\documentclass{article} 
\usepackage{iclr2027_conference,times}

\usepackage{amsmath,amsfonts,bm}

\def\eqref#1{equation~\ref{#1}}
\def\Eqref#1{Equation~\ref{#1}}
\def\plaineqref#1{\ref{#1}}

\def\1{\bm{1}}

\DeclareMathAlphabet{\mathsfit}{\encodingdefault}{\sfdefault}{m}{sl}
\SetMathAlphabet{\mathsfit}{bold}{\encodingdefault}{\sfdefault}{bx}{n}

\DeclareMathOperator*{\argmin}{arg\,min}

\usepackage{hyperref}
\usepackage{url}
\usepackage{comment}
\usepackage{amsmath, amssymb}
\usepackage{pifont}
\usepackage{graphicx}
\usepackage{booktabs}
\usepackage{wrapfig}
\usepackage{xcolor}
\usepackage{multirow}
\usepackage{wrapfig}
\usepackage{subcaption}

\newcommand\scalemath[2]{\scalebox{#1}{\mbox{\ensuremath{\displaystyle #2}}}}
\newcommand{\cmark}{\ding{51}}
\newcommand{\xmark}{\ding{55}}

\title{Learn the Dynamics, Enforce the Physics: Generative Latent Operators for High-Dimensional Beam Evolution}

\author{
Haohan Zou$^{1}$ \quad
Alexander Scheinker$^{2}$ \quad
Yuanyuan Shi$^{1}$ \\
$^{1}$University of California San Diego 
$^{2}$Los Alamos National Laboratory \\
\texttt{\{hazou, yyshi\}@ucsd.edu} \quad
\texttt{ascheink@lanl.gov}
}

\iclrfinalcopy 
\begin{document}

\maketitle

\begin{abstract}
    Particle-in-cell (PIC) simulation is indispensable for modeling beam dynamics in particle accelerators, but its computational cost limits many-query workflows, such as configuration exploration, virtual diagnostics, and controller design. We introduce \emph{GLOBE}, a physics-constrained \textbf{G}enerative \textbf{L}atent \textbf{O}perator for \textbf{B}eam \textbf{E}volution, modeling the nonlinear, control-dependent, multiscale evolution of 3D beam--field dynamics covering $10^7$ spatial cells. GLOBE compresses beam--field states into a compact latent space and factors their evolution into two components. A control-conditioned neural operator learns the dominant dynamics, while an amplitude-normalized generative model captures the residual distribution, recovering fine-scale structures and conditional variations. GLOBE enforces field constraints by regime-specific, parameter-free operators: a Poisson solve reconstructs the electric field in the electrostatic regime, and a spectral projection imposes Gauss's law and a divergence-free magnetic field in the electromagnetic regime. On two high-resolution 3D particle-accelerator benchmarks, bunched-beam transport and plasma-wakefield acceleration, GLOBE lowers relative-$L_2$ error by 9\% and 13\%, low-frequency spectral error by 69\% and 51\%, and mid-frequency spectral error by 54\% and 20\% relative to the strongest baseline (DPOT), satisfies field constraints, and reduces hours of PIC simulation to a minute per rollout.
\end{abstract}

\section{Introduction}
Particle accelerators are foundational instruments across high-energy physics, materials science, and medical therapy. Their beam dynamics are governed by the Vlasov--Maxwell equations, where charged particles evolve collectively with their self-generated electromagnetic fields, producing nonlinear phenomena such as emittance growth, instabilities, and particle loss. As direct experimentation on the physical machine is constrained, high-fidelity simulation is indispensable to accelerator design and operation. The particle-in-cell (PIC) method~\citep{birdsall2018plasma, hockney2021computer} is the standard tool, advancing millions of macroparticles together with their electromagnetic fields resolved on a spatial mesh. At high resolution and over long horizons, however, each simulation costs hours of GPU time, prohibitive for many-query workflows such as configuration exploration, virtual diagnostics, and controller design~\citep{emma2018machine, edelen2020machine,arpaia2021machine}. This computational burden has motivated efficient data-driven surrogates for particle accelerators. Existing works map machine settings to low-dimensional beam quantities~\citep{edelen2016neural,edelen2020machine}, reconstruct phase-space projections from indirect measurements~\citep{roussel2023phase, scheinker2024cdvae, scheinker2024conditional,  scheinker2025physics, scheinker2026phaseflow4D}, or forecast beam evolution across successive linac modules~\citep{rautela2024conditional}. Yet efficient long-horizon prediction for the 3D beam--field evolution on the simulation mesh remains largely unexplored.

Advances in scientific machine learning have established two prominent paradigms for PDE modeling. Neural operators learn function-to-function mappings for PDE solution operators and achieve strong performance across various physical systems~\citep{lu2021learning, li2021fourier, cao2021choose, li2023transformer, hao2023gnot, li2023scalable, hao2024dpot,wu2024transolver, rahman2024pretraining, wang2025cvit}. But they exhibit spectral bias, yielding over-smoothed predictions with unresolved fine-scale structures~\citep{rahaman2019on, xu2025understanding}. Generative models offer a complementary capability in recovering fine-scale structures and representing predictive variability, with successes in partially observed PDEs, turbulent-flow modeling, and probabilistic weather forecasting~\citep{shysheya2024conditional,huang2024diffusionpde,lienen2024from,du2024conditional,li2024generative,price2025probabilistic}. But evolving the forecast generatively is challenging, as the model must jointly learn temporal dynamics and conditional variations. These complementary strengths have motivated hybrid surrogates that combine deterministic forecast with generative refinement, either to resolve spectral content~\citep{oommen2024integrating, khodakarami2025mitigating} or to model correction distributions around deterministic predictions~\citep{doi:10.1126/sciadv.adx2372,mardani2025residual}. However, these hybrid designs scale poorly to high-resolution 3D systems with heterogeneous field amplitudes.

\begin{figure*}[t!]
    \centering
    \includegraphics[width=0.91\textwidth]{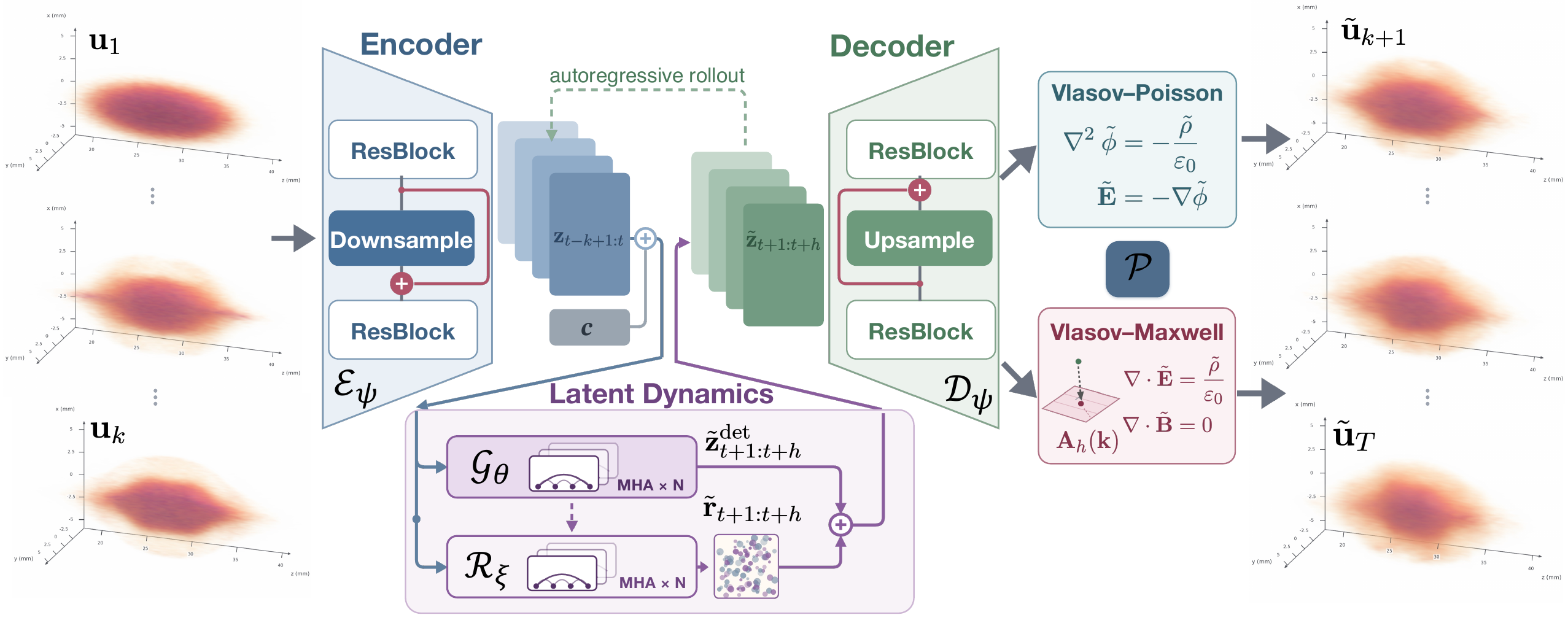}
    \caption{\small \textbf{Framework Overview:} The encoder $\mathcal{E}_\psi$ compresses initial beam--field states $\mathbf{u}_{1:k}$ into latent $\mathbf{z}_{1: k}$. At each rollout step $t$, the latest $k$ latents $\mathbf{z}_{t-k+1: t}$ and machine operation controls $\mathbf{c}$ are given to the latent dynamics models. The neural operator $\mathcal{G}_\theta$ predicts the dominant evolution $\tilde{\mathbf{z}}^{\text{det}}_{t+1:t+h}$ and the flow-matching model $\mathcal{R}_\xi$ samples the residual $\tilde{\mathbf{r}}_{t+1:t+h}$. Their sum $\tilde{\mathbf{z}}_{t+1:t+h}$ is appended to the history and the prediction advances. Finally, $\mathcal{D}_\psi$ reconstructs the mesh-resolved states and $\mathcal{P}$ enforces regime-specific field constraints.} 
    \label{fig:framework} 
\end{figure*}

Motivated by these gaps, we propose \emph{GLOBE}, a physics-constrained \textbf{G}enerative \textbf{L}atent \textbf{O}perator for \textbf{B}eam \textbf{E}volution, extending hybrid PDE modeling to high-resolution 3D beam--field evolution over $10^7$ spatial cells. For tractability, GLOBE compresses full-resolution beam--field states into a compact latent space. In this latent space, a control-conditioned neural operator captures the dominant dynamics, while an amplitude-normalized conditional generative model learns the residual distribution to recover fine-scale structures and sampling-induced conditional variability. Regime-specific, parameter-free operators enforce field constraints by construction, separating dynamics learning from constraint satisfaction. Our main contributions are summarized as follows:
\begin{itemize}
    \item We design GLOBE, to our knowledge, the first hybrid surrogate for 3D beam--field evolution accommodating $10^7$ spatial cells, factoring the latent forecast into a control-conditioned deterministic operator and an amplitude-normalized generative residual. 
    
    \item We introduce regime-specific, parameter-free operators that enforce field constraints by construction: a Poisson solve for electric-field reconstruction in the electrostatic regime and a spectral Maxwell projection in the electromagnetic regime.
    
    \item We release two high-resolution 3D particle-accelerator benchmarks. On both, GLOBE attains lower relative-$L_2$ error by 9\% and 13\%, low-frequency spectral error by 69\% and 51\%, and mid-frequency spectral error by 54\% and 20\% relative to the strongest baseline (DPOT), satisfies field constraints, and reduces hours of simulation to a minute per rollout.

\end{itemize}

\section{Literature Review}
\subsection{Deep Learning for PDE Modeling}

Neural operators learn mappings between function spaces, enabling one model to generalize across families of initial conditions, boundary conditions, and system parameters. Representative architectures include DeepONet~\citep{lu2021learning}, Fourier Neural Operator~\citep{li2021fourier, kovachki2021universal, guibas2022efficient, hao2024dpot}, and attention-based operators~\citep{cao2021choose,li2023transformer, hao2023gnot, li2023scalable, wu2024transolver, rahman2024pretraining,mccabe2024multiple, wang2025cvit,holzschuh2026pd}. To reduce cost in high dimensions, latent operator methods evolve dynamics in compressed latent spaces~\citep{wu2022learning,wang2024latent,alkin2024universal,serrano2024aroma}. However, neural operators exhibit spectral bias, yielding over-smoothed predictions missing fine-scale features and sharp gradients~\citep{rahaman2019on, xu2025understanding}. 

In parallel, generative models learn distributions over PDE solutions and dynamical states. Diffusion-, score-, and flow-based models have enabled conditional generation, inverse reconstruction, and spatiotemporal prediction of PDE systems~\citep{shysheya2024conditional,huang2024diffusionpde,li2025videopde}, with applications to turbulent-flow modeling and probabilistic weather forecasting~\citep{lienen2024from,du2024conditional, li2024generative, andrae2025continuous, price2025probabilistic, kohl2026benchmarking}. However, modeling the complete evolution and its conditional structure generatively is challenging, and repeated denoising or numerical integration at each forecast increases rollout cost. 

To combine neural operators' modeling capability with generative models' probabilistic representation, recent studies explore hybrid deterministic--generative formulations.~\citet{oommen2024integrating, khodakarami2025mitigating} couple neural operators with conditional diffusion to recover high-frequency structures missed by deterministic forecasts, mitigating spectral bias in turbulent and multiscale systems.~\citet{doi:10.1126/sciadv.adx2372, mardani2025residual} learn generative corrections around deterministic predictions for high-resolution downscaling and probabilistic weather forecasting. These studies demonstrate the effectiveness of hybrid decomposition across turbulence and weather applications, but its extension to high-resolution 3D systems remains challenging: the cost of state-space model scales with the grid size, and generative training is biased by heterogeneous field amplitudes.

\subsection{Learning-based Methods for Particle Accelerators}

Machine learning is emerging as a promising approach in accelerator physics, helping address the challenges of high-dimensional configuration spaces and expensive simulations~\citep{edelen2018opportunities,arpaia2021machine}. Early methods learn mappings from accelerator configurations to beam-quality metrics for rapid modeling and design~\citep{edelen2016neural, edelen2020machine}, and pursue data-driven tuning and control via adaptive feedback~\citep{scheinker2018demonstration,scheinker2019model,leemann2019demonstration}, Bayesian optimization~\citep{duris2020bayesian,roussel2024bayesian}, and reinforcement learning~\citep{kain2020sample,kaiser2024reinforcement}. Recent works reconstruct the beam's phase-space distribution from indirect diagnostics. Approaches include differentiable beam-dynamics simulations fit to measurements~\citep{roussel2023phase, roussel2024efficient}, adaptive latent-space tuning that tracks time-varying distributions~\citep{scheinker2021adaptive6D, scheinker2025physics}, and conditional generative models that map machine signals to phase-space projections~\citep{scheinker2024cdvae, scheinker2024conditional, scheinker2026phaseflow4D}. Beyond reconstruction, \citet{rautela2024conditional} forecast downstream phase-space projections across successive linac modules with a latent autoregressive generative model. Despite this progress, efficient long-horizon prediction of 3D mesh-resolved, self-consistent beam--field states remains underexplored.

\section{Preliminary: Beam Dynamics}
\label{sec:preliminary}

\textbf{Vlasov--Maxwell Equations.}
The collective beam dynamics in particle accelerators are governed by the relativistic Vlasov--Maxwell equations~\citep{glassey1996cauchy,davidson2001physics}. For particles of charge $q$ and mass $m$, let $f(\mathbf r,\mathbf p,t)$ be the phase-space distribution of beam population at time $t$, position $\mathbf r=(x,y,z)\in\mathbb R^3$, and momentum $\mathbf p = \gamma m\mathbf v \in\mathbb R^3$, where velocity $\mathbf{v}=\dot{\mathbf{r}}$, relativistic factor $\gamma=\sqrt{1+|\mathbf p|^2/(m^2c^2)}$, and speed of light $c$. It evolves according to the Vlasov equation,
\vspace{-2pt}
\begin{equation}
\label{eq:vlasov-maxwell}
\frac{\partial f}{\partial t} +
\mathbf{v}\cdot\nabla_{\mathbf{r}}f + \mathbf{F}\cdot\nabla_{\mathbf{p}}f = 0, \qquad
\mathbf{F} = \frac{\mathrm{d}\mathbf{p}}{\mathrm{d}t} = q\left( \mathbf{E} + \mathbf{v}\times\mathbf{B} \right),
\vspace{-2pt}
\end{equation}
where $\mathbf{F}$ is the Lorentz force, $\mathbf{E}$ and $\mathbf{B}$ are the electric and magnetic fields. These fields consist of externally applied fields and self-fields generated by the beam: $\mathbf{E} = \mathbf E_{\mathrm{ext}}+\mathbf E_{\mathrm{beam}}$ and $\mathbf B=\mathbf B_{\mathrm{ext}}+\mathbf B_{\mathrm{beam}}$. External fields are generated by accelerator components such as magnets and RF cavities, while self-fields are generated by beam charge density $\rho$ and current density $\mathbf{J}$, given by
\vspace{-2pt}
\begin{equation*}
    \rho(\mathbf{r},t) =  q\int_{\mathbb{R}^{3}} f(\mathbf{r},\mathbf{p},t) \,\mathrm{d}\mathbf{p}, \quad
    \mathbf{J}(\mathbf{r},t) =  q\int_{\mathbb{R}^{3}} \mathbf{v}(\mathbf{p}) f(\mathbf{r}, \mathbf{p},t)\, \mathrm{d}\mathbf{p}.
\vspace{-2pt}
\end{equation*}
The beam self-fields satisfy Maxwell's equations: 
\vspace{-2pt}
\begin{equation}
\label{eq: maxwell equations}
    \scalemath{0.94}{\nabla_{\mathbf{r}}\cdot\mathbf{E}_{\mathrm{beam}} = \frac{\rho}{\varepsilon_0},
    \nabla_{\mathbf{r}}\cdot\mathbf{B}_{\mathrm{beam}} = 0,
    \frac{\partial\mathbf{B}_{\mathrm{beam}}}{\partial t}= -\nabla_{\mathbf{r}}\times\mathbf{E}_{\mathrm{beam}},
    \frac{\partial\mathbf{E}_{\mathrm{beam}}}{\partial t} = c^{2}\,\nabla_{\mathbf{r}}\times\mathbf{B}_{\mathrm{beam}} - \frac{\mathbf{J}}{\varepsilon_0}},
\end{equation}
where $\varepsilon_0$ is the vacuum permittivity. Equations~\plaineqref{eq:vlasov-maxwell}--\plaineqref{eq: maxwell equations} form a nonlinear self-consistent system: $f$ determines $(\rho,\mathbf J)$ and the beam self-fields, which in turn govern $f$ through the Lorentz force $\mathbf{F}$.

\textbf{From Vlasov--Maxwell to Vlasov--Poisson.}
In the low-energy regime, the magnetic and inductive effects associated with the beam self-fields are negligible. Therefore we take the time-varying beam magnetic field $\mathbf{B}_{\mathrm{beam}} \approx 0$. By Faraday's law, this gives $\nabla_\mathbf{r} \times \mathbf{E}_{\mathrm{beam}} \approx 0$, so the beam electric field is approximately curl-free and can be expressed as $\mathbf E_{\mathrm{beam}}=-\nabla_r \phi$, where  $\phi(\mathbf{r}, t)$ is the electrostatic scalar potential. The Vlasov--Maxwell equation then reduces to the Vlasov--Poisson equation, 
\vspace{-2pt}
\begin{equation}
\label{eq:vlasov-poisson}
    \frac{\partial f}{\partial t}
    + \mathbf{v}\cdot\nabla_{\mathbf{r}}f
    + q\left(\mathbf{E}_{\mathrm{ext}} + \mathbf{v}\times\mathbf{B}_{\mathrm{ext}} - \nabla_{\mathbf{r}}\phi\right)\cdot\nabla_{\mathbf{p}}f = 0,
    \quad
    -\nabla_{\mathbf{r}}^{2}\phi = \frac{\rho}{\varepsilon_0} = \frac{q}{\varepsilon_0}\int_{\mathbb{R}^{3}} f\,\mathrm{d}\mathbf{p}.
\vspace{-2pt}
\end{equation}
Unlike the full electromagnetic case, the electrostatic beam field has no independent temporal evolution. Given the instantaneous charge density $\rho$ and corresponding boundary conditions, $\phi$ is determined by the Poisson equation, and $\mathbf{E}_{\mathrm{beam}}$ follows directly from $-\nabla_\mathbf{r}\phi$~\citep{davidson2001physics}. 

\section{Proposed Framework}
This section introduces GLOBE, our physics-constrained Generative Latent Operator for Beam Evolution. As shown in Figure~\ref{fig:framework}, GLOBE has three modules: (i) a compressive autoencoder $(\mathcal{E}_\psi, \mathcal{D}_\psi)$ learning a compact representation of beam--field states, (ii) a latent dynamics model combining a deterministic neural operator $\mathcal{G}_\theta$ with a flow-matching generative residual $\mathcal{R}_\xi$, and (iii) regime-specific, parameter-free operators $\mathcal{P}$ enforcing instantaneous field constraints. We detail each module below. 

\textbf{Problem Setup.}
Let $\mathbf{u}_t \in \mathbb{R}^{C \times Z \times Y \times X}$ be a beam--field state at time $t$ on a $Z\times Y\times X$ grid with $C$ variables. Given the observed history $\mathbf{u}_{1:k}$ and machine operation controls $\mathbf{c}$, GLOBE models the conditional distribution of future states $p\!\left(\mathbf{u}_{k+1:T}\mid\mathbf{u}_{1:k},\mathbf c\right)$, rolled out autoregressively in blocks of $h$ steps. We denote $\hat{\mathbf{u}}(\mathbf{k}) =  \int_{\mathbb{R}^3} \mathbf{u}(\mathbf{x}) e^{-i\mathbf{k}\cdot \mathbf{x}} \mathrm{d} \mathbf{x}$ as the Fourier transform of $\mathbf{u}$ at frequency $\mathbf{k} \in \mathbb{R}^3$.

\subsection{Compressive Autoencoder}
\label{sec:autoencoder}

\begin{figure*}[h]
    \centering
    \includegraphics[width=0.87\textwidth]{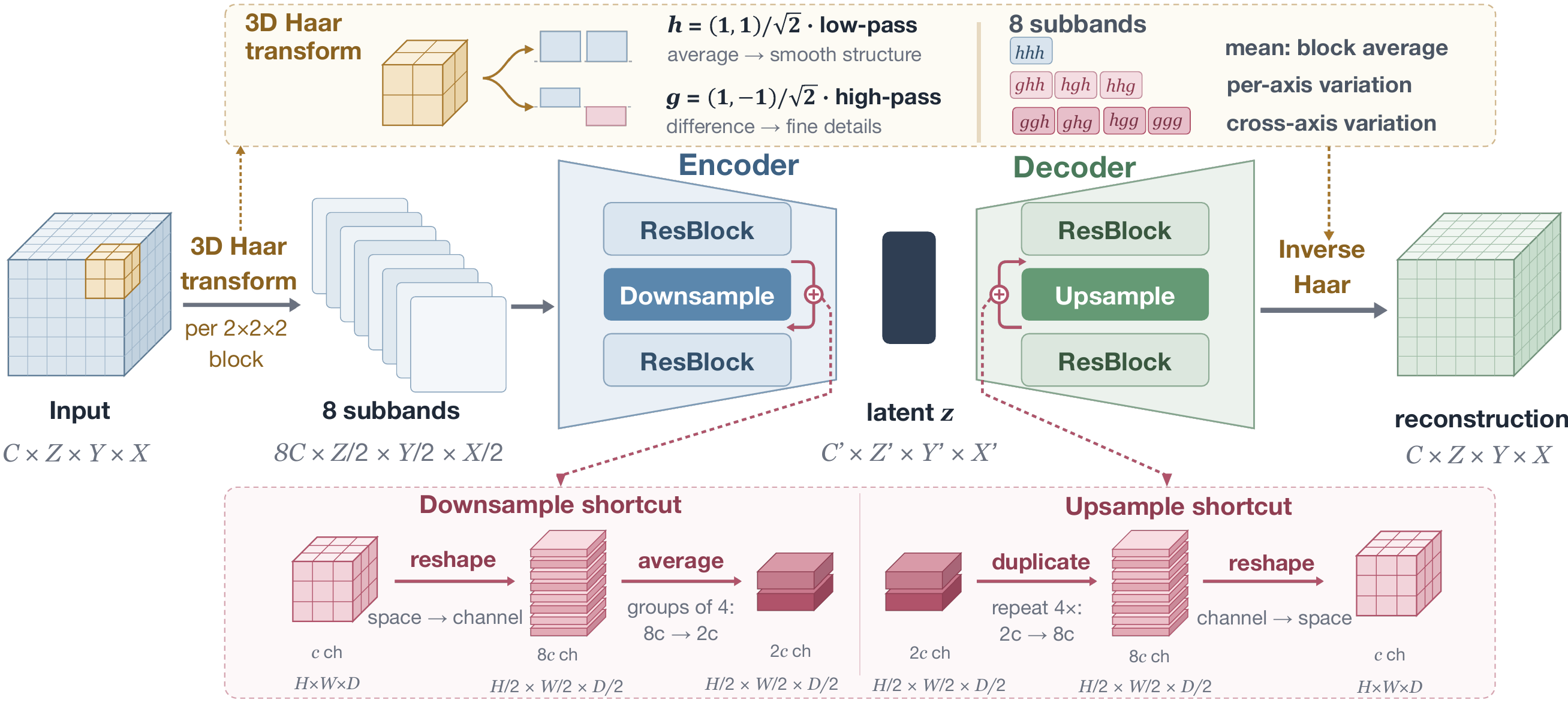}
    \caption{\small \textbf{Compressive Autoencoder}: For each 3D input, every $2\times2\times2$ block is split by the low-/high-pass pair $(\mathbf{h}, \mathbf{g})$ into $8$ wavelet subbands, then the encoder $\mathcal{E}_\psi$ compresses it to a grid latent $\mathbf{z}$. The decoder $\mathcal{D}_\psi$ mirrors the hierarchy, closing with the inverse Haar transform. Around every down-/up-sampling module, a parameter-free space--channel shortcut ($\oplus$) provides references across resolution by channel averaging and duplication with reshaping, so the corresponding convolutional branch fits only a residual correction. }
    \label{fig:AE}
\end{figure*}

GLOBE maps the high-dimensional 3D accelerator beam--field states into a compact latent representation, in which subsequent dynamics are modeled. As illustrated in Figure~\ref{fig:AE}, we employ a convolutional autoencoder $(\mathcal{E}_\psi, \mathcal{D}_\psi)$ to learn this representation. The encoder compresses each beam--field state $\mathbf{u}_t$ into latent $\mathbf z_t = \mathcal E_\psi(\mathbf u_t) \in \mathbb{R}^{C' \times Z' \times Y' \times X'}$ at reduced resolution, from which the decoder reconstructs $\tilde{\mathbf{u}}_t = \mathcal{D}_\psi(\mathbf{z}_t)$. The compression substantially reduces the number of spatial tokens from $10^7$ to $\sim 10^4 - 10^5$ while retaining a structured 3D grid for latent dynamics modeling. 

\vspace{4pt}

\textbf{Autoencoder Backbone.} 
The encoder composes a sequence of residual blocks and down-sampling modules, while the decoder mirrors this structure with up-sampling modules and residual blocks, without encoder–decoder skip connections. Each residual block operates at a fixed resolution for progressive feature refinement, whereas the down- and up-sampling modules perform the transitions between resolution scales. To preserve fine-scale structures and stabilize training under strong compression, we augment the backbone with two parameter-free designs.

\textbf{Haar Transform and Multiscale Resolution Shortcuts.}  
First, the 3D Haar transform decomposes each non-overlapping $2\times2\times2$ spatial block into eight wavelet subbands: one local average, three per-axis variations, and four cross-axis variations. It is applied before the encoder and inverted after the decoder. By separating local variations from smooth background before learned compression, the transform helps preserve fine beam structures~\citep{dai2026physics}. Second, each learned down- and up-sampling module is paired with a space--channel shortcut that maps features between resolutions~\citep{chen2025deep}. During down-sampling, the shortcut reshapes spatial blocks into channels and adjusts the channel width through group averaging. During up-sampling, channel features are duplicated and rearranged back into the spatial dimensions. This multiscale path provides a reference across resolutions, allowing the learned resolution-changing modules to focus only on the correction needed for accurate reconstruction, which eases optimization at high compression ratios.

\textbf{Noise-robust Reconstruction Objective.}
The autoencoder is trained to accurately reconstruct the beam and field quantities while preserving the spatial structures with three complementary losses:
\vspace{-2pt}
\begin{equation}
\label{eq:rec}
\mathcal{L}_{\text{rec}}(\mathbf{u}_t,\tilde{\mathbf{u}}_t)
= \underbrace{\ell(\mathbf{u}_t,\tilde{\mathbf{u}}_t)}_{\mathcal{L}_{\text{rel}}}
+ \lambda_{\text{fd}}\underbrace{\textstyle\sum_{j=1}^{3}\ell\big(\partial_{x_j}\mathbf{u}_t,\partial_{x_j}\tilde{\mathbf{u}}_t\big)}_{\mathcal{L}_{\text{fd}}}
+ \lambda_{\text{spec}}\underbrace{\ell\big(\mathbf{w}\odot\hat{\mathbf{u}}_t,\,\mathbf{w}\odot\hat{\tilde{\mathbf{u}}}_t\big)}_{\mathcal{L}_{\text{spec}}},
\vspace{-2pt}
\end{equation}
where $\ell(\mathbf{a},\mathbf{b})=\frac{\|\mathbf{a}-\mathbf{b}\|_2}{\max(\|\mathbf{a}\|_2,\epsilon)}$ is the relative-$L_2$ error with $\epsilon\in \mathbb{R}^{+}$ for numerical stability, $\odot$ is element-wise multiplication, and $\lambda_{\text{spec}}, \lambda_{\text{fd}}\in \mathbb{R}^{+}$ are loss weights. $\mathcal{L}_\text{rel}$ measures the overall relative error between the original and reconstructed states. $\mathcal{L}_{\text{fd}}$ preserves local gradient consistency, retaining sharp spatial variations such as beam edges that are under-weighted by relative reconstruction loss alone. $\mathcal{L}_{\text{spec}}$ promotes consistency of spectral content with $\mathbf{w} = (w(\mathbf{k}))_{\mathbf{k}}$, a vector of spectral weights over all Fourier modes that grow with $|\mathbf{k}|$ and additionally emphasize the mid-frequency band. It encourages coherent mid-frequency structure while reducing the dominance of low-frequency energy and suppressing high-frequency noise. To make the decoder robust to imperfect predicted latent states during inference, we apply $\mathcal{L}_{\text{rec}}$ to both the original latent $\mathbf{z}_t$ and the noise-perturbed latent. Together, we formulate the noise-robust reconstruction objective as:
\vspace{-2pt}
\begin{equation*}
    \mathcal{L}_{\text{AE}}(\mathbf{u}_t, \tilde{\mathbf{u}}_t) = \mathcal{L}_{\text{rec}} \left(\mathbf{u}_t, \tilde{\mathbf{u}}_t\right) + \lambda_{\text{noise}}\mathcal{L}_{\text{rec}}\Big(\mathbf{u}_t, \mathcal{D}_\psi\big(\mathbf{z}_t +\sigma \mathrm{std}(\mathbf{z}_t)\boldsymbol{\epsilon}\big)\Big),\:
    \sigma \sim \mathcal{U}(0, \sigma_{\max}),\:
    \boldsymbol{\epsilon} \sim \mathcal{N}(\mathbf{0}, \mathbf{I}).
\vspace{-2pt}
\end{equation*}
This objective maintains reconstruction fidelity while improving decoder's robustness to latent error.

\subsection{Latent Dynamics: Deterministic Anchor and Generative Residual}
\label{sec:latent_dynamics}

Deterministic neural operators can model dynamical evolution accurately, but tend to attenuate weaker fine-scale structures and average over realization-dependent variations. Fully generative dynamics, in contrast, burden the generative model with learning evolution that is largely predictable. GLOBE therefore uses a deterministic forecast as the anchor and pairs it with a conditional generative residual that refines unresolved fine-scale structures and captures sampling-induced variations.

\textbf{Deterministic Latent Neural Operator. }
Given the latest $k$-step latent histories and prescribed machine operation controls $\mathbf{c}$, the deterministic operator $\mathcal{G}_\theta$ predicts the next $h$-step latent states, trained against the encoded ground truth with a relative-$L_2$ objective,
\vspace{-2pt}
\begin{equation*}
    \tilde{\mathbf{z}}_{t+1:t+h}^{\text{det}}= \mathcal{G}_{\theta}(
        \mathbf z_{t-k+1:t},\mathbf{c}), \quad
    \mathcal{L}_{\mathcal{G}_\theta} = \frac{\|\mathbf{z}_{t+1:t+h} - \tilde{\mathbf{z}}_{t+1:t+h}\|_2}{\max(\|\mathbf{z}_{t+1:t+h}\|_2, \epsilon)}, \quad \epsilon \in \mathbb{R}^+.
\vspace{-2pt}
\end{equation*}
The prediction $\tilde{\mathbf{z}}^{\text{det}}$ captures the dominant temporal evolution, and serves as a dynamical anchor providing the reference around which the residual distribution is modeled. Since autoregressive rollout feeds the operator its imperfect predictions rather than the exact latent histories seen in training, we use two techniques to limit error accumulation. First, Gaussian perturbations are added to the input latent histories during training, exposing the operator to small deviations from the encoded trajectory~\citep{sanchez2020learning}. In addition, we employ pushforward training, feeding the operator its own prediction for a subsequent supervised forecast~\citep{brandstetter2022message}.

\textbf{Generative Latent Residual.}
We define the discrepancy around the deterministic anchor as $\mathbf{r}_{t+1:t+h} = \mathbf{z}_{t+1:t+h} - \tilde{\mathbf{z}}_{t+1:t+h}^{\text{det}}$ and model it with a generative model $\mathcal{R}_\xi$. Because its magnitude varies substantially across physical variables, forecast stages, and spatial locations, modeling the raw residual would place disproportionate weight on large-amplitude components. We therefore normalize it before training by a conditional amplitude factorized at three granularities,
\begin{equation*}
    \mathbf{s}_\omega(\mathbf{p}) = \boldsymbol{\sigma}_r \odot s_{\mathrm{win}}(\mathbf{p}) \odot s_{\mathrm{tok}}\left(\mathbf{p}\right), \quad \bar{\mathbf{r}}_{t+1:t+h} = \mathbf{r}_{t+1:t+h} \oslash \mathbf{s}_\omega(\mathbf{p})\,, \quad \mathbf{p} = (\mathbf{z}_{t-k+1:t}, \tilde{\mathbf{z}}_{t+1:t+h}^{\text{det}}),
\end{equation*}
where $\odot$ and $\oslash$ denote element-wise multiplication and division. $\boldsymbol{\sigma}_r \in \mathbb{R}^{1 \times C' \times 1 \times 1 \times 1}$ is the per-channel root-mean-square (RMS) of the residual over the training set, $s_{\mathrm{win}} > 0$ adjusts the amplitude of each forecast block, and $s_{\mathrm{tok}} \in \mathbb{R}^{1 \times 1 \times Z' \times Y' \times X'}$ captures spatial amplitude variation shared across channels and steps. The latter two normalizers are predicted sequentially by small networks from the conditioning context $\mathbf{p}$. Each network minimizes the zero-mean negative Gaussian log-likelihood $\frac{m}{s(\mathbf{p})^2} + 2 \log \big(s(\mathbf{p})\big)$, where $m$ is the mean-square residual after the preceding normalization. For $s_{\mathrm{win}}$, $m$ is averaged over the full forecast block after channel normalization, and for $s_{\mathrm{tok}}$, $m$ is averaged over steps and channels at each spatial token after channel and window normalization. The conditional optimum satisfies $s(\mathbf p)^2=\mathbb E[m\mid\mathbf p]$, so each network predicts the conditional RMS at its granularity, yielding an approximately unit-RMS normalized residual $\bar{\mathbf r}$.

We learn the normalized residual distribution with conditional flow matching~\citep{liu2023flow,lipman2023flow}. Given Gaussian noise $\boldsymbol{\epsilon}\sim\mathcal N(\mathbf0,\mathbf I)$ and a logit-normal flow time~\citep{esser2024scaling} $\tau=\operatorname{sigmoid}(\eta), \;\eta\sim\mathcal N(0,1)$, we construct the linear interpolant  
$\mathbf{x}_\tau = (1-\tau) \boldsymbol{\epsilon} + \tau \bar{\mathbf{r}}$ and train the conditional velocity field $\mathbf{v}_\xi$ to regress its constant velocity $(\bar{\mathbf{r}} -\boldsymbol{\epsilon})$,
\begin{equation*}
    \mathcal{L}_{\mathcal{R}_\xi} = \mathbb{E}_{\tau, \boldsymbol{\epsilon}, \bar{\mathbf{r}}} \left[\|\mathbf{v}_\xi (\mathbf{x}_\tau; \mathbf{p}, \mathbf{c}, \tau) - (\bar{\mathbf{r}} -\mathbf{\boldsymbol{\epsilon}})\|_2^2 \right]+ \lambda_{\text{ES}} \mathcal{L}_{\text{ES}}, \quad \lambda_{\text{ES}} \in \mathbb{R}^{+},
\end{equation*}
where $\mathcal{L}_{\text{ES}}$ is an energy-score regularizer~\citep{Pacchiardi2024Probabilistic} that keeps generated samples close to target while preserving ensemble dispersion. At inference, $\mathcal R_\xi$ maps Gaussian noise into a normalized residual by integrating $\mathbf{v}_\xi$ with Euler's method. Rescaling  by $\mathbf s_\omega (\mathbf{p})$, the predicted residual $\tilde{\mathbf{r}}_{t+1:t+h}$ is added to the deterministic forecast, yielding the final latent prediction $\tilde{\mathbf{z}}_{t+1:t+h}$.

\textbf{Backbone Architecture.}
Transformer~\citep{vaswani2017attention} with self-attention over spatial latent tokens is used for the deterministic neural operator $\mathcal{G}_\theta$ and generative residual model $\mathcal{R}_\xi$. As self-attention is permutation-equivariant, sinusoidal Fourier features of the normalized 3D coordinates are used for positional embeddings~\citep{tancik2020fourier}. Machine operation controls $\mathbf{c}$, together with flow time $\tau$ for $\mathcal{R}_\xi$, are injected into each block via adaLN-Zero~\citep{Peebles_2023_ICCV}.

\subsection{Physics-constrained Operators}
\label{sec:physics_constraint} 
To preserve the instantaneous field constraints, GLOBE applies a parameter-free, regime-specific operator $\mathcal{P}$ after decoding, enforcing instantaneous field constraint satisfaction by construction.

\textbf{Vlasov--Maxwell Constraint Enforcement.}
In the electromagnetic regime, the charge density and electromagnetic fields carry independent dynamical information and are therefore predicted jointly. By \Eqref{eq: maxwell equations}, the decoded prediction $\tilde{\mathbf{u}} = (\tilde{\rho}, \tilde{\mathbf{E}}, \tilde{\mathbf{B}})$ should satisfy
\vspace{-5pt}
\begin{equation}
\label{eq:maxwell-constraint}
\text{Gauss's Law: }\nabla\cdot\mathbf E - \frac{\rho}{\varepsilon_0} = 0,
\qquad
\text{Divergence-free magnetic field: }\nabla\cdot\mathbf B =0.
\vspace{-5pt}
\end{equation}
These constraints are linear differential relations. Under the Fourier transform, spatial differentiation becomes multiplication by the corresponding frequency component, $\partial_{x_j} \to \ ik_j$. Constraints in \Eqref{eq:maxwell-constraint} therefore become, independently at each Fourier mode $\mathbf{k}$,
\vspace{-3pt}
\begin{equation*}
i\mathbf k^{\mathsf T}\hat{\mathbf E}(\mathbf k)
-\frac{\hat{\rho}(\mathbf k)}{\varepsilon_0}=0,
\qquad
i\mathbf k^{\mathsf T}\hat{\mathbf B} (\mathbf k)=0.
\vspace{-3pt}
\end{equation*}
As a result, the admissible state values form a linear subspace that can be written as $\mathbf {A}(\mathbf k)\hat{\mathbf u}(\mathbf k)=\mathbf 0$. 
We define $\mathcal{P}_\mathbf{k}$ as a weighted equality-constraint optimization solution operator that projects the decoded prediction $\hat{\tilde{\mathbf{u}}} (\mathbf{k})$ to its closest constraint-satisfying coefficient $\hat{\mathbf{v}}^\star(\mathbf{k})$:
\vspace{-4pt}
\begin{equation*}
    \scalemath{0.94}{\hat{\mathbf{v}}^\star(\mathbf{k}) :=
    \mathcal{P}_{\mathbf{k}} \big(\hat{\tilde{\mathbf{u}}}(\mathbf{k})\big) = 
    \argmin_{\mathbf{A}(\mathbf{k}) \hat{\mathbf{v}}(\mathbf{k}) = 0} \|\hat{\mathbf{v}}(\mathbf{k}) - \hat{\tilde{\mathbf{u}}}(\mathbf{k})\|_{\mathbf{W}^{-1}} = \Big[\mathbf{I} - \mathbf{W} \mathbf{A}(\mathbf{k})^{\mathsf H} \big(\mathbf{A}(\mathbf{k}) \mathbf{W} \mathbf{A}(\mathbf{k})^{\mathsf{H}}\big)^{\dagger}\mathbf{A}(\mathbf{k})\Big] \hat{\tilde{\mathbf{u}}}(\mathbf{k})},
\vspace{-4pt}
\end{equation*}
where $\mathbf W\succ0$ accounts for value-scale differences across variables. $(\cdot)^{\mathsf{H}}$ and $(\cdot)^\dagger$ denote the Hermitian transpose and pseudo-inverse. Applying the projection to each Fourier mode and transforming the corrected coefficients back to the physical space yields a state $\mathbf{v}^\star$ satisfying the instantaneous Maxwell constraints by construction. In implementation, derivatives are evaluated as finite differences, acting in Fourier space as multiplication by a modified frequency, e.g. $i \sin(k_j h)/h$ for a central difference of spacing $h$. Thus we build $\mathbf{A}_h(\mathbf{k})$ from these multipliers so the projected state satisfies the constraints in simulation's own discretization. Details are in Appendix~\ref{app:spectral_operator}.

\textbf{Vlasov--Poisson Field Reconstruction.}
As established in Section~\ref{sec:preliminary}, in the electrostatic regime, the beam field is uniquely determined by the charge density and the boundary condition:
\begin{equation*}
    \text{Poisson Equation: } -\nabla_{\mathbf r}^2\phi=\frac{\rho}{\varepsilon_0}, \qquad \mathbf{E} = -\nabla_{\mathbf{r}} \phi.
\end{equation*}
Therefore, GLOBE predicts only the charge density $\tilde{\rho}$, solves the Poisson equation for the predicted scalar potential $\tilde{\phi}$ with an iterative Krylov solver (Jacobi-preconditioned BiCGStab~\citep{doi:10.1137/0913035}), and obtains the predicted beam electric field $\tilde{\mathbf{E}}$ by applying the negative discrete gradient to $\tilde{\phi}$. This enforces the Poisson equation, Gauss's law, and the curl-free condition by construction.

\section{Experiments}

We evaluate GLOBE on two high-resolution 3D particle-accelerator benchmarks generated with WarpX~\citep{WarpX}, one per regime in Section~\ref{sec:preliminary}. (1)~\emph{Bunched-beam transport (electrostatic regime)}: a $750\, \mathrm{keV}$ proton bunch propagates through a periodic quadrupole--RF focusing channel. The quadrupole and RF strengths are tuned with bunch charge to maintain beam matching, as mismatch drives envelope oscillations, emittance growth, and halo formation that degrade downstream beam quality. The benchmark varies focusing strength, bunch charge, and initial mismatch across 464 simulations (52 for testing), each with $T=401$ charge-density frames on a $256\times336\times336$ grid over $80$ focusing periods, requiring $\sim6600$s GPU per trajectory. (2)~\emph{Plasma-wakefield acceleration (PWFA; electromagnetic regime)}: an ultra-relativistic electron driver excites a nonlinear plasma wake with multi-GV/m gradients, accelerating a trailing witness bunch. Driver charge, energy, and size, and witness charge and position, determine the energy gain and witness beam quality. The benchmark varies these parameters across 504 simulations (56 for testing), each with $T=60$ frames over ten variables, four species-resolved charge densities and ($\mathbf{E}$, $\mathbf{B}$), on a $336\times216\times216$ grid over $19.6 \,\mathrm{mm}$ of propagation, requiring $\sim4130$s GPU per trajectory. Details are in Appendix~\ref{app:dataset}.

\textbf{Baselines:} 
We compare against four representative baselines at matched model size: DPOT~\citep{hao2024dpot}, UPT~\citep{alkin2024universal}, MPP~\citep{mccabe2024multiple}, and P3D~\citep{holzschuh2026pd}. For P3D, both deterministic (P3D-D) and generative formulation (P3D-G) are compared with. We report three GLOBE variants sharing the same autoencoder: GLOBE-D, the deterministic operator $\mathcal{G}_\theta$ alone, GLOBE-C, the full model with both deterministic operator $\mathcal{G}_\theta$ and generative residual $\mathcal{R}_\xi$, and GLOBE-G, an ablation that forecasts the latent directly with a single flow-matching model.

\textbf{Evaluation Metrics:} 
We evaluate predictive quality with relative-$L_2$ error (Rel-$L_2$) and spectral error $L_{\mathcal{F}}$, averaged over rollout steps and test trajectories. Following~\citet{khodakarami2025mitigating}, we split spectrum into low ($|\mathbf{k}|\leq0.34k_{\mathrm{Nyq}}$), mid ($0.34k_{\mathrm{Nyq}}<|\mathbf{k}|\leq0.49k_{\mathrm{Nyq}}$), and high ($|\mathbf{k}|>0.49k_{\mathrm{Nyq}}$) frequency bands, with $k_{\mathrm{Nyq}}$ the per-axis Nyquist frequency. $L_{\mathcal F}$ is the relative error between the radially averaged power spectra of prediction and reference in each band. For generative models we report the mean error over 8 samples. Inference time per trajectory is  also reported.

\subsection{Quantitative Comparison with Baselines}

\begin{table*}[b]
\vspace{2pt}
\centering
\captionsetup{skip=2pt}
\caption{\small Quantitative comparison of GLOBE and baselines on bunched-beam transport and PWFA.}
\label{tab:comparison}
\resizebox{\textwidth}{!}{
\begin{tabular}{lcccccccc}
\toprule
& \multicolumn{4}{c}{\textbf{Bunched-Beam Transport}}
& \multicolumn{4}{c}{\textbf{PWFA}} \\
\cmidrule(lr){2-5}\cmidrule(lr){6-9}
\textbf{Method} & Params & Rel-$L_2$ $\downarrow$ & $L_{\mathcal{F}}$ (low / mid / high) $\downarrow$ & Time $\downarrow$
& Params & Rel-$L_2$ $\downarrow$ & $L_{\mathcal{F}}$ (low / mid / high) $\downarrow$ & Time $\downarrow$ \\
\midrule
DPOT  & 91.4M & 0.177 & 0.112 / 0.986 / \textbf{0.954} & \textbf{7.55}s & 109.4M & 0.224 & 0.156 / 0.377 / 0.634 & \textbf{3.04}s \\
UPT   & 92.6M & 0.261 & 0.147 / 0.999 / 1.000 & 1384s & 107.7M & 0.378 & 0.251 / 0.640 / 2.585 & 114.5s \\
MPP   & 90.0M    & 0.347    & 0.384 / 0.991 / 0.978         & 943s    & 110.9M     & 0.325    &0.443 / 1.279 / 3.212       & 73.5s \\
P3D-D & 91.4M & 0.329 & 0.197 / 0.961 / 2.160 & 91.3s & 110.9M & 0.520 & 0.314 / 0.789 / 19.3  & 17.9s \\
P3D-G & 91.4M & 0.490 & 0.437 / \textbf{0.323} / 1.115 & 1490s & 111.0M & 1.219 & 0.794 / 151 / 6002 & 309s \\
\midrule
GLOBE-D (Ours) & 63.6M & \textbf{0.144} & 0.039 / 0.994 / 1.003 & 16.3s & 82.4M & \textbf{0.174} & 0.083 / 0.350 / \textbf{0.592} & 5.4s \\
GLOBE-G (Ours) & 63.9M & 0.251 & 0.153 / 0.545 / 0.967 & 73.1s & 83.1M & 1.954 & 11.3 / 225 / 909 & 23.6s \\
GLOBE-C (Ours) & 91.4M & 0.161 & \textbf{0.035} / 0.457 / 0.973 & 64.7s & 111.8M & 0.194 & \textbf{0.077} / \textbf{0.300} / 0.618 & 21.8s \\
\bottomrule
\end{tabular}
}
\end{table*}

Table~\ref{tab:comparison} compares the proposed GLOBE models with four representative baselines on both benchmarks. The deterministic baselines include DPOT, UPT, MPP, and P3D-D. Within this group, GLOBE-D achieves the lowest Rel-$L_2$ error, reaching $0.144$ on bunched-beam transport and $0.174$ on PWFA, while using fewer parameters than the baselines. We attribute this improvement to full attention in the compressed latent space, which enables global spatial interactions that would be intractable in the raw 3D space. However, like other deterministic models, GLOBE-D has much larger errors at mid-frequency than at low-frequency. Since higher frequencies correspond to smaller spatial scales, this indicates reduced accuracy in capturing finer-scale spatial structure. 

Adding the generative residual model yields GLOBE-C, which preserves the overall accuracy of GLOBE-D while better capturing finer-scale beam structures. Notably, it reduces mid-frequency error from $0.994$ to $0.457$ on bunched-beam transport and from $0.350$ to $0.300$ on PWFA. In contrast, directly modeling the full dynamics via generative models performs poorly. Both P3D-G and GLOBE-G exhibit much larger Rel-$L_2$ and spectral errors, especially on PWFA. These results support our staged design, where the deterministic model captures the dominant beam--field evolution and the generative model predicts the residual to recover unresolved fine-scale structures.

GLOBE is efficient at inference, as modeling occurs in the compressed latent space. Among deterministic baselines, GLOBE-D uses only $16.3$s / $5.4$s per rollout. UPT ($1384$s / $114.5$s) incurs query-wise cross-attention during decoding, MPP ($943$s / $73.5$s) attends over spatio-temporal tokens of the full history, and P3D-D ($91.3$s / $17.9$\,s) repeatedly processes full 3D states with convolutional and windowed-attention blocks. DPOT remains fastest ($7.55$s / $3.04$s), operating on patchified states with lightweight input/output mappings, whereas GLOBE decodes latent forecasts via 3D convolutional blocks. GLOBE-C avoids full-state generative rollout by integrating flows in latent space and decoding only the final forecasts, achieving $64.7$s / $21.8$s versus P3D-G $1490$ / $309$s, whose iterative generation operates on full 3D states. Relative to WarpX, GLOBE-C gains $102\times$ / $189\times$ speedups.

\begin{figure*}[t]
    \centering
    \includegraphics[width=0.9\textwidth]{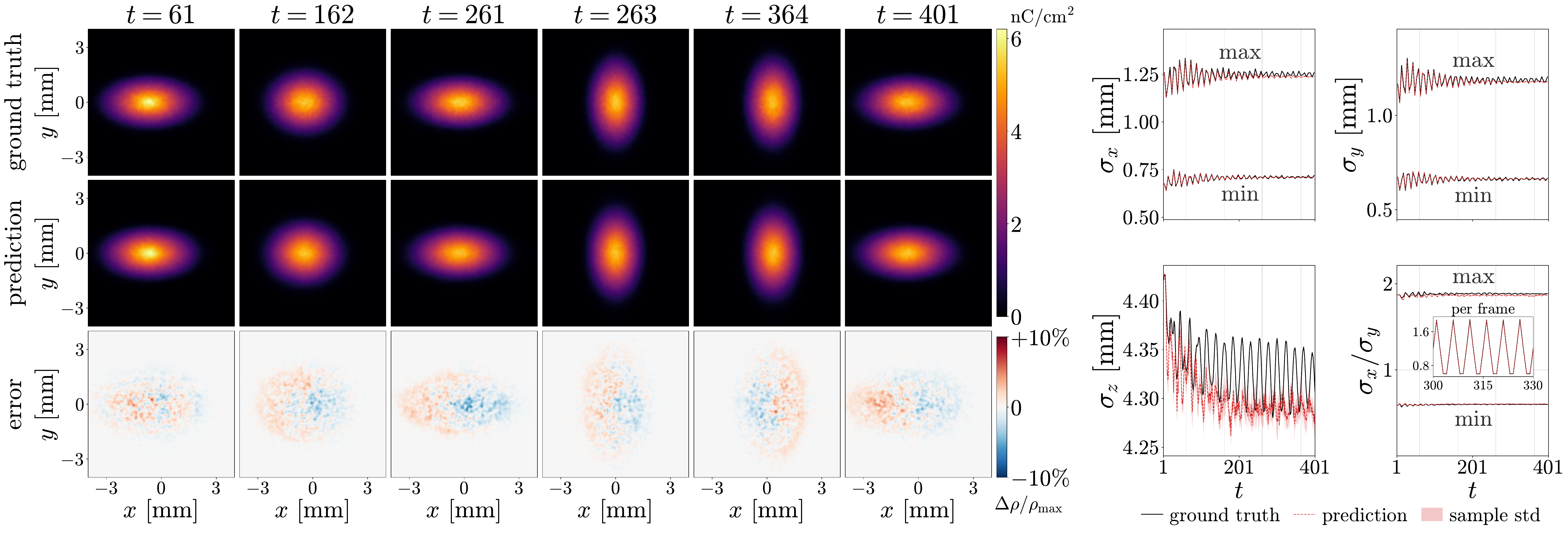}
    \caption{\footnotesize \textbf{Bunched-beam transport.} Left: Transverse charge projection $\int \rho  \,\mathrm{d}z$, top: ground truth, bottom: one prediction sample from GLOBE-C. Right: Evolution of $\sigma_x$, $\sigma_y$, $\sigma_z$, and $\sigma_x/\sigma_y$. Due to rapid oscillation, the main curves of transverse quantities show period-wise extrema and the insets show frame-wise evolution. Prediction curve is the mean of eight GLOBE-C samples, and shading indicates the sample standard deviation.}
    \label{fig:bunched_main}
\end{figure*}

\subsection{Long-Horizon Prediction and Visualization}
We visualize long-horizon predictions on both benchmarks.
Figure~\ref{fig:bunched_main} shows a held-out test trajectory for bunched-beam transport, conditioned on the initial eight frames and operation controls, followed by a 393-frame prediction with forecast window $h = 8$. In the left panel, we observe that GLOBE-C closely reproduces the time-varying transverse beam shape, the gradual change in beam size, and the halo effect, i.e., the diffuse particle population extending beyond the dense beam core, throughout the trajectory. The right panel further quantifies the beam-envelope evolution through the transverse beam widths $\sigma_x$ and $\sigma_y$, longitudinal bunch size $\sigma_z$, and transverse aspect ratio $\sigma_x/\sigma_y$. All four predicted quantities closely track the ground truth without noticeable error accumulation.

\begin{figure*}[h!]
    \centering
    \includegraphics[width=0.9\textwidth]{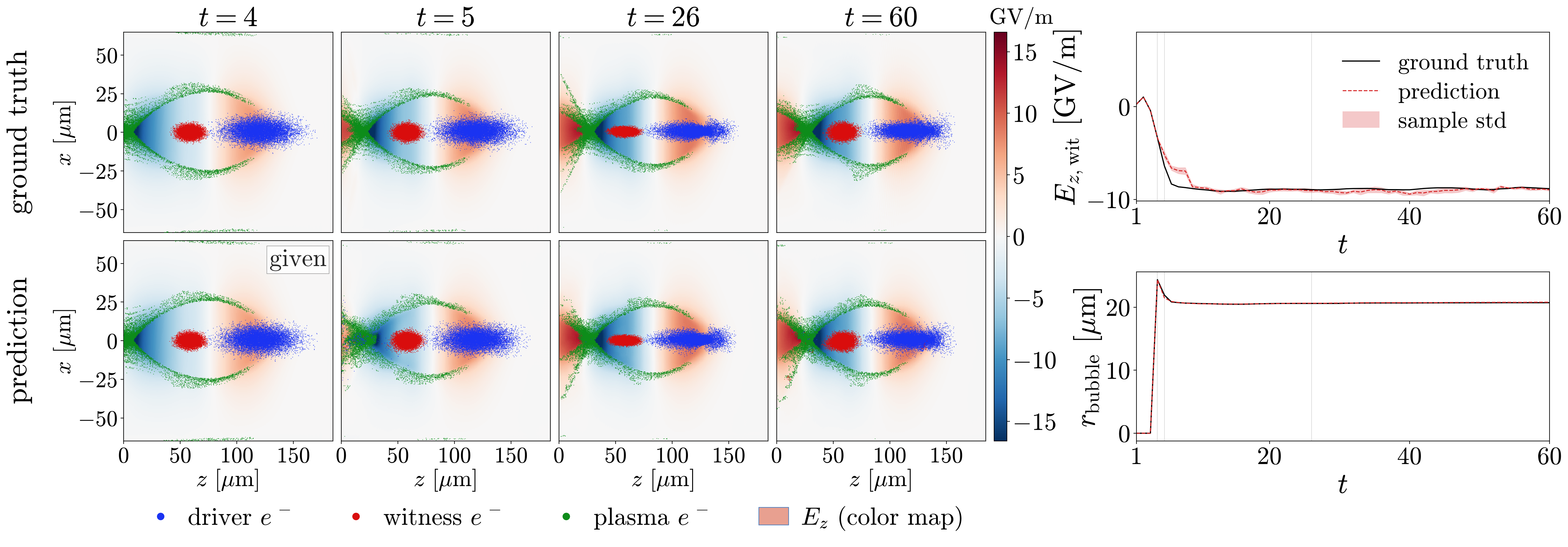}
    \caption{\footnotesize \textbf{PWFA.} Left: longitudinal electric field $E_z$ in the $x$--$z$ mid-plane, overlaid with pseudo-particles of driver, witness, and plasma electrons sampled from charge density fields of WarpX (top) and GLOBE-C (bottom). First column is the last input frame; the rest show the forecast. Right: evolution of $E_{z, \text{wit}}$ and $r_{\text{bubble}}$.}
    \label{fig:pwfa_main}
\end{figure*}
Figure~\ref{fig:pwfa_main} shows a held-out PWFA test trajectory, using the first four frames as input and predicting the subsequent 56 frames with forecast window $h = 4$. In the left panels, the driver bunch (blue) expels plasma electrons (green), creating an ion cavity enclosed by the curved green electron sheath and a longitudinal wakefield $E_z$. The witness bunch (red) is located where $E_z$ provides an accelerating force, increasing its energy as it propagates. GLOBE-C closely reproduces the cavity shape, surrounding electron sheath, and witness-bunch evolution in the ground truth. The right panel quantifies this evolution by the accelerating field at the witness centroid $E_{z,\mathrm{wit}}$ and the maximum cavity radius $r_{\mathrm{bubble}}$. Both predictions closely track the ground truth throughout the rollout.

\subsection{Ablation Studies}

\begin{wraptable}{r}{0.48\textwidth}
\vspace{-3pt}
\centering
\captionsetup{skip=2pt}
\caption{\footnotesize Ablation of two designs in our autoencoder: Haar transform (Haar) and multiscale resolution shortcut (Shortcut). The last row combines both designs and corresponds to the proposed autoencoder.}
\label{table:ab_ae}

\resizebox{\linewidth}{!}{
\begin{tabular}{@{}cccccc@{}}
\toprule
Haar & Shortcut & $\mathcal L_{\mathrm{rel}}\!\downarrow$
& $\mathcal L_{\mathrm{spec}}\!\downarrow$
& $\mathcal L_{\mathrm{fd}}\!\downarrow$
& Runtime (s) $\downarrow$ \\
\midrule
\xmark & \xmark & 0.0891 & 0.1070 & 0.697 & 6.04 \\
\cmark & \xmark & 0.0547 & 0.0740 & 0.566 & \textbf{0.97} \\
\cmark & \cmark & \textbf{0.0457} & \textbf{0.0685} & \textbf{0.489} & 1.00 \\
\bottomrule
\end{tabular}
}
\vspace{-6pt}
\end{wraptable}
\textbf{Compressive Autoencoder Design Ablation.}
Table~\ref{table:ab_ae} isolates the benefits of compressive autoencoder's two designs in Section \ref{sec:autoencoder}: the Haar transform and multiscale resolution shortcut. The Haar transform maps each $2 \times 2 \times 2$ block into eight subbands, reducing the spatial grid by $8\times$ while separating local variations from the smooth background. This reduces training runtime by $6.2\times$ and lowers the three losses in \Eqref{eq:rec} by $39\%$, $31\%$, and $19\%$. Adding the shortcut further reduces losses by $16\%$, $7\%$, and $14\%$ with nearly unchanged runtime. These results show that the Haar transform provides the main efficiency gain, while the shortcut provides complementary accuracy gains via fixed multiscale references for the learned convolutional branch.

\begin{wraptable}{r}{0.49\textwidth}
\vspace{-4pt}
\centering
\captionsetup{skip=2pt}
\caption{\footnotesize Ablation of the physics-constrained operator on PWFA. Constraint residuals are normalized RMS values averaged over frames and trajectories.}
\label{table:ab_constraint}
\resizebox{\linewidth}{!}{
\begin{tabular}{@{}lccc@{}}
\toprule
& Rel-$L_2$ $\downarrow$
& $\nabla\!\cdot\!\mathbf E-\rho/\varepsilon_0$ $\downarrow$
& $\nabla\!\cdot\!\mathbf B$ $\downarrow$ \\
\midrule
w/o $\mathcal P$
& \textbf{0.191}
& $2.72\times10^{-2}$
& $6.85\times10^{-3}$ \\
w/ $\mathcal P$
& 0.194
& $\mathbf{2.77\times10^{-7}}$
& $\mathbf{2.78\times10^{-7}}$ \\
\bottomrule
\end{tabular}
}
\vspace{0pt}
\end{wraptable}
\textbf{Physics-constrained Operator Ablation.} Table~\ref{table:ab_constraint} evaluates the parameter-free physics-constrained decoding in Section~\ref{sec:physics_constraint} on the PWFA dataset. We compare GLOBE-C without the physics-constrained operator (denoted as w/o $\mathcal{P}$) against GLOBE-C with $\mathcal{P}$ applied to enforce the field constraints (denoted as w/ $\mathcal{P}$). Applying $\mathcal P$ maintains comparable prediction accuracy (Rel-$L_2$ $0.191$ vs.\ $0.194$) while reducing violations of Gauss's law and divergence-free magnetic-field constraint from $10^{-2}$--$10^{-3}$ to $10^{-7}$, improving physical consistency by four to five orders of magnitude with negligible accuracy loss.

\begin{wraptable}{r}{0.55\textwidth}
\vspace{-4pt}
\centering
\captionsetup{skip=2pt}
\caption{\footnotesize Ablation of latent Transformer backbone vs. HFS.}
\label{tab:backbone}

\resizebox{\linewidth}{!}{
\begin{tabular}{@{}lcccc@{}}
\toprule
& \multicolumn{2}{c}{\textbf{Bunched-Beam Transport}}
& \multicolumn{2}{c}{\textbf{PWFA}} \\
\cmidrule(lr){2-3}\cmidrule(lr){4-5}
Method
& Rel-$L_2$ $\downarrow$ & $L_{\mathcal F}$ $\downarrow$
& Rel-$L_2$ $\downarrow$ & $L_{\mathcal F}$ $\downarrow$ \\
\midrule
HFS
& 0.163
& 0.090 / 0.995 / 1.004
& 0.835
& 2.04 / 6.44 / 258 \\

GLOBE-D (Ours)
& \textbf{0.144}
& \textbf{0.039} / \textbf{0.994} / \textbf{1.003}
& \textbf{0.174}
& \textbf{0.083} / \textbf{0.350} / \textbf{0.592} \\
\midrule

HFS+DM
& 0.173
& 0.088 / 0.682 / 1.034
& 3.258
& 46.8 / 1301 / 2446 \\

GLOBE-C (Ours)
& \textbf{0.161}
& \textbf{0.035} / \textbf{0.457} / \textbf{0.973}
& \textbf{0.194}
& \textbf{0.077} / \textbf{0.300} / \textbf{0.618} \\
\bottomrule
\end{tabular}
}
\vspace{-6pt}
\end{wraptable}
\textbf{Latent Backbone Ablation.}
Table~\ref{tab:backbone} compares GLOBE's global self-attention latent Transformer in Section \ref{sec:latent_dynamics} against HFS~\citep{khodakarami2025mitigating}, a U-Net based neural operator with diffusion refinement. As the original HFS does not scale to our high-resolution 3D grids, we place it in the same learned latent space as GLOBE with a matched model size. HFS is competitive with GLOBE-D on the single-channel bunched-beam transport (Rel-$L_2$ $0.163$ vs. $0.144$), and diffusion refinement (HFS + DM) reduces its mid-frequency error from $0.995$ to $0.682$. On the multi-channel PWFA task, however, HFS degrades substantially (Rel-$L_2$ $0.835$ vs.\ $0.174$), with particularly high spectral prediction errors, and HFS+DM further diverges. The larger gap in PWFA highlights the benefit of global attention for strongly coupled multi-channel dynamics.  Overall, the results support GLOBE's hybrid design: a robust deterministic operator first captures the dominant dynamics, leaving the generative model to refine the remaining variability.

\vspace{-4pt}
\section{Conclusion}
\vspace{-4pt}
This work introduced GLOBE, a hybrid learning framework for high-resolution 3D beam dynamics in particle accelerators. GLOBE combines a compressed latent representation, a control-conditioned Transformer for dominant dynamics, a generative residual model for fine-scale refinement, and parameter-free operators imposing regime-specific field constraints. Across two particle accelerator benchmarks, GLOBE achieves the best prediction accuracy among the evaluated baselines, satisfies instantaneous field constraints, and reduces hours of PIC simulation to about one minute per rollout. Future work will extend GLOBE toward a foundation model for particle accelerator simulation across machines, control settings, and initial/boundary conditions. An alternative direction is to leverage the learned dynamics for beam-state estimation and control, enabling accurate virtual diagnostics and real-time decision-making across diverse accelerator applications.

\subsection*{AI use statement}

In this work, we used generative AI tools to assist with coding and implementation of methods designed by the authors, and to polish parts of the manuscript for readability. We did not use generative AI tools to generate synthetic datasets, develop theoretical models or conceptual frameworks, formulate mathematical claims, propose or refine hypotheses, design or provide feedback on research methodology or experiments, clean or reformat datasets, or interpret results. Providing ingredients for or writing proofs, assisting with translation, and supporting qualitative or thematic data analysis are not applicable to this work. AI-assisted implementations were checked and validated by the authors, and all AI-assisted writing was reviewed for accuracy against the underlying methods and results. The authors take full responsibility for the final content of this work, including text, claims or artifacts produced with the aid of generative AI.

\subsubsection*{Acknowledgments}
H. Zou and Y. Shi gratefully acknowledge support from the Michael R. Anastasio LANL–UC Early Career Faculty Fellowship, DOE ASCR Award DE-SC0025495, the Schmidt Sciences AI2050 Early Career Fellowship, and an NVIDIA Academic Grant. A. Scheinker gratefully acknowledges support from the Directed Research project 20270020DR "Adaptive Physics-Constrained Generative Diffusion for Time-Varying Dynamic Systems" funded by Los Alamos National Laboratory which is operated by Triad National Security, LLC, for the National Nuclear Security Administration of the U.S. Department of Energy under Contract No.~89233218CNA000001.

\bibliography{iclr2027_conference}
\bibliographystyle{iclr2027_conference}

\newpage
\appendix

\section*{Contents of Supplementary Materials}

\subsection*{Dataset:}
\begin{itemize}
    \item Appendix \ref{app:dataset_1}: Dataset I: Space-charge-dominated Bunched-beam Transport
    \item Appendix \ref{app:dataset_2}: Dataset II: Plasma-Wakefield Acceleration (PWFA)
\end{itemize}

\subsection*{Framework Details:}
\begin{itemize}
    \item Appendix \ref{app:autoencoder}: Compressive Autoencoder
    \item Appendix \ref{app:deterministic_latent}: Deterministic Latent Neural Operator
    \item Appendix \ref{app:Generative_Latent_Residual}: Generative Latent Residual
    \item Appendix \ref{app:spectral_operator}: Physics-Constrained Operators 
\end{itemize}

\subsection*{Baseline Descriptions:}
\begin{itemize}
    \item Appendix \ref{app:baseline_dpot}: Denoising Operator Transformer (DPOT)
    \item Appendix \ref{app:baseline_upt}: Universal Physics Transformer (UPT)
    \item Appendix \ref{app:baseline_mpp}: Multiple Physics Pretraining (MPP)
    \item Appendix \ref{app:baseline_p3d}: P3D
\end{itemize}

\subsection*{Supplementary Experiment Results:}
\begin{itemize}
    \item Appendix \ref{app:amplitude-normalization-ablation}: Amplitude-normalization Ablation
    \item Appendix \ref{app:bunched-beam-visualization}: Space-charge-dominated Bunched-beam Transport
    \item Appendix \ref{app:PWFA-visualization}: Plasma Wakefield Acceleration
\end{itemize}

\newpage
\textbf{Computation Resources}: All experiments in this work were performed on a server with Nvidia RTX Pro 6000 Blackwell Server Edition 96 GB GPU.

\section{Dataset Generation}
\label{app:dataset}

We construct two complementary 3D particle accelerator benchmarks generated with WarpX~\citep{WarpX}. The first one models space-charge-dominated bunched-beam transport in the electrostatic Vlasov-Poisson regime, while the second models plasma-wakefield acceleration (PWFA) in the fully electromagnetic Vlasov-Maxwell regime. Together, these datasets span two distinct accelerator settings that require different physical scales, simulation geometries, control variables, state representations, and field treatments. 

\subsection{Space-Charge-Dominated Bunched-Beam Transport}
\label{app:dataset_1}

The first benchmark is inspired by the low-energy, high-current proton transport conditions at the entrance to the first drift-tube-linac (DTL) tank of the Los Alamos Neutron Science Center (LANSCE). The reference beam energy of $750\, \mathrm{keV}$, RF frequency $201.25\, \mathrm{MHz}$, and $7.5\, \mathrm{mm}$ bore radius follow LANSCE Tank-1 entrance settings.  The benchmark is not intended as a cell-by-cell digital twin of the LANSCE lattice. Instead, it retains the essential ingredients of intense linac injection, strong space charge, alternating-gradient transverse focusing, longitudinal RF confinement, and controlled mismatch and centroid perturbations, that drive envelope oscillation, emittance growth, halo formation, and particle loss.

At this energy level, the beam self-field is well described in the electrostatic Vlasov--Poisson regime. We simulate a proton bunch in an idealized periodic focusing channel. In the beam co-moving frame, the longitudinal domain spans one RF wavelength, $L_z \approx 59.5 \,\mathrm{mm}$, while the transverse computational box is $16 \,\mathrm{mm} \times 16 \, \mathrm{mm}$. A grounded cylindrical conductor of radius $7.5 \,\mathrm{mm}$ models the accelerator aperture. The longitudinal domain is taken to be periodic, and the applied lattice combines alternating-gradient quadrupole focusing with a periodic RF bucket. The state for model learning is the beam charge density only, from which the electrostatic self-field can be reconstructed directly. The applied external fields are prescribed by machine configuration.

\paragraph{Design Parameters.}
Each trajectory corresponds to a different beam and accelerator configuration. We sample five physically interpretable dimensionless beam-dynamics quantities and control parameters:
\begin{align*}
    \sigma_{0t} \in [60^\circ, 85^\circ], \quad
    \frac{\sigma_{0l}}{\sigma_{0t}} \in [0.5, 1.2], \quad
    \eta_t \in [0.4, 0.9], \quad
    M \in [1.0, 1.4], \quad
    \frac{x_{\mathrm{off}}}{a} \in [0,0.15]. 
\end{align*}
Here, $\sigma_{0t}$ denotes the zero-current transverse phase advance per focusing period, $\frac{\sigma_{0l}}{\sigma_{0t}}$ controls the relative longitudinal focusing strength, and $\eta_t$ is the transverse tune depression and determines the strength of the space-charge interaction. The remaining two describe the initial beam state: $M$ controls the initial transverse mismatch, and $\frac{x_{\mathrm{off}}}{a}$ specifies the normalized horizontal beam offset. The synchronous RF phase is fixed at $\phi_s = -30^\circ$.

For each sampled design, an optics-inversion procedure maps the dimensionless design variables to the physical input parameters required by WarpX, including the quadrupole strength $E_{Q0}$, RF-field amplitude $E_{0T}$, total bunch charge $q_{\mathrm{tot}}$, and matched beam dimensions. These quantities play two different roles in the PIC simulation. The accelerator settings $(E_{Q0}, E_{0T}, q_{\mathrm{tot}})$ are deployment-known machine controls and are provided explicitly to the surrogate as control parameters $\mathbf{c}$. In contrast, the mismatch $M$ and centroid offset $x_{\mathrm{off}}$ define the initial beam condition and are not supplied as additional control variables. Feasibility and stability checks are applied before simulation so that the retained configurations represent physically meaningful transport conditions rather than failed numerical or optical setups.

\paragraph{Dataset Construction.}
The main design space is sampled using maximin latin-hypercube points. We supplement these samples with nine representative anchor configurations and two additional high-phase-advance cases with $\sigma_{0t} = 87^\circ, 89^\circ$, which probe dynamics near the envelope-instability regime. PIC simulations contain stochastic fluctuations arising from finite-particle sampling. To quantify this variability, we select $16$ feasible physical configurations distributed throughout the design space and simulate each using four independent seeds. The two high-phase-advance cases are also evaluated using four independent seeds. All replicate simulations corresponding to the same physical configuration are assigned to the same data split to prevent leakage. In total, the dataset contains $420$ distinct physically feasible configurations with $474$ simulation runs. After simulations, we removed ten simulations where the beam loss is significant ($> 20\%$ of injected charge scraped on the bore wall), resulting in a dataset of $464$ simulations used in our experiments. The training, validation, and test sets contain $364$, $48$, and $52$ simulations, respectively.

\paragraph{Numerical Configuration and Data Storage.}
We simulate the trajectory using approximately $8 \times 10^5$ proton macroparticles on a Cartesian grid of size $256 \times 336\;\times 336 \; (Z \times Y \times X)$. One alternating-gradient focusing period is resolved using $50$ PIC time steps. Each simulation is evolved for $4000$ steps, corresponding to $80$ focusing periods. Mesh states and beam diagnostics are stored every $10$ steps, resulting in $T = 401$ frames per trajectory. For the standard trajectories, only charge density $\rho$ is stored in the compressed dataset. For a subset of audit trajectories, we additionally retain scalar potential $\phi$ and electric fields $\mathbf{E}$, which provides independent field-level ground truth for evaluating the Poisson-based field reconstruction. Physically meaningful beam diagnostics that can be derived directly from the predicted charge density, including the beam centroid, rms beam size, transmitted charge, and aperture loss, are also recorded for each simulation. These quantities can be calculated analytically from $\rho$, allowing the learned surrogate to be evaluated not only through pointwise field errors but also through accelerator-relevant physical observables. The total computational cost of this dataset is about $900$ GPU hours.

\subsection{Vlasov-Maxwell Dataset: Plasma-Wakefield Acceleration (PWFA)}
\label{app:dataset_2}

The second benchmark models three-dimensional beam-driven plasma-wakefield acceleration (PWFA), an advanced accelerator concept in which the accelerating structure is created dynamically by a plasma rather than conventional RF cavity. An ultra-relativistic electron driver bunch is injected into a preformed plasma and displaces the plasma electrons, exciting a co-moving electromagnetic wake. A trailing witness bunch is placed at an accelerating phase of this wake and gains energy from the longitudinal wakefield. Thus, the accelerating field is generated by the collective plasma response to the injected driver bunch, not by an externally prescribed accelerating cavity. Because the driver, witness, plasma electrons, and electromagnetic fields evolve self-consistently, this benchmark exercises the full Vlasov--Maxwell regime. 

The simulation contains four particle species: an electron driver bunch, an electron witness bunch, plasma electrons, and stationary hydrogen ions. The driver and witness are represented by approximately $10^6$ and $2.5 \times 10^5$ macroparticles, respectively. The background plasma uses two electron macroparticles per cell and one ion macroparticle per cell, with nominal density $n_0 = 10^{23} \mathrm{m}^{-3} = 10^{17} \mathrm{cm}^{-3}$. At the plasma entrance, the density rises smoothly from zero to $n_0$ over a $2\,\mathrm{mm}$ up-ramp and is then held constant.

\paragraph{Design Parameters.}
The PWFA campaign is constructed by sampling six dimensionless physical parameters that span distinct wakefield and beam-loading regimes:
\begin{align*}
    &\Lambda \in [0.05, 1.2], \quad \gamma_d \in [1500, 6000], \quad k_p\sigma_z \in [0.30, 1.19], \\ &M \in [0.5, 2.5], \quad \frac{x_{\mathrm{off}}}{\sigma_x} \in [0,0.5], \quad \eta_{\mathrm{load}} \in [0.2, 1.5].
\end{align*}
Here, $\Lambda$ controls the normalized driver strength and spans weak excitation through strongly nonlinear wake formation. $\gamma_d$ sets the driver energy. $k_p\sigma_z$ determines the bunch duration relative to plasma scale. $M$ controls the transverse driver size relative to its matched value. $\frac{x_{\mathrm{off}}}{\sigma_x}$ introduces transverse symmetry breaking, and $\eta_{\mathrm{load}}$ controls witness beam loading. 

For each sampled design, these dimensionless quantities are converted into the physical parameters required to initialize the WarpX simulation containing
\begin{equation*}
    (Q_d,\,
    \gamma_d,\,
    \sigma_{z,d},\,
    \sigma_{r,d},\,
    x_{d,\mathrm{off}},\,
    Q_w,\,
    \Delta z_{dw},\,
    I_w).
\end{equation*}
$Q_d$ is the driver charge, $\gamma_d$ is the driver Lorentz factor, $\sigma_{z,d}$ and $\sigma_{r,d}$ are its longitudinal and transverse rms sizes, $x_{d, \mathrm{off}}$ is its transverse centroid offset, $Q_w$ is the witness charge, $\Delta {z}_{dw}$ is the driver--witness separation, and $I_w$ indicates whether a witness bunch is present. These physical parameters are provided to the surrogate as control parameters $\mathbf{c}$.

\paragraph{Dataset Construction.}
The main design space is sampled using maximin latin-hypercube points together with nine representative anchor configurations. To probe additional physically important effects, we supplement the main sweep with several controlled subsets. First, 24 \emph{imperfection twins} are created from selected baseline simulations. These simulations preserve the underlying particle realization while introducing controlled imperfections such as head--tail tilt, transverse emittance asymmetry, correlated energy chirp, and witness offset. Such paired simulations isolate the influence of physical perturbations from ordinary PIC sampling noise. Second, $40$ \emph{driver-only twins} are generated by removing the witness bunch from selected configurations while retaining the same driver and plasma setup. These pairs enable controlled comparison of loaded and unloaded plasma wakes. Finally, six hand-selected configurations probe more extreme wake and depletion regimes and are reserved for evaluation. Similar to the Vlasov-Poisson dataset, we additionally select $16$ representative configurations and repeat them using four independent particle seeds to estimate the stochastic variability introduced by finite-particle PIC sampling. In total, the PWFA dataset contains $456$ distinct physically feasible configurations with $504$ simulations. The training set, validation set, and test sets contain $399$, $49$, and $56$ simulations, respectively

\paragraph{Numerical Configuration and Data Storage.}
We simulate the trajectory using a Cartesian grid of size $336 \times 216 \times 216 \;(Z \times Y \times X)$, covering a transverse region of approximately $129.6 \,\mu \mathrm{m} \times 129.6 \,\mu \mathrm{m}$ and a longitudinal moving window of approximately $184.8 \,\mu \mathrm{m}$. The corresponding spatial resolutions are approximately $\Delta x = \Delta y = 0.6 \,\mu \mathrm{m}, \Delta z = 0.55 \,\mu \mathrm{m}$. Each simulation evolves $36000$ PIC steps, corresponding to approximately $19.6 \,\mathrm{mm}$ of plasma propagation. Three-dimensional fields are stored every $600$ time steps, producing $61$ diagnostic frames per trajectory. The initial frame contains a short numerical initialization transient and is retained for auditing but excluded from model training; consequently, frames $1 - 60$ form the learning trajectory. Each trajectory stores ten three-dimensional channels:
\begin{equation*}
    \left(\rho_{\mathrm{driver}},\rho_{\mathrm{witness}}, \rho_{\mathrm{plasma}\text{-}e}, \rho_{\mathrm{plasma}\text{-}i}\right), \quad 
    \left(E_x, E_y, E_z\right), \quad
    \left(B_x, B_y, B_z\right).
\end{equation*}
The species-resolved charge-density channels describe the evolution of the particle populations, while the six electromagnetic-field components characterize the plasma wake. The total computational cost of this dataset is about $600$ GPU hours.

\newpage
\section{Framework Details}
\label{app:framework}

\subsection{Compressive autoencoder}
\label{app:autoencoder}

We employ a deterministic 3D convolutional autoencoder $(\mathcal{E}_\psi, \mathcal{D}_\psi)$ to map each mesh-resolved high-dimensional PIC state $\mathbf{u}_t$ to a spatial grid latent $\mathbf{z}_t = \mathcal{E}_\psi (\mathbf{u}_t)$ and reconstruct it as $\tilde{\mathbf{u}}_t = \mathcal{D}_\psi(\mathbf{z}_t)$. The encoder and decoder consist of a sequence of residual blocks and resolution-changing layers. Each residual block applies GroupNorm, SiLU activation, and a 3D convolution twice in succession, with the block input added to the output through an identity residual connection. Each downsampling layer uses a strided 3D convolution to jointly reduce the selected spatial dimensions and transform the feature channels. The decoder mirrors this hierarchy in reverse: each upsampling layer first restores the spatial resolution by trilinear interpolation and then applies a 3D convolution to transform the feature channels. A pointwise channel projection maps the final encoder feature map to the grid latent, with a corresponding projection at the beginning of the decoder. No encoder--decoder skip connections bypass the latent bottleneck.

\paragraph{Latent Dimensions.}
Both datasets use a $16 \times$ reduction in total tensor elements. For the space-charge-dominated bunched-beam transport dataset, the encoder receives the single charge density field $\mathbf{u}_t \in \mathbb{R}^{1 \times 256 \times 336 \times 336}$. The 3D Haar transform first maps the single input channel to eight subband channels while halving each spatial dimension,
\begin{equation*}
    1 \times 256 \times 336 \times 336 \rightarrow 8 \times 128 \times 168 \times 168.
\end{equation*}
A learned convolution projects the 8 Haar channels to $32$ feature channels. The subsequent encoder stage changes the channel widths as $32 \to 64 \to 128 \to 128$, while the spatial grid evolves as
\begin{equation*}
    128 \times 168 \times 168 \rightarrow 64 \times 84 \times 84 \rightarrow 32 \times 42 \times 42 \rightarrow 16 \times 42 \times 42. 
\end{equation*}
A final projection maps the $128$-channel tensor to the $64$-channel latent, so that 
\begin{equation*}
    \mathbf{z}_t \in \mathbb{R}^{64 \times 16 \times 42 \times 42}.
\end{equation*}

For the PWFA dataset, the encoder receives $10$-channel state $\mathbf{u}_t \in \mathbb{R}^{10 \times 336 \times 216 \times 216}$. The Haar transform first maps each channel to $8$ subbands resulting in $80$ channels at resolution $168 \times 108 \times 108$. These channels are projected to $64$ learned features and passed through the $64 \to 128 \to 256 \to 256$ hierarchy, with spatial resolutions evolving as
\begin{equation*}
    168 \times 108 \times 108 \rightarrow 84 \times 54 \times 54 \rightarrow 42 \times 27 \times 27 \rightarrow 42 \times 27 \times 27. 
\end{equation*}
A final projection maps the $256$-channel tensor to the $320$-channel latent, producing the latent 
\begin{equation*}
    \mathbf{z}_t \in \mathbb{R}^{320 \times 42 \times 27 \times 27}.
\end{equation*}

\paragraph{3D Haar Transform.}

Inspired by~\citet{dai2026physics}, we apply a fixed orthonormal 3D Haar transform before the learned encoder. Let 
\begin{equation*}
    h = \frac{1}{\sqrt{2}}[1,1], \quad g = \frac{1}{\sqrt{2}}[1,-1]
\end{equation*}
denote the one-dimensional low-pass and high-pass Haar filters, respectively. The low-pass filter computes the local average over two neighboring samples, whereas the high-pass filter computes their local difference. Taking their tensor products along the three spatial axes gives the eight separable filters:
\begin{equation*}
    (hhh), \quad
    (ghh, hgh, \ hhg),\quad
    (ggh,  ghg, hgg, \ ggg).
\end{equation*}
The $hhh$ component is the three dimensional low-pass subband and represents the local block average. The three subbands containing a single high-pass factor, $ghh$, $hgh$, $hhg$, represent directional variations along the three spatial axes. The remaining four subbands contain high-pass factors along two or three axes and represent mixed or cross-directional variations within the block. Thus, the transform separates local average and multiaxial detail information before learned processing. Defining 
\begin{equation*}
    H = \frac{1}{\sqrt 2}  \begin{bmatrix}
        1 & 1 \\
        1 & -1
    \end{bmatrix}, \quad
    M = H \otimes H \otimes H \in \mathbb{R}^{8 \times 8},
\end{equation*}
where $\otimes$ is the Kronecker product. The Haar transform of each non-overlapping $2\times2\times2$ block, flattened as $\mathbf{v} \in \mathbb{R}^8$, is computed as $\mathbf{f} = M \mathbf{v}$, where the eight entries of $\mathbf{f}$ are the corresponding Haar subband coefficients. The transform is applied independently to every input channel with stride $2$, reshaping the input shape as:
\begin{equation*}
    C \times Z\times Y \times X \rightarrow 8C \times \frac{Z}{2} \times \frac{Y}{2} \times \frac{X}{2}.
\end{equation*}
Because $H$ is orthogonal, the 3D Haar transform is invertible. The decoder predicts the eight subbands for each output channel and applies the fixed inverse Haar transform to recover the original spatial resolution.

\paragraph{Space--channel Residual Shortcuts.}
Following~\citet{chen2025deep}, each learned down- and up-sampling block is paired with a parameter-free shortcut that transfers information across different spatial scales. For a downsampling layer with input $\mathbf{x} \in \mathbb{R}^{C \times Z \times Y \times X}$ and stride $\mathbf{s} = (s_z, s_y, s_x)$. Let $r = s_zs_ys_x$. The shortcut first applies a space-to-channel rearrangement, which folds the samples within each non-overlapping $s_z \times s_y \times s_x$ spatial block into the channel dimension:
\begin{equation*}
     C \times Z \times Y \times X \longrightarrow rC \times \frac{Z}{s_z} \times \frac{Y}{s_y} \times \frac{X}{s_x}.
\end{equation*}
This operation is pure reshaping and permutation, therefore introducing no interpolation or learned parameters. The resulting $rC$ channels are then resized to the target width $C'$. Consecutive groups of $rC/C'$ channels are averaged so that the shortcut output has the same shape as the learned strided-convolution branch. The shortcut output is then added to the learned strided-convolution branch. 

The upsampling shortcut applies the reverse sequence. For an upsampling factor $\mathbf{s} = (s_z, s_y, s_x)$, the input channels $C'$ are first repeated to obtain $rC$ channels, after which a channel-to-space rearrangement distributes each group of $r$ channels over an $s_z \times s_y \times s_x$ spatial block:
\begin{equation*}
    rC \times \frac{Z}{s_z} \times \frac{Y}{s_y} \times \frac{X}{s_x}\rightarrow C \times Z \times Y \times X. 
\end{equation*}
The shortcut output tensor is added to the learned upsampling branch, which performs trilinear interpolation followed by a 3D convolution. Hence, the shortcut provides a parameter-free shape-matched pathway across every resolution change, while the learned branch supplies the residual transformation.

\paragraph{Reconstruction Objectives.}
The autoencoder is trained to reconstruct each 3D PIC state using a composite objective:
\begin{equation*}
    \mathcal{L}_{\text{rec}}(\mathbf{u}_t, \tilde{\mathbf{u}}_t) = \mathcal{L}_{\text{rel}}(\mathbf{u}_t, \tilde{\mathbf{u}}_t) + \lambda_{\text{spec}}\mathcal{L}_{\text{spec}}(\mathbf{u}_t, \tilde{\mathbf{u}}_t) + \lambda_{\text{fd}}\mathcal{L}_{\text{fd}}(\mathbf{u}_t, \tilde{\mathbf{u}}_t). 
\end{equation*}

The first and dominant term is the relative-$L_2$ loss,
\begin{equation*}
    \mathcal{L}_{\text{rel}}(\mathbf{u}_t, \tilde{\mathbf{u}}_t) = \Bigg(\frac{\|\mathbf{u}_t - \tilde{\mathbf{u}}_t\|_2^2}{\max(\|\mathbf{u}_t\|_2^2 , \epsilon)}\Bigg)^{\frac{1}{2}},
\end{equation*}
which measures the normalized pixel-wise reconstruction error and provides a scale-normalized measurement of the overall state reconstruction accuracy.

We measure reconstruction accuracy in local spatial variation using a gradient loss 
\begin{equation*}
    \mathcal{L}_{\text{fd}}(\mathbf{u}_t, \tilde{\mathbf{u}}_t) = 
    \sum_{j=1}^3 \Bigg({\frac{ \|\partial_{x_j} \mathbf{u}_t - \partial_{x_j} \tilde{\mathbf{u}}_t \|_2^2}{\max( \|\partial_{x_j} \mathbf{u}_t\|_2^2, \epsilon)}\Bigg)^{\frac{1}{2}}}.
\end{equation*}
The spatial derivatives are evaluated using finite differences along the three coordinate directions. Whereas the relative-$L_2$ term primarily measures agreement in field values, $\mathcal{L}_{\text{fd}}$ directly penalizes errors in their spatial variation, placing additional emphasis on sharp gradients, beam boundaries, and other localized structures that can contribute only weakly to the pointwise reconstruction error.

To account for reconstruction error across spatial frequencies, we additionally use a weighted spectral loss, 
\begin{equation*}
    \mathcal{L}_{\text{spec}}(\tilde{\mathbf{u}}_t, \mathbf{u}_t) = \Bigg({\frac{\| \mathbf{w} \odot \hat{\mathbf{u}}_t(\mathbf{k}) - \mathbf{w} \odot\hat{\tilde{\mathbf{u}}}_t(\mathbf{k})\|_2^2}{\max(\ \|\mathbf{w} \odot \hat{{\mathbf{u}}}_t(\mathbf{k})\|_2^2 , \epsilon_2)}\Bigg)^{\frac{1}{2}}},
\end{equation*}
where $\mathbf{w} = (w(\mathbf{k}))_{\mathbf{k}}$ assigns different contributions to different wavenumber ranges designed as
\begin{equation*}
    w_0(\mathbf{k}) = 1 + \frac{|\mathbf{k}|}{k_{\max}}, \quad
    w^2(\mathbf{k}) = \begin{cases}
        5 w_0(\mathbf{k}),  & \frac{1}{3} k_{\max} < |\mathbf{k}| < \frac{2}{3} k_{\max} \\
        w_0(\mathbf{k}), & \text{otherwise}.
    \end{cases}
\end{equation*}
This weighting reduces the dominance of low-wavenumber modes, whose large energy contribution is already well captured by the relative-$L_2$ loss, while placing additional emphasis on intermediate wavenumbers that contain coherent small-scale beam structure. The highest-wavenumber band is not additionally boosted, since it is largely dominated by finite-particle PIC sampling noise and would otherwise encourage the model to overfit realization-specific fluctuations.

Since the decoder receives predicted rather than exact encoded latents during dynamics rollout, we additionally train our decoder to tolerate latent perturbations. Given $\mathbf{z} = \mathcal{E}_\psi(\mathbf{u})$, we construct
\begin{equation*}
    \mathbf{z} + \sigma \mathrm{std}(\mathbf{z}) \boldsymbol{\epsilon}, \quad \sigma \sim \mathcal{U}(0,0.15), \quad \boldsymbol{\epsilon} \sim \mathcal{N}(\mathbf{0}, \mathbf{I}),
\end{equation*}
where $\mathrm{std}(\mathbf{z})$ scales the perturbation relative to the magnitude of the latent representation. The autoencoder is then trained to reconstruct the clean target from both the encoded and perturbed latents, 
\begin{equation*}
     \mathcal{L}_{\text{AE}}(\tilde{\mathbf{u}}_t, \mathbf{u}_t) = \mathcal{L}_{\text{rec}} \left(\tilde{\mathbf{u}}_t, \mathbf{u}_t\right) + \lambda_{\text{noise}}\mathcal{L}_{\text{rec}}\big(\mathcal{D}_\psi(\mathbf{z}_t +\sigma \mathrm{std}(\mathbf{z}_t)\boldsymbol{\epsilon}), \mathbf{u}_t\big).
\end{equation*}
This perturbation exposes the decoder to off-center latent states during training, improving robustness to latent prediction errors encountered during autoregressive rollout.

\newpage
\subsection{Deterministic Latent Neural Operator.}
\label{app:deterministic_latent}

\paragraph{Forecast Representation.}
Given a $k$-step latent history $\mathbf z_{t-k+1:t}\in \mathbb R^{k\times C'\times Z'\times Y'\times X'}$ and machine controls $\mathbf c$, the deterministic operator $\mathcal G_\theta$ predicts the next $h$ latent states,
$\tilde{\mathbf z}^{\mathrm{det}}_{t+1:t+h}\in \mathbb R^{h\times C'\times Z'\times Y'\times X'}$. Let $N=Z'Y'X'$ denote the number of spatial tokens in the compressed latent grid. We first flatten the spatial dimensions of each latent frame,
\begin{equation*}
    \mathbf z_i
    \in \mathbb R^{C'\times Z'\times Y'\times X'}
    \;\longrightarrow\;
    \mathbf Z_i
    =\operatorname{Flat}(\mathbf z_i)
    \in\mathbb R^{N\times C'} .
\end{equation*}
The $k$ flattened frames are then concatenated along the feature dimension at each spatial location,
\begin{equation*}
    \mathbf Z_{\mathrm{hist}} = \left[\mathbf Z_{t-k+1},\ldots,\mathbf Z_t \right]
    \in\mathbb R^{N\times kC'}.
\end{equation*}
Thus, the token dimension indexes the $N$ spatial tokens, while the temporal history is encoded in the $kC'$ feature dimension. Finally, a linear lifting layer maps these features to the Transformer hidden dimension $d$.

\paragraph{Spatial Positional Encoding.}
Flattening the latent grid removes its explicit three-dimensional
arrangement. We therefore associate each token $i$ with its normalized
coordinate
$\mathbf x_i=(x_i,y_i,z_i)\in[-1,1]^3$
and encode it using sinusoidal Fourier features
~\citep{tancik2020fourier},
\begin{equation*}
    \boldsymbol{\phi}(\mathbf x_i)
    =
    \left[
        \sin(2\pi\mathbf B\mathbf x_i),
        \cos(2\pi\mathbf B\mathbf x_i)
    \right],
    \qquad
    \mathbf e_i^{\mathrm{pos}}
    =
    \mathbf W_{\mathrm{pos}}
    \boldsymbol{\phi}(\mathbf x_i),
\end{equation*}
where $\mathbf B \in \mathbb{R}^{M \times 3}$ denotes the prescribed frequency bank, and $\mathbf{W} \in \mathbb{R}^{d \times 2M}$ is a learned linear projection that maps the Fourier-feature vector $\boldsymbol(\phi)(\mathbf{x}_i)$ into the Transformer hidden dimension $d$. Stacking $\mathbf e_i^{\mathrm{pos}}$ over all $N$ spatial tokens gives $\mathbf E_{\mathrm{pos}}\in\mathbb R^{N\times d}$, which is added to the lifted latent tokens before the Transformer. This preserves the physical identity of each spatial token while retaining the global interaction enabled by full attention.

\paragraph{Full-Attention Backbone.}
We process the resulting $N$ spatial tokens using a Transformer with dense self-attention~\citep{vaswani2017attention}. Full attention allows every latent location to interact directly with all other locations, providing global spatial coupling across the compressed 3D domain. Because attention is applied after spatial compression, this global interaction is tractable despite the high resolution of the original PIC state. After the final block, the spatial token is projected from hidden dimension $d$ directly to dimension $hC'$, which corresponds to the $C'$ channel features over the $h$-step latent forecast. The flattened spatial token is reshaped back to the original latent grid,
\begin{equation*}
    \mathbb R^{N\times hC'}
    \;\longrightarrow\;
    \mathbb R^{h\times N\times C'}
    \;\longrightarrow\;
    \mathbb R^{h\times C'\times Z'\times Y'\times X'},
\end{equation*}
yielding $\tilde{\mathbf z}^{\mathrm{det}}_{t+1:t+h}$. The complete $h$-step forecast is therefore predicted jointly rather than autoregressively within each forecast block.

\paragraph{Control Conditioning with adaLN-Zero.}
The machine controls $\mathbf c$ are first mapped to a conditioning
embedding
\begin{equation*}
    \mathbf e_c=\operatorname{MLP}_c(\mathbf c)\in\mathbb R^d .
\end{equation*}
Following adaLN-Zero~\citep{Peebles_2023_ICCV}, the control embedding $\mathbf e_c$ is injected into each Transformer block through adaptive layer normalization. For each token $\mathbf h\in\mathbb R^d$, layer normalization
standardizes its hidden features,
\begin{equation*}
    \operatorname{LN}(\mathbf h)
    =
    \frac{\mathbf h-\mu(\mathbf h)}
    {\sqrt{\sigma^2(\mathbf h)+\epsilon}},
\end{equation*}
where $\mu(\mathbf h)$ and $\sigma^2(\mathbf h)$ are the mean and variance
over the $d$ hidden dimensions. A modulation network maps $\mathbf e_c$ to
channel-wise scale and shift vectors
$\boldsymbol{\gamma},\boldsymbol{\beta}\in\mathbb R^d$, giving
\begin{equation*}
    \operatorname{adaLN}(\mathbf h;\mathbf e_c)
    =
    \bigl(1+\boldsymbol{\gamma}(\mathbf e_c)\bigr)
    \odot\operatorname{LN}(\mathbf h)
    +
    \boldsymbol{\beta}(\mathbf e_c).
\end{equation*}
Thus, $\boldsymbol{\gamma}$ rescales and $\boldsymbol{\beta}$ shifts each
hidden feature according to the machine controls.

AdaLN-Zero predicts separate $(\boldsymbol{\gamma},\boldsymbol{\beta},\boldsymbol{\alpha})$ for the attention and feed-forward branches, where $\boldsymbol{\alpha}\in\mathbb R^d$ gates the corresponding residual update:
\begin{align*}
    \mathbf H'
    &=
    \mathbf H+
    \boldsymbol{\alpha}_{\mathrm{attn}}(\mathbf e_c)
    \odot
    \operatorname{Attn}\!\left(
        \operatorname{adaLN}_{\mathrm{attn}}(\mathbf H;\mathbf e_c)
    \right),\\
    \mathbf H^{+}
    &=
    \mathbf H'+
    \boldsymbol{\alpha}_{\mathrm{ffn}}(\mathbf e_c)
    \odot
    \operatorname{FFN}\!\left(
        \operatorname{adaLN}_{\mathrm{ffn}}(\mathbf H';\mathbf e_c)
    \right).
\end{align*}

The modulation network therefore produces six $d$-dimensional vectors per
block: scale, shift, and gate parameters for both attention and feed-forward
branches. The gates $\boldsymbol{\alpha}$ are initialized to zero, so each
block initially behaves as an identity mapping and gradually learns
control-dependent updates.

\newpage
\subsection{Generative Latent Residual.}
\label{app:Generative_Latent_Residual}

\paragraph{Amplitude-Normalization.} 

The latent residual
$\mathbf r_{t+1:t+h} = \mathbf z_{t+1:t+h} - \tilde{\mathbf z}^{\mathrm{det}}_{t+1:t+h}
\in \mathbb R^{h\times C'\times Z'\times Y'\times X'}$
inherits the strong amplitude heterogeneity of beam dynamics across latent channels, forecast blocks, and spatial regions. Training flow matching directly on these raw residuals biases the objective toward large-amplitude components, making weaker structures difficult to learn consistently. We instead normalize the residual to have approximately unit RMS before generative modeling, while predicting a separate conditional amplitude factor that restores its original scale at inference.

A fixed dataset-level normalization is insufficient because the residual magnitude varies with the current beam state and deterministic forecast, while the true future residual is unavailable at deployment. We therefore predict the scale from
\[
\mathbf p
=
\left(
\mathbf z_{t-k+1:t},
\tilde{\mathbf z}^{\mathrm{det}}_{t+1:t+h}
\right),
\]
which is available at inference, and factor it across three granularities:
\begin{equation*}
\mathbf s_\omega(\mathbf p)
=
\boldsymbol{\sigma}_r
\odot
s_{\mathrm{win}}(\mathbf p)
\odot
s_{\mathrm{tok}}(\mathbf p),
\qquad
\bar{\mathbf r}_{t+1:t+h}
=
\mathbf r_{t+1:t+h}
\oslash
\mathbf s_\omega(\mathbf p).
\end{equation*}
Here,
$\boldsymbol{\sigma}_r
\in
\mathbb R^{1\times C'\times1\times1\times1}$
is a global per-channel RMS,
$s_{\mathrm{win}}(\mathbf p)\in\mathbb R_{+}$
is a conditional normalizer for the whole forecast block, and
$s_{\mathrm{tok}}(\mathbf p)
\in
\mathbb R^{1\times1\times Z'\times Y'\times X'}$
is a conditional spatial normalizer shared across channels and forecast steps.

The design follows the RMS normalization principle. For residual values
$\{q_i\}_{i=1}^{N}$ with empirical mean square
\begin{equation*}
    m=\frac{1}{N}\sum_{i=1}^{N}q_i^2,
\end{equation*}
normalizing by $s=\sqrt{m}$ gives unit RMS,
\begin{equation*}
    \frac{1}{N}\sum_{i=1}^{N} \left(\frac{q_i}{s}\right)^2 =1.
\end{equation*}
Because this target RMS depends on the unknown future residual, we instead predict $s(\mathbf p)$ from the available conditioning information. Using the zero-centered Gaussian scale objective
\begin{equation*}
\mathcal L_{\mathrm{scale}}
\big(s(\mathbf p)\big)
=
\frac{m}{s(\mathbf p)^2}
+
2\log s(\mathbf p),
\end{equation*}
which is used only for scale estimation and does not assume that the normalized residual itself is Gaussian. For a fixed conditioning context $\mathbf p$, the expected objective is
\begin{equation*}
\mathbb E\!\left[
    \mathcal L_{\mathrm{scale}}
    \big(s(\mathbf p);m\big)
    \mid \mathbf p
\right]
=
\frac{\mathbb E[m\mid\mathbf p]}{s(\mathbf p)^2}
+
2\log s(\mathbf p).
\end{equation*}
Minimizing this objective with respect to $s(\mathbf p)$ gives
\begin{equation*}
s^\star(\mathbf p)^2
=
\mathbb E[m\mid\mathbf p],
\qquad
s^\star(\mathbf p)
=
\sqrt{\mathbb E[m\mid\mathbf p]}.
\end{equation*}
Thus, when optimized over the training distribution, the scale predictor
estimates the conditional RMS of the residual at the corresponding granularity.

We estimate the factors sequentially. First, the global channel scale
$\boldsymbol{\sigma}_r$
is computed from the training residuals and used to form
$\mathbf r_{\mathrm{ch}}
=
\mathbf r\oslash\boldsymbol{\sigma}_r$.
For each $h$-step forecast block, we define
\begin{equation*}
m_{\mathrm{win}}
=
\frac{1}{hC'Z'Y'X'}
\sum_{j,c,z,y,x}
r_{\mathrm{ch},j,c,z,y,x}^{\,2},
\end{equation*}
and train a network conditioned on $\mathbf p$ to predict
$s_{\mathrm{win}}(\mathbf p)$ optimized with $\mathcal L_{\mathrm{scale}}$.
After freezing this network, we form
$\mathbf r_{\mathrm{win}}
=
\mathbf r_{\mathrm{ch}}
/
s_{\mathrm{win}}(\mathbf p)$.
The remaining spatial variation is modeled token-wise using
\begin{equation*}
m_{\mathrm{tok}}(z,y,x)
=
\frac{1}{hC'}
\sum_{j,c}
r_{\mathrm{win},j,c,z,y,x}^{\,2},
\end{equation*}
from which a second network predicts
$s_{\mathrm{tok}}(\mathbf p)$
with the same scale objective. The resulting factorized normalizer
$\mathbf s_\omega
=
\boldsymbol{\sigma}_r
\odot s_{\mathrm{win}}
\odot s_{\mathrm{tok}}$
places residual components on comparable scales for flow matching, while the same predicted factors restore their amplitude at inference.

\paragraph{Flow-matching Training.}

After freezing the residual normalizer, the generative model $\mathcal{R}_\xi$ learns the conditional distribution of normalized residual $\bar{\mathbf{r}}_{t+1:t+h}$ using conditional flow matching. For each training example, we sample $\boldsymbol{\epsilon} \sim \mathcal{N}(\mathbf{0},\mathbf{I})$ and $\eta \sim \mathcal{N}(0,1)$, set $\tau = \operatorname{sigmoid}(\eta)$, and construct the linear interpolant 
\begin{equation*}
    \mathbf{x}_\tau = (1 - \tau) \boldsymbol{\epsilon} + \tau \bar{\mathbf{r}}_{t+1:t+h}.
\end{equation*}
The velocity network $\mathbf{v}_\xi$ takes $\mathbf{x}_\tau$, the latent history $\mathbf{z}_{t-k+1:t}$ and deterministic forecast $\tilde{\mathbf{z}}_{t+1:t+h}^{\text{det}}$, controls $\mathbf{c}$, and flow time $\tau$ as input, and predicts transport velocity along the linear path. The model is trained with
\begin{equation*}
    \mathcal{L}_{\mathcal{R}_\xi} = \mathbb{E}_{\tau, \boldsymbol{\epsilon}, \bar{\mathbf{r}}} \|\mathbf{v}_\xi (\mathbf{x}_\tau; \mathbf{z}_{t-k+1:t}, \tilde{\mathbf{z}}_{t+1:t+h}, \mathbf{c}, \tau) - (\bar{\mathbf{r}} -\mathbf{\boldsymbol{\epsilon}})\|_2^2 + \lambda_{\text{ES}} \mathcal{L}_{\text{ES}}.
\end{equation*}
The $\mathcal{L}_{\text{ES}}$ is a two-sample energy-score regularizer on generated residuals. Given two independent noise samples $\boldsymbol{\epsilon}^{(i)} \sim \mathcal{N}(\mathbf{0}, \mathbf{I})$, through one Euler step from $\tau = 0$,
\begin{equation*}
    \tilde{\mathbf{r}}^{(i)} = \boldsymbol{\epsilon}^{(i)} + \mathbf{v}_\xi(\boldsymbol{\epsilon}^{(i)}; \mathbf{z}_{t-k+1:t}, \tilde{\mathbf{z}}_{t+1:t+h}, \mathbf{c},0),\quad i = 1, 2
\end{equation*}
The energy-score term is calculated as
\begin{equation*}
    \mathcal{L}_{\text{ES}} = \left[ \frac{\|\tilde{\mathbf{r}}^{(1)} - \bar{\mathbf{r}}\|_2 + \|\tilde{\mathbf{r}}^{(2)} - \bar{\mathbf{r}}\|_2}{2} - \frac{\|\tilde{\mathbf{r}}^{(1)} - \tilde{\mathbf{r}}^{(2)}\|_2}{2}\right]. 
\end{equation*}
The energy-score term complements the pointwise flow-matching loss by encouraging generated residuals to remain consistent with the target while discouraging an under-dispersed conditional ensemble.

At inference, we sample $\mathbf{x}_0 \sim \mathcal{N}(\mathbf{0}, \mathbf{I})$ and integrate
\begin{equation*}
    \frac{\mathrm{d} \mathbf{x}_\tau}{\mathrm{d} \tau} = \mathbf{v}_\xi (\mathbf{x}_\tau; \mathbf{z}_{t-k+1:t}, \tilde{\mathbf{z}}_{t+1:t+h}, \mathbf{c}, \tau),
\end{equation*}
from $\tau = 0$ to $1$ by $16$-step Euler integration. The terminal state $\mathbf{x}_1$ is the normalized sample, which is rescaled as $\tilde{\mathbf{r}}_{t+1:t+h}$ and added to the latent forecast.

\paragraph{Backbone Architecture.}

\begin{figure*}[h]
    \centering
    \includegraphics[width=0.51\textwidth]{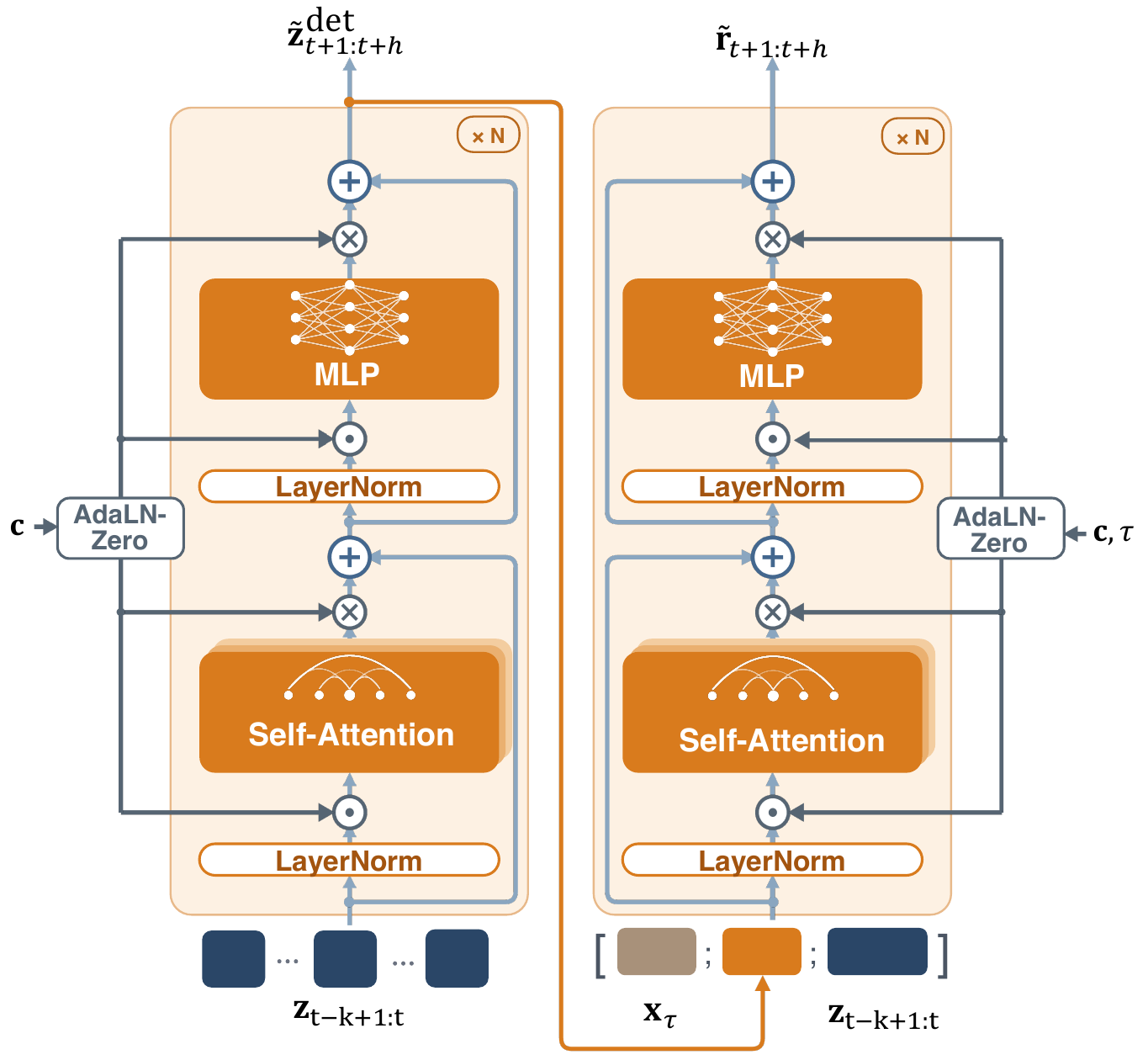}
    \caption{ Visualization of the Transformer backbone used by deterministic neural operator $\mathcal{G}_\theta$ and generative residual model $\mathcal{R}_\xi$.}
    \label{fig:transformer_backbone}
\end{figure*}

Both the deterministic operator and generative flow matching employ the transformer backbone, as visualized in Figure~\ref{fig:transformer_backbone}. The flow time $\tau$ is injected to each transformer block by AdaLN-Zero, same as the control parameters $\mathbf{c}$. For GLOBE-C, the deterministic neural operator $\mathcal{G}_\theta$ has $12$ Transformer blocks with hidden dimension $d = 512$. The generative residual model has $10$ Transformer blocks with hidden dimension $d = 384$. For the GLOBE-G ablation variant, we keep the hidden dimension and number of Transformer blocks the same as in \(G_\theta\). All Transformer blocks use eight attention heads and an MLP ratio of 4.

\newpage
\subsection{Physics-constrained Operator}
\label{app:spectral_operator}

\paragraph{Vlasov--Poisson Regime.}
Given the decoded charge density $\tilde{\rho}$, the predicted electrostatic potential $\tilde{\phi}$ is obtained by solving the discrete Poisson equation
\begin{equation*}
    -\Delta_h \tilde{\phi} = \frac{\tilde{\rho}}{\varepsilon_0} \quad \text{in } \Omega, \qquad \tilde{\phi}|_{\Gamma_a} = 0, \qquad \tilde{\phi}(x,y,z + L_z) = \tilde{\phi}(x,y,z), 
\end{equation*}
where $\Gamma_a$ denotes the grounded cylindrical beam-pipe boundary and $L_z$ is one RF wavelength in the beam co-moving frame. Longitudinal periodicity is incorporated directly into the discrete Poisson operator by wrapping the finite-difference stencil across the two $z$ faces. The grounded Dirichlet condition fixes the potential reference, so the potential is uniquely determined by $\tilde{\rho}$. In implementation, we use an iterative Krylov solver (Jacobi-preconditioned BiCGStab~\citep{doi:10.1137/0913035}) to solve $\tilde{\phi}$, with numerical tolerance $10^{-8}$.

The electric self-field is subsequently reconstructed on the same collocated grid using the discrete gradient,
\begin{equation*}
    \tilde{\mathbf{E}} = -\nabla_h \tilde{\phi}.
\end{equation*}
Away from the embedded boundary, $\nabla_h$ uses second-order centered differences, for example,
\begin{equation*}
    (D_j^h \phi)_{\mathbf{n}} =  \frac{\phi_{\mathbf{n} + \mathbf{e}_j} - \phi_{\mathbf{n} - \mathbf{e}_j}}{2\Delta_j}, \quad j \in \{x, y, z\},
\end{equation*}
where $\mathbf{e}_j$ denotes one grid step along direction $j$. Periodic wrapping is used in the longitudinal direction, while boundary-adjacent transverse differences are evaluated consistently with the grounded embedded-boundary discretization.

\paragraph{Vlasov--Maxwell Regime.}

Unlike the electrostatic regime, in the Vlasov--Maxwell system, the charge density and electromagnetic fields carry independent dynamical information and are therefore predicted jointly. The instantaneous Maxwell constraints are formulated as
\begin{equation}
\label{eq:vm-constraint}
    \nabla_h^E \cdot \mathbf{E} - \frac{\rho}{\varepsilon_0} = 0, \qquad \nabla_h^B \cdot \mathbf{B} = 0.
\end{equation}

Both discrete divergence operators, $\nabla_h^E$ and $\nabla_h^B$, are composed of translation-invariant finite-difference and averaging stencils and therefore diagonalized under the discrete Fourier transform.

Let
\begin{equation*}
    k_j = \frac{2\pi n_j}{L_j}, \qquad n_j = 0, \cdots, N_j-1, \qquad j \in \{x, y, z\},
\end{equation*}
represent the discrete Fourier wavenumber along direction $j$, where $L_j = N_j \Delta_j$. The electric-field divergence uses the wide centered difference
\begin{equation*}
    (D_j^h f)_{\mathbf{n}} = \frac{f_{\mathbf{n} + \mathbf{e}_j} -f_{\mathbf{n} - \mathbf{e}_j} }{2\Delta_j},
\end{equation*}
whose Fourier symbol is
\begin{equation*}
    d_j^E(\mathbf{k}) = \frac{i \sin (k_j \Delta_j)}{\Delta_j}. 
\end{equation*}

The stored magnetic field follows the centering convention of the CKC/Yee field solver, so its matched discrete divergence has a different symbol. Defining the forward-difference and averaging symbols
\begin{equation*}
    d_j^+(\mathbf{k}) = \frac{e^{i k_j \Delta_j} -1}{\Delta_j}, \qquad a_j(\mathbf{k}) = \frac{1+e^{i k_j\Delta_j}}{2},
\end{equation*}
the magnetic-divergence symbol along component $j$ is
\begin{equation*}
    d_j^B (\mathbf{k}) = d_j^+(\mathbf{k}) a_{t_1}(\mathbf{k}) a_{t_2}(\mathbf{k})T_j(\mathbf{k}),
\end{equation*}
where $t_1, t_2$ denote the two directions transverse to $j$, and $T_j$ is the fixed CKC transverse-smoothing symbol,
\begin{equation*}
    T_j = 1 - 2\beta \Big[r_{t_1}\big(1-\cos(k_{t_1}\Delta_{t_1})\big) + r_{t_2} \big(1-\cos(k_{t_2}\Delta_{t_2})\big) \Big] - 4 \gamma_j \big(1 - \cos(k_{t_1}\Delta_{t_1}) \cos(k_{t_2}\Delta_{t_2})\big)
\end{equation*}
where $r_{t_j}, \beta, \gamma_j$ are fixed by CKC stencil and the grid spacings. Thus, $\nabla_j^E$ and $\nabla_j^B$ deliberately use different symbols because they must match the discrete representation of the stored electric and magnetic fields.

For each Fourier mode, define
\begin{equation*}
    \hat{\mathbf{u}} = \begin{bmatrix}
        \hat{\rho} & \hat{E}_x & \hat{E}_y &  \hat{E}_z &  \hat{B}_x & \hat{B}_y & \hat{B}_z
    \end{bmatrix}^\top
\end{equation*}
\Eqref{eq:vm-constraint} then becomes the homogeneous linear system
\begin{equation*}
    \mathbf{A}_h(\mathbf{k}) \hat{\mathbf{u}}(\mathbf{k}) = 0,
\end{equation*}
where
\begin{equation*}
    \mathbf{A}_h(\mathbf{k}) = \begin{bmatrix}
        -\frac{1}{\varepsilon_0} & d_x^E & d_y^E & d_z^E & 0 & 0 & 0 \\
        0 & 0 & 0 & 0 & d_x^B & d_y^B & d_z^B
    \end{bmatrix}. 
\end{equation*}
$\mathbf{A}_h$ is the Fourier-space matrix representation of Gauss's law and magnetic divergence constraint under the matched finite difference operators.

Independently at each Fourier mode $\mathbf{k}$, given an unconstrained decoder prediction $\hat{\tilde{\mathbf{u}}}(\mathbf{k})$, we seek the smallest weighted correction that satisfies \Eqref{eq:vm-constraint},
\begin{equation*}
    \hat{\mathbf{u}}^\star(\mathbf{k}) = \argmin_{\hat{\mathbf{u}}(\mathbf{k}): \mathbf{A}_h(\mathbf{k})\hat{\mathbf{u}}(\mathbf{k}) = 0} \|\hat{\mathbf{u}}(\mathbf{k}) - \hat{\tilde{\mathbf{u}}}(\mathbf{k})\|_{\mathbf{W}^{-1}}^2,
\end{equation*}
where $\mathbf{W}$ is a diagonal matrix whose diagonal entries contain the per-channel scales used by the reconstruction metrics. The closed-form solution is
\begin{equation*}
    \hat{\mathbf{u}}^\star(\mathbf{k}) = \hat{\tilde{\mathbf{u}}}(\mathbf{k}) - \mathbf{W} \mathbf{A}_h^\mathsf{H}(\mathbf{k})\Big(\mathbf{A}_h(\mathbf{k})\mathbf{W} \mathbf{A}_h^\mathsf{H}(\mathbf{k})\Big)^\dagger\mathbf{A}_h(\mathbf{k}) \hat{\tilde{\mathbf{u}}}.
\end{equation*}
where $\cdot^\mathsf{H}$ and $\cdot^\dagger$ denote the Hermitian transpose and Moore--Penrose pseudoinverse. After applying this projection independently to all Fourier modes, the corrected coefficients are assembled and mapped back to the computational grid using the inverse discrete Fourier transform. 

The spectral projection is implemented using a periodic 3D FFT of the matched discrete difference stencils. Periodicity is used only to diagonalize the operators in Fourier space. The stored simulation window itself is non-periodic. Directly removing the full periodic Gauss residual would couple opposite faces and impose an artificial periodic boundary condition. We therefore compute
\begin{equation*}
    D_G =
    \nabla_h^E\!\cdot\tilde{\mathbf E}
    - \tilde{\rho}/\varepsilon_0
\end{equation*}
in real space and form the masked residual
\begin{equation*}
    D_G^{\mathrm{proj}} = m D_G,
\end{equation*}
where $m=0$ within two cells of each face and $m=1$ in the interior. The Fourier transform of $D_G^{\mathrm{proj}}$, rather than the full
$\mathbf A_h\hat{\tilde{\mathbf u}}$, is supplied to the Gauss row of the minimum-norm correction above. Consequently, the corrected state satisfies Gauss's law exactly in the interior, while the boundary shell remains unconstrained.

Let $\Omega_{\mathrm{int}}$ denote the unmasked region. By the divergence theorem,
\begin{equation*}
    \Phi_E
    =
    \oint_{\partial\Omega_{\mathrm{int}}}
    \mathbf E\cdot\mathbf n\,dA
    =
    \int_{\Omega_{\mathrm{int}}}
    \nabla\!\cdot\mathbf E\,dV ,
\end{equation*}
so Gauss's law requires
\begin{equation*}
    \varepsilon_0\Phi_E = Q_{\mathrm{int}},
    \qquad
    Q_{\mathrm{int}}
    =
    \int_{\Omega_{\mathrm{int}}}\rho\,dV .
\end{equation*}
Thus, a nonzero enclosed charge requires a corresponding electric flux through the boundary of the interior region. A fully periodic projection cannot represent this relation because the net flux over a periodic box is identically zero. Leaving the boundary shell unconstrained allows the required flux to pass through the interior boundary while enforcing Gauss's law exactly inside.

This also clarifies the $\mathbf k=0$ mode. For nonzero mode, $\mathbf{E}$ and $\rho$ are corrected simultaneously with respect to Gauss's law in the interior. However, for $\mathbf{k}= 0$, $d_j^E(\mathbf 0)=0$, the electric field cannot change the zero-frequency component of the Gauss residual. The zero mode of the masked residual therefore measures the global mismatch between the electric flux through $\partial\Omega_{\mathrm{int}}$ and the enclosed net charge,
\begin{equation*}
    D_G(\mathbf 0)
    \propto
    \Phi_E-\frac{Q_{\mathrm{int}}}{\varepsilon_0}.
\end{equation*}
Because $\rho$ is included jointly with $\mathbf E$ in $\mathbf A_h$, the Gauss row remains nonsingular at $\mathbf k=0$: at this
mode the minimum-norm correction acts through the charge-density component, corresponding to a spatially uniform shift of $\rho$. The corrected state therefore satisfies
\begin{equation*}
    Q_{\mathrm{int}}^\star
    =
    \varepsilon_0\Phi_E,
\end{equation*}
without forcing the simulation window to be charge neutral. At nonzero modes, $\rho$ and $\mathbf E$ are corrected jointly according to the same projection defined above.

\newpage
\section{Baseline Descriptions}
\label{app:baseline}

\subsection{Denoising Operator Transformer (DPOT)}
\label{app:baseline_dpot}
DPOT~\citep{hao2024dpot} is a Transformer-based PDE foundation model designed
for autoregressive prediction across heterogeneous physical systems. Its
architecture first patchifies the input fields and projects each spatial patch
into a hidden representation. Information from multiple input time steps is
combined through a temporal aggregation module, followed by a stack of Fourier
attention blocks based on the Adaptive Fourier Neural Operator (AFNO). A
lightweight output projection maps the processed patch representations back to
the predicted physical fields. DPOT is trained using a relative-$L_2$ objective. In our experiments, we use the DPOT architecture at a matched model size and adapt its input and output channels to the corresponding beam--field states. In both benchmarks, DPOT is trained to predict $h$-step future states given $k$-step histories. 

\subsection{Universal Physics Transformer (UPT)}
\label{app:baseline_upt}

UPT~\citep{alkin2024universal} is a latent neural operator consisting of an
encoder, a latent dynamics approximator, and a query-based decoder. The encoder
compresses the input physical state into a fixed set of latent tokens using
local aggregation and Perceiver-style cross-attention. A Transformer
approximator then propagates these tokens forward in latent space. To recover
the physical field, the decoder embeds the coordinates of the desired output
locations as queries and cross-attends to the propagated latent tokens,
allowing the latent representation to be evaluated at arbitrary spatial
positions. UPT is trained with a next-step prediction loss together with
inverse decoding and inverse encoding losses, all implemented with mean-square
error, to promote consistency of the latent representation during rollout. In
our setting, the full 3D mesh locations are used as decoder queries to recover
the beam--field state during inference. During training, due to the design of the model architecture, UPT is trained to predict the next single future state given the past $k$-step histories. For training tractability, we uniformly select $150528$ (bunched-beam transport) and $163296$ (PWFA) query points among the spatial grids for loss function evaluation on both datasets.

\subsection{Multiple Physics Pretraining (MPP)}
\label{app:baseline_mpp}

MPP~\citep{mccabe2024multiple} is an autoregressive Transformer framework
designed to learn dynamics shared across heterogeneous physical systems. It
first normalizes and projects different physical fields into a common embedding
space, followed by spatial patchification. Its backbone is an Axial Vision
Transformer (AViT), which factorizes self-attention across the temporal and
individual spatial axes to reduce the cost of processing high-dimensional
spatiotemporal fields. The processed tokens are projected back to the physical
variables to predict the next state. MPP is trained with a
normalized mean-square error. For our comparison, we instantiate the MPP backbone at a model size comparable to GLOBE and adapt the field embedding and output layers to each benchmark. Since MPP's output head decodes a single time slot into one frame, training predicts only the next state given the $k$-step histories.

\subsection{P3D}
\label{app:baseline_p3d}

P3D~\citep{holzschuh2026pd} is a scalable backbone developed specifically for
high-resolution 3D physics simulations. It combines convolutional blocks for
local feature extraction with windowed self-attention in a hierarchical
U-shaped encoder--decoder architecture, enabling direct processing of
full-resolution 3D states while limiting the cost of attention. We evaluate
the same backbone under two training formulations.

\textbf{P3D-D.}
For the deterministic baseline, P3D directly maps the input state and
conditioning variables to the future physical state and is trained with the
supervised mean-square error proposed in the original work. Predictions are
then rolled out autoregressively in the full 3D state space.

\textbf{P3D-G.}
For the generative baseline, we follow the probabilistic formulation of P3D
and train the same 3D backbone with conditional flow matching. Starting from
Gaussian noise, P3D-G learns a velocity field transporting the noise
distribution toward the conditional distribution of the target physical state.
At inference, the corresponding ODE is integrated iteratively to generate each
forecast. Unlike GLOBE, which performs flow integration on a compressed latent
grid, P3D-G applies the generative backbone directly to the full 3D state.

\newpage
\section{Supplementary Experiment Results}

\subsection{Amplitude-normalization Ablation}
\label{app:amplitude-normalization-ablation}

Table~\ref{tab:norm} evaluates the amplitude normalization introduced in Section~\ref{sec:latent_dynamics} by replacing the factorized normalizer $s_\omega=\boldsymbol{\sigma}_r\odot s_{\mathrm{win}}\odot s_{\mathrm{tok}}$ with the global per-channel scale $\boldsymbol{\sigma}_r$ alone. On bunched-beam transport, the $\boldsymbol{\sigma}_r$-only variant increases the mid-band spectral error from $0.457$ to $0.894$, approaching the deterministic GLOBE-D baseline ($0.994$). Without window- and token-level scaling, the flow-matching objective is dominated by larger-amplitude residuals, while weaker fine-scale variations, particularly in the low-density regions near the bunch boundary, receive insufficient weight. Consequently, the residual model provides only limited refinement over the deterministic prediction.

The effect is more pronounced on PWFA, where residual magnitudes vary by orders of magnitude across field channels, forecast windows, and spatial locations. Using only $\boldsymbol{\sigma}_r$ increases Rel-$L_2$ from $0.194$ to $0.266$ and the low/mid/high spectral errors from $0.077/0.300/0.618$ to $0.231/4.32/6.66$. The factorized normalizer compensates for these heterogeneous amplitudes, bringing residual targets to comparable, approximately unit-RMS scales across channels, windows, and tokens. This substantially improves the conditioning of the flow-matching objective and enables stable and effective generative refinement on both benchmarks.

\begin{table*}[t]
\centering
\caption{Ablation of the amplitude normalization for the generative residual. GLOBE-C uses the factorized normalizer $s_\omega=\boldsymbol{\sigma}_r\odot s_{\mathrm{win}}\odot s_{\mathrm{tok}}$; the variant keeps only the global per-channel scale $\boldsymbol{\sigma}_r$. GLOBE-D (deterministic anchor, no residual) is shown for reference.}
\label{tab:norm}
\resizebox{\textwidth}{!}{
\begin{tabular}{lcccc}
\toprule
& \multicolumn{2}{c}{\textbf{Bunched-Beam Transport}}
& \multicolumn{2}{c}{\textbf{PWFA}} \\
\cmidrule(lr){2-3}\cmidrule(lr){4-5}
\textbf{Method} & Rel-$L_2$ $\downarrow$ & $L_{\mathcal{F}}$ (low / mid / high) $\downarrow$
& Rel-$L_2$ $\downarrow$ & $L_{\mathcal{F}}$ (low / mid / high) $\downarrow$ \\
\midrule
GLOBE-D (no residual)                   & \textbf{0.144} & 0.039 / 0.994 / 1.003 & \textbf{0.174} & 0.083 / 0.350 / \textbf{0.592} \\
\midrule
GLOBE-C ($\boldsymbol{\sigma}_r$ only) & 0.147 & 0.039 / 0.894 / 0.986 & 0.266 & 0.231 / 4.32 / 6.66 \\
GLOBE-C (Ours)                          & 0.161 & \textbf{0.035} / \textbf{0.457} / \textbf{0.973} & 0.194 & \textbf{0.077} / \textbf{0.300} / 0.618 \\
\bottomrule
\end{tabular}
}
\vspace{8pt}
\end{table*}

\subsection{Space-charge-dominated Bunched-Beam Transport}
\label{app:bunched-beam-visualization}

We further visualize representative bunched-beam trajectories from the test set spanning different charges, focusing strengths, initial mismatches, and space-charge regimes. For each trajectory, we report three complementary views. The transverse projection tracks the beam cross section together with the second-order moments $\sigma_x$, $\sigma_y$, $\sigma_z$, and the aspect ratio $\sigma_x/\sigma_y$. The longitudinal projection visualizes the coupled transverse--longitudinal evolution. Finally, an $x$--$z$ slice at $y=0$ examines fine-scale spatial structure. For the latter, we additionally show the mid-band component obtained by a 2-D Fourier band-pass retaining $0.168<|\mathbf{k}|\leq0.246$ cycles/cell, normalized by the same peak density $\rho_{\max}$ and displayed on a symmetric scale. This isolates the spatial scales at which deterministic regression tends to smooth the PIC realization and highlights the structure recovered by generative residual refinement.

\begin{figure*}[h]
    \centering
    \includegraphics[width=\textwidth]{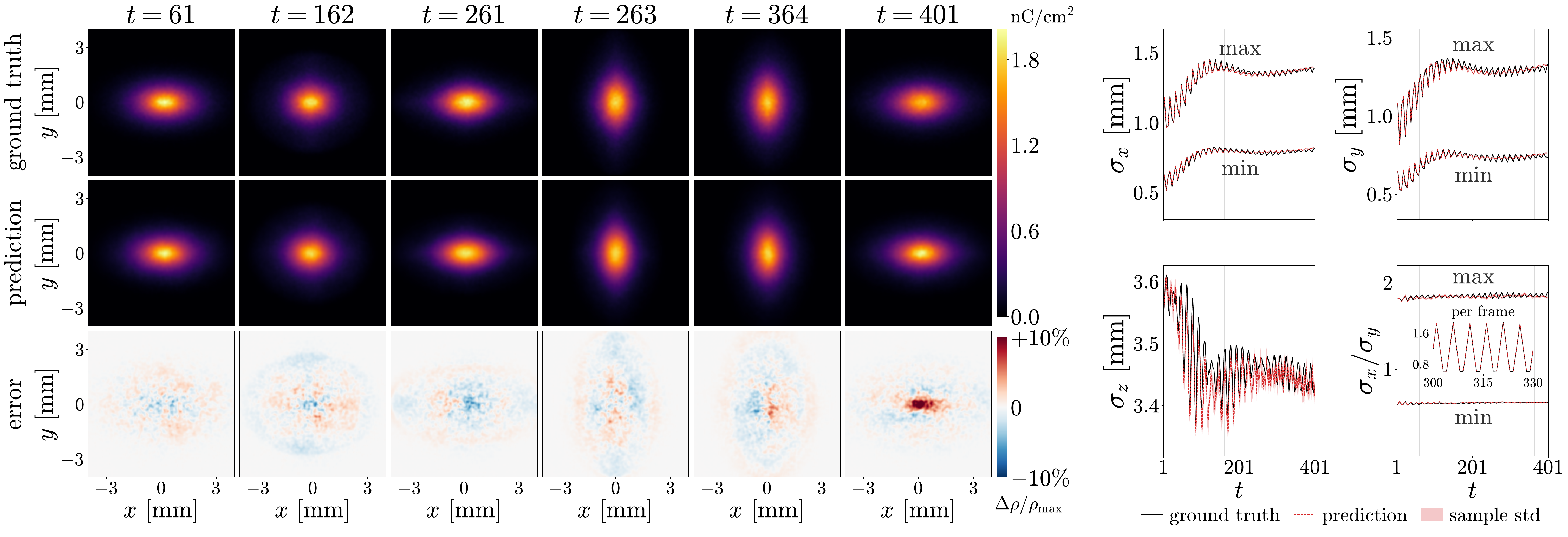}
    \caption{\small
    Bunched-beam Transport Run 392: transverse beam evolution. Ground-truth and predicted density projections are shown together with the evolution of
    $\sigma_x$, $\sigma_y$, $\sigma_z$, and $\sigma_x/\sigma_y$.
    }
    \label{fig:bunched_run392_transverse}
\end{figure*}

\begin{figure*}[h]
    \centering
    \includegraphics[width=0.85\textwidth]{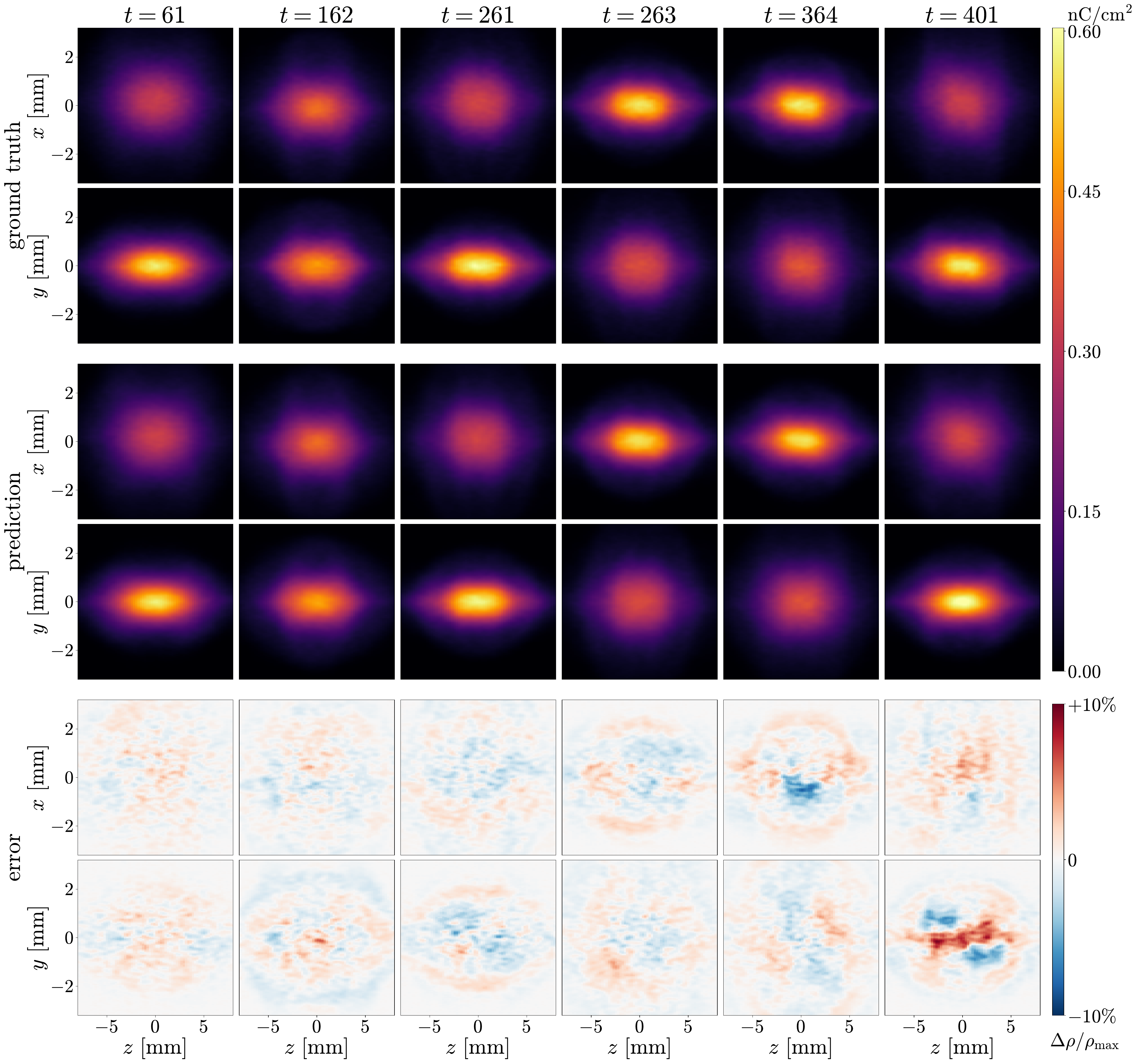}
    \caption{\small
    Bunched-beam Transport Run 392: longitudinal beam evolution. Ground-truth and predicted
    longitudinal projections are compared over the rollout, with relative density errors shown below.}
    \label{fig:bunched_run392_longitudinal}
\end{figure*}

\begin{figure*}[h]
    \centering
    \includegraphics[width=0.85\textwidth]{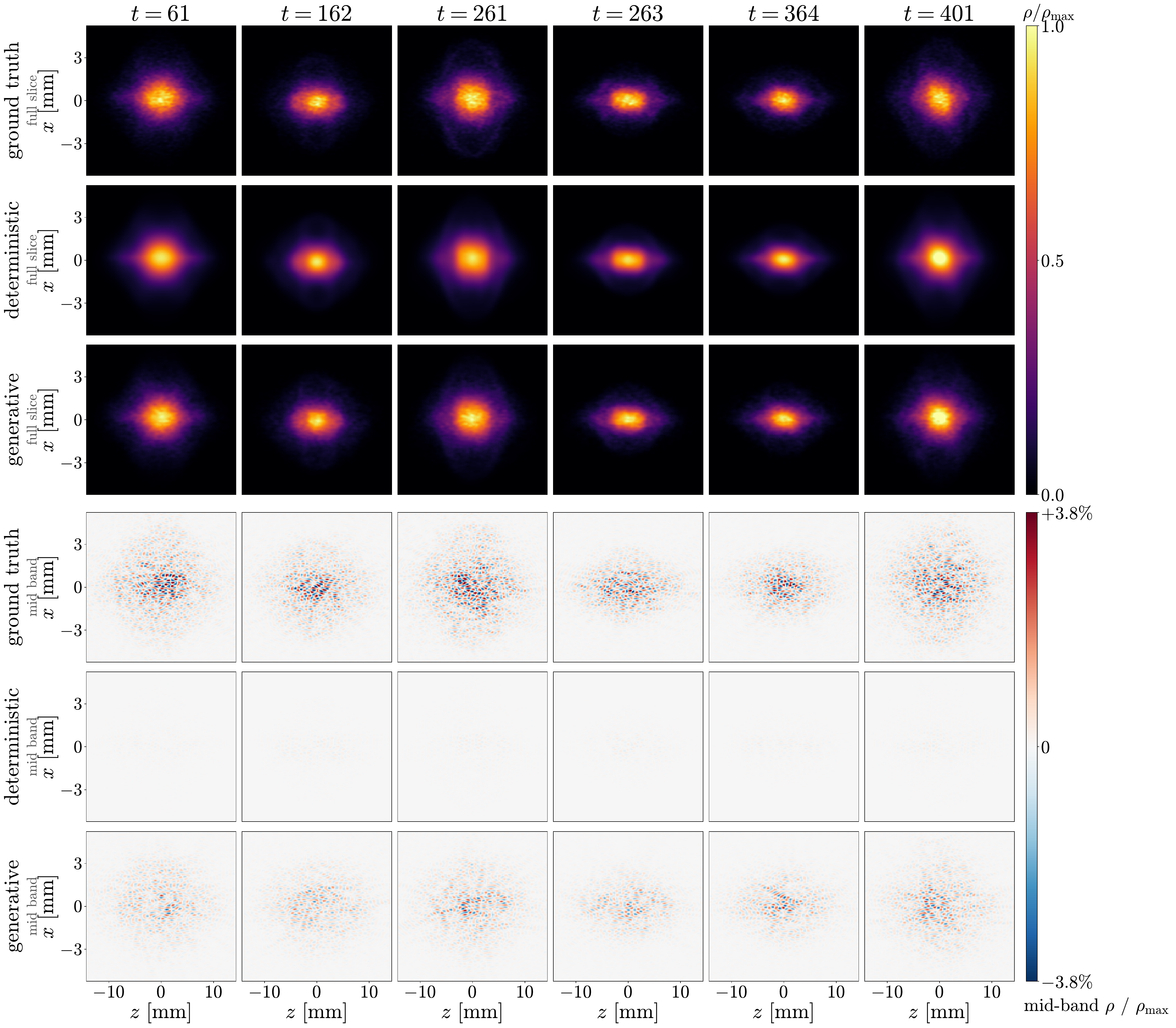}
    \caption{\small
    Bunched-beam Transport Run 392: fine-scale structure in the $y=0$ longitudinal slice.
    The deterministic prediction captures the bulk density evolution,
    while generative refinement recovers the realization-dependent
    mid-band structure observed in WarpX.}
    \label{fig:bunched_run392_texture}
\end{figure*}

\paragraph{Run 392: mismatched case.}
Run 392, visualized in Figures~\ref{fig:bunched_run392_transverse},~\ref{fig:bunched_run392_longitudinal}, and~\ref{fig:bunched_run392_texture} is the representative trajectory with relatively weak space charge but a $20\%$ initial envelope mismatch. The mismatch produces sustained transverse envelope oscillations, with $\sigma_x$ increasing by approximately $42\%$ over the rollout while $\sigma_z$ contracts by about $5\%$. GLOBE accurately follows the alternating transverse deformation and long-term envelope evolution. The deterministic prediction captures the bulk beam shape but smooths much of the realization-dependent mid-band structure. Generative residual refinement restores this texture while preserving the predicted envelope dynamics.

\begin{figure*}[h]
    \centering
    \includegraphics[width=\textwidth]{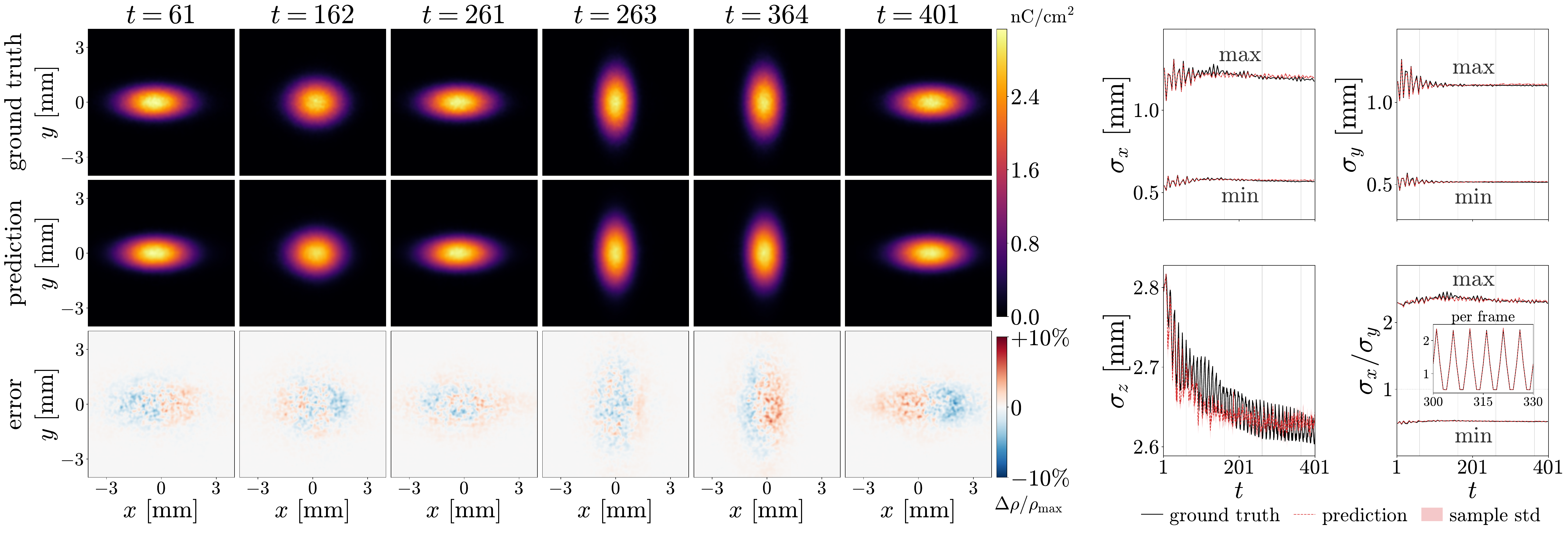}
    \caption{\small
    Bunched-beam Transport Run 244: transverse beam evolution. Ground-truth and predicted density projections are shown together with the evolution of
    $\sigma_x$, $\sigma_y$, $\sigma_z$, and $\sigma_x/\sigma_y$.
    }
    \label{fig:bunched_run244_transverse}
\end{figure*}

\begin{figure*}[h]
    \centering
    \includegraphics[width=0.85\textwidth]{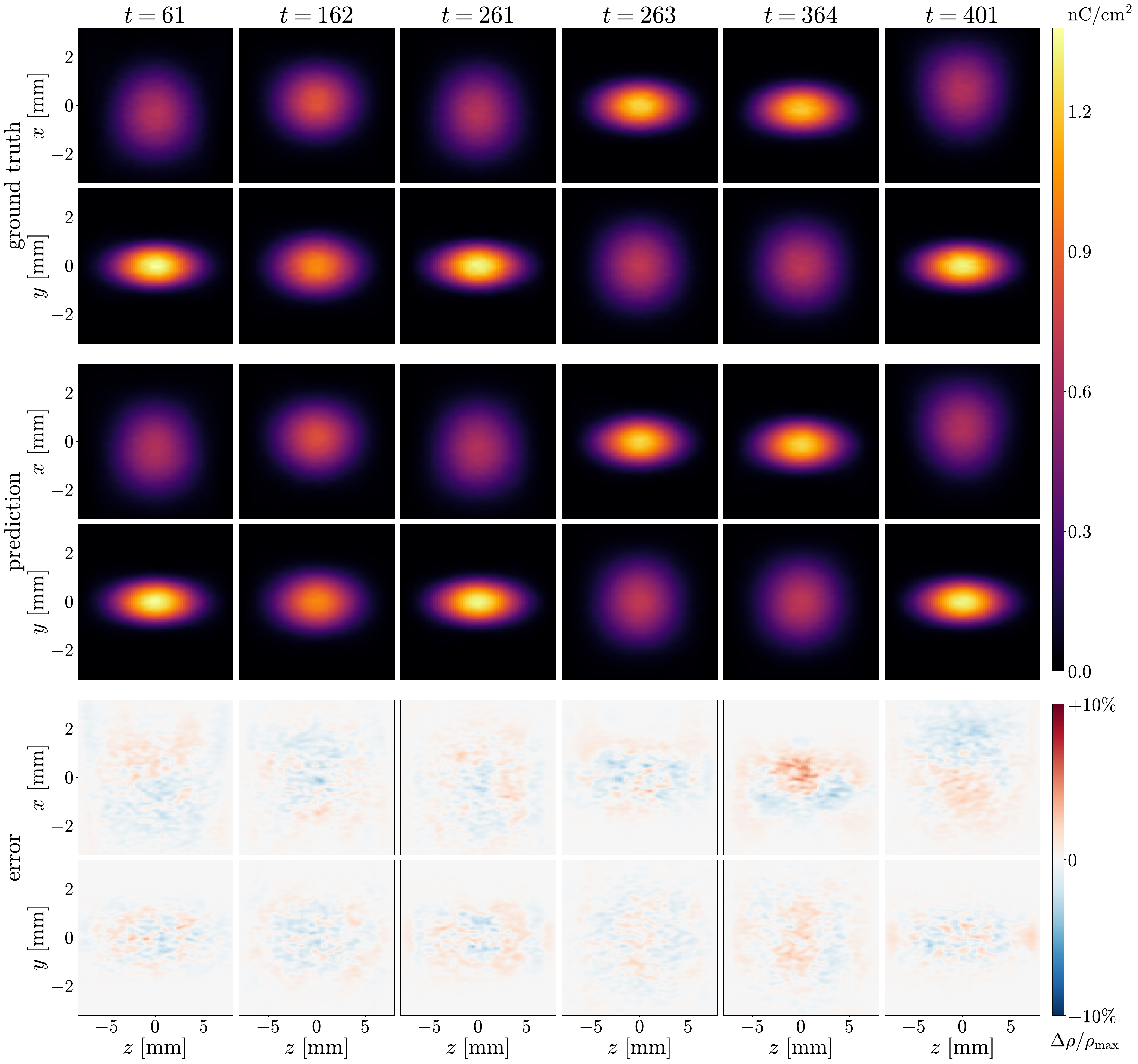}
    \caption{\small
    Bunched-beam Transport Run 244: longitudinal beam evolution. Ground-truth and predicted
    longitudinal projections are compared over the rollout, with relative density errors shown below.}
    \label{fig:bunched_run244_longitudinal}
\end{figure*}

\begin{figure*}[h]
    \centering
    \includegraphics[width=0.85\textwidth]{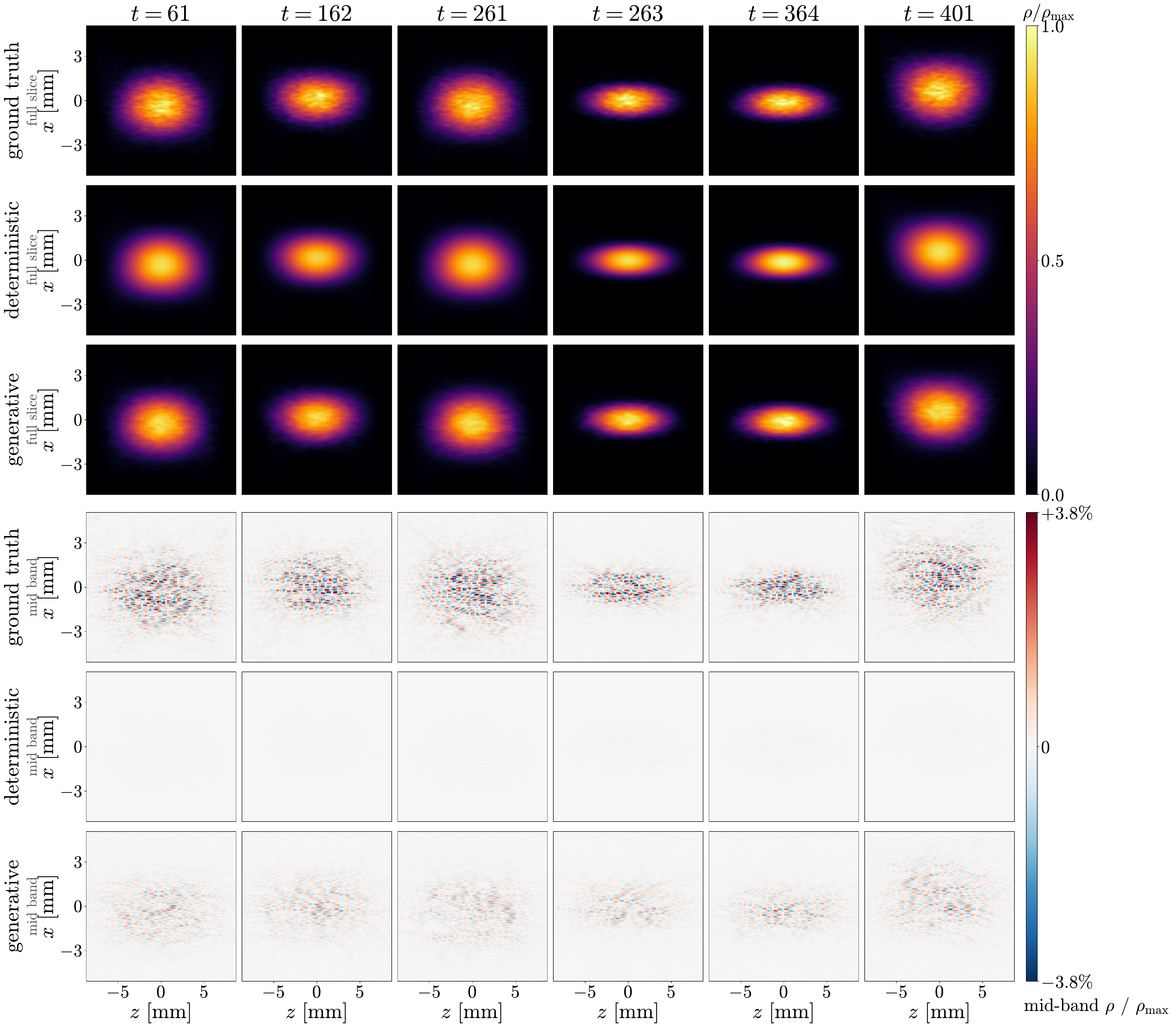}
    \caption{\small
    Bunched-beam Transport Run 244: fine-scale structure in the $y=0$ longitudinal slice.
    The deterministic prediction captures the bulk density evolution,
    while generative refinement recovers the realization-dependent
    mid-band structure observed in WarpX.}
    \label{fig:bunched_run244_texture}
\end{figure*}

\paragraph{Run 244: low-charge, strongly focused case.}
Run 244, visualized in Figures~\ref{fig:bunched_run244_transverse},~\ref{fig:bunched_run244_longitudinal}, and~\ref{fig:bunched_run244_texture} has the lowest bunch charge and strongest focusing among the selected cases, together with the largest initial transverse offset. The beam remains well confined while undergoing a pronounced exchange of transverse aspect ratio, with $\sigma_x/\sigma_y$ varying from approximately $2.3$ to $0.5$. GLOBE closely tracks this repeated $x$--$y$ deformation and the longitudinal beam size. The spatial comparisons show that the deterministic forecast recovers the smooth envelope but largely suppresses the mid-band PIC texture, whereas the generative residual restores fluctuations with spatial extent and amplitude similar to the reference.

\begin{figure*}[h]
    \centering
    \includegraphics[width=\textwidth]{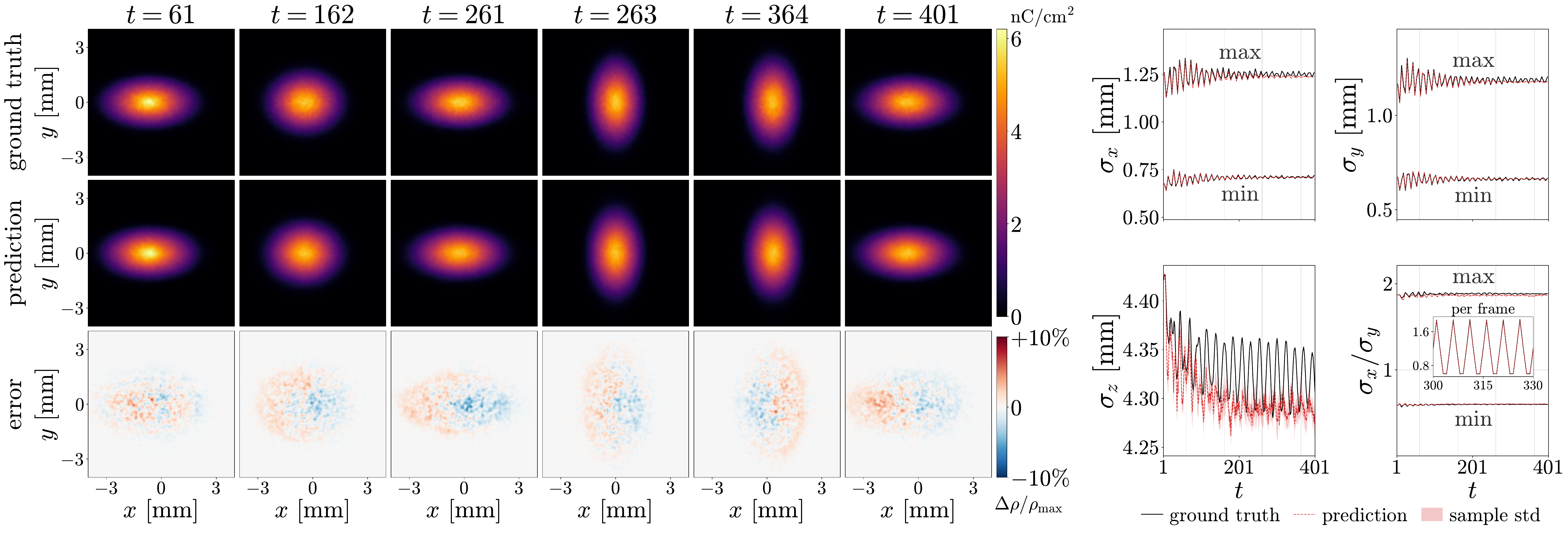}
    \caption{\small
    Bunched-beam Transport Run 389: transverse beam evolution. Ground-truth and predicted density projections are shown together with the evolution of
    $\sigma_x$, $\sigma_y$, $\sigma_z$, and $\sigma_x/\sigma_y$.
    }
    \label{fig:bunched_run389_transverse}
\end{figure*}

\begin{figure*}[h]
    \centering
    \includegraphics[width=0.85\textwidth]{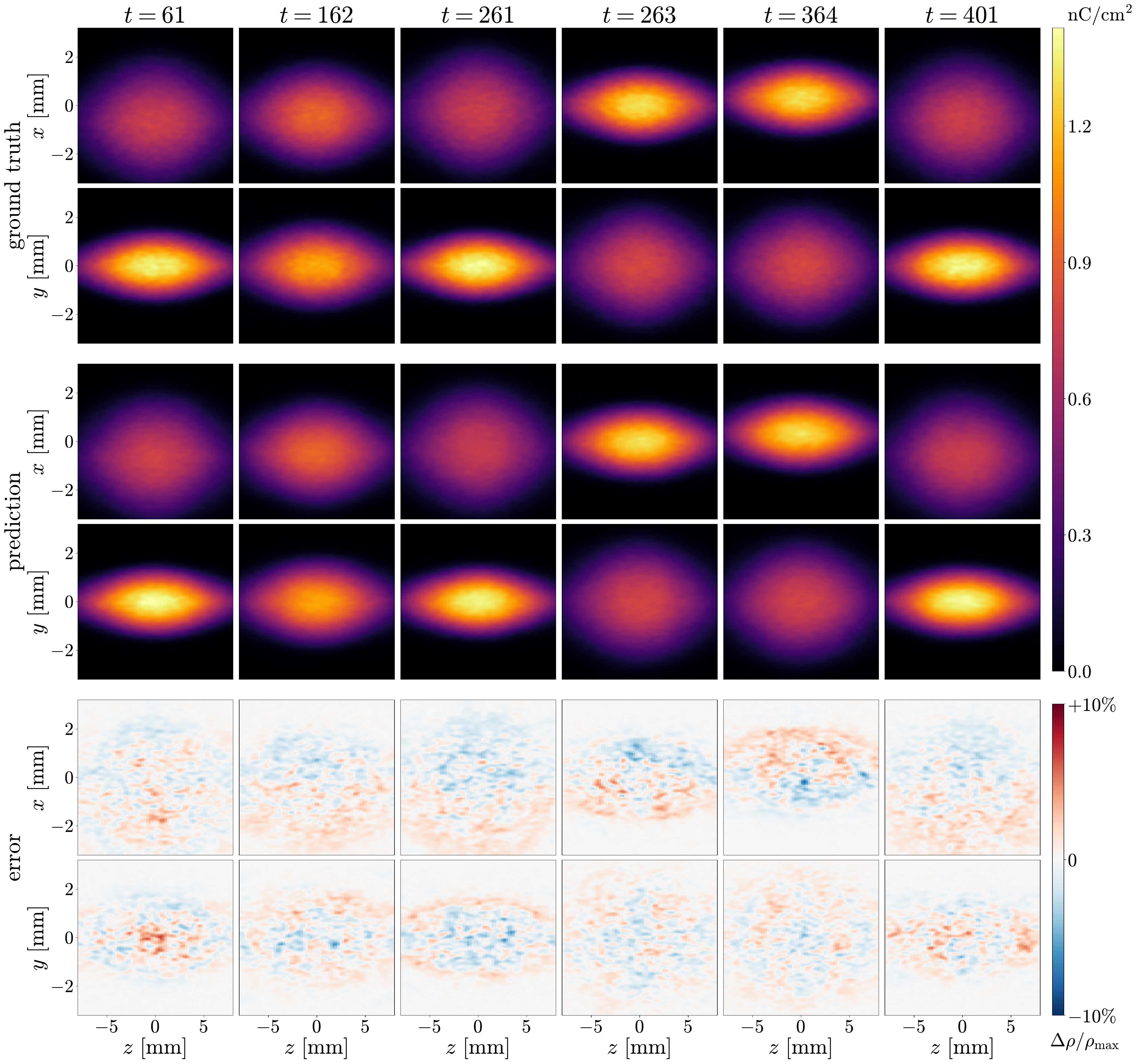}
    \caption{\small
    Bunched-beam Transport Run 389: longitudinal beam evolution. Ground-truth and predicted
    longitudinal projections are compared over the rollout, with relative density errors shown below.}
    \label{fig:bunched_run389_longitudinal}
\end{figure*}

\begin{figure*}[h]
    \centering
    \includegraphics[width=0.85\textwidth]{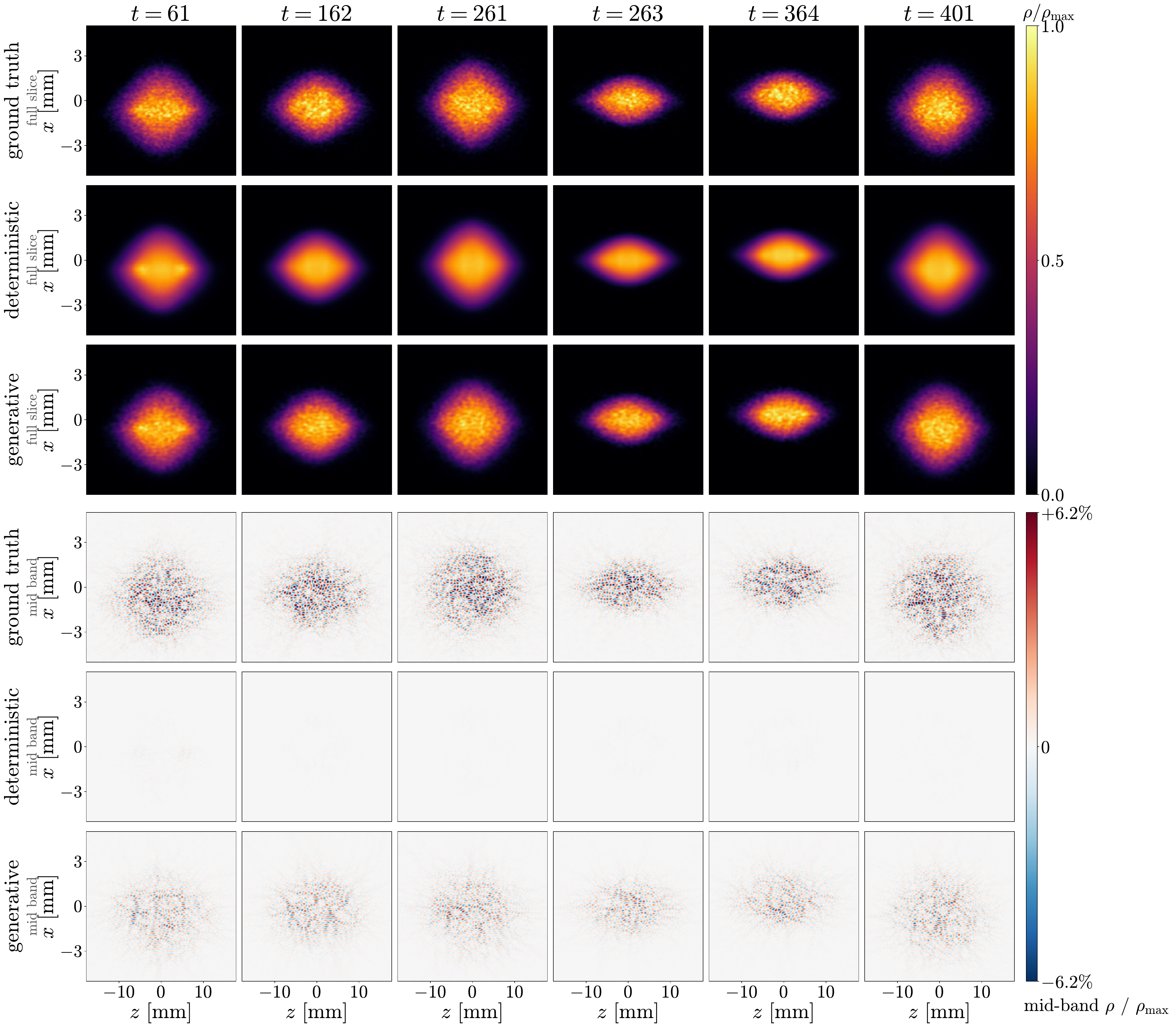}
    \caption{\small
    Bunched-beam Transport Run 389: fine-scale structure in the $y=0$ longitudinal slice.
    The deterministic prediction captures the bulk density evolution,
    while generative refinement recovers the realization-dependent
    mid-band structure observed in WarpX.}
    \label{fig:bunched_run389_texture}
\end{figure*}

\paragraph{Run 389: near-matched intermediate case.}
Run 389, visualized in Figures~\ref{fig:bunched_run389_transverse},~\ref{fig:bunched_run389_longitudinal}, and~\ref{fig:bunched_run389_texture} is the representative trajectory discussed in the main text, operating with intermediate space charge and a nearly matched initial beam. Its envelope changes only moderately, with roughly $5\%$ transverse growth over the full trajectory. GLOBE reproduces both the recurring transverse shape changes and the nearly stationary longitudinal extent. Although the deterministic prediction already captures the macroscopic evolution accurately, its density field is visibly smoother than the WarpX realization. The generative residual recovers much of the missing mid-band texture without altering the underlying envelope dynamics.

\begin{figure*}[h]
    \centering
    \includegraphics[width=\textwidth]{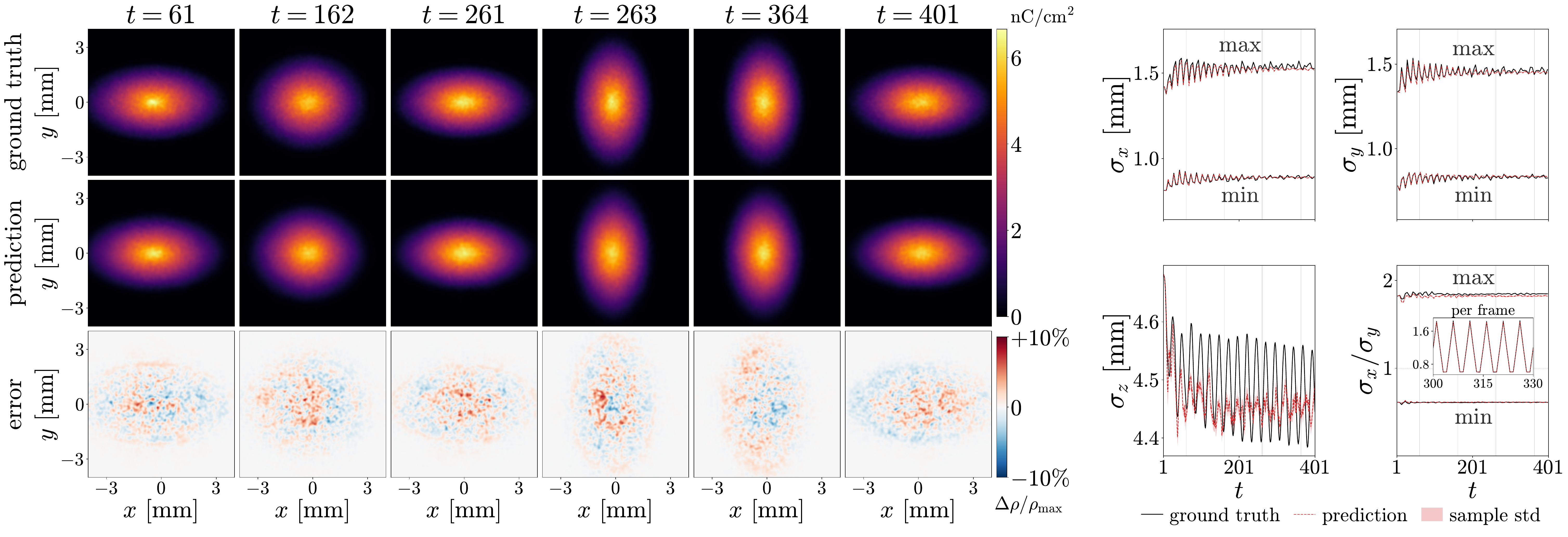}
    \caption{\small
    Bunched-beam Transport Run 87: transverse beam evolution. Ground-truth and predicted density projections are shown together with the evolution of
    $\sigma_x$, $\sigma_y$, $\sigma_z$, and $\sigma_x/\sigma_y$.
    }
    \label{fig:bunched_run87_transverse}
\end{figure*}

\begin{figure*}[h]
    \centering
    \includegraphics[width=0.85\textwidth]{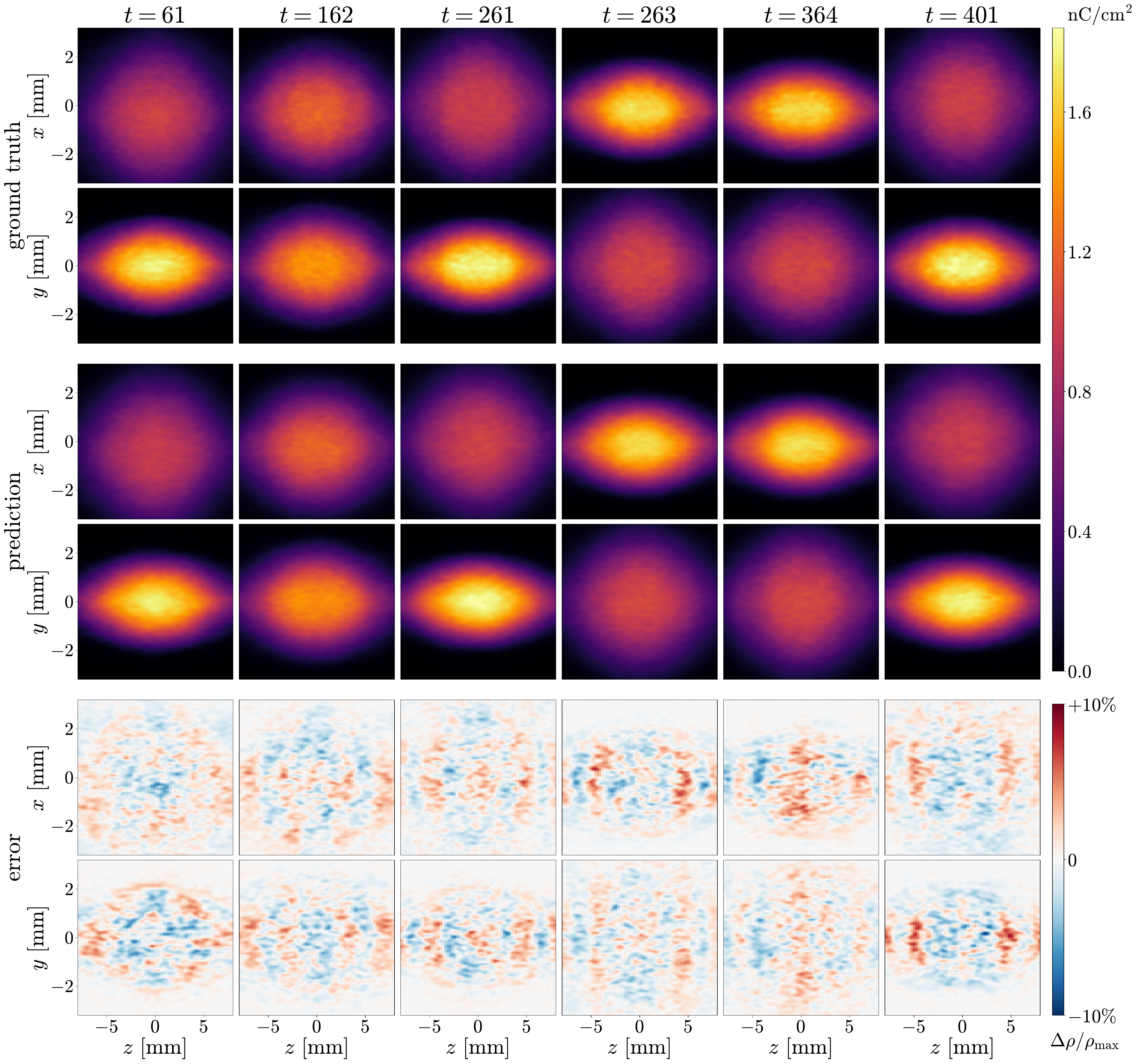}
    \caption{\small
    Bunched-beam Transport Run 87: longitudinal beam evolution. Ground-truth and predicted
    longitudinal projections are compared over the rollout, with relative density errors shown below.}
    \label{fig:bunched_run87_longitudinal}
\end{figure*}

\begin{figure*}[h]
    \centering
    \includegraphics[width=0.85\textwidth]{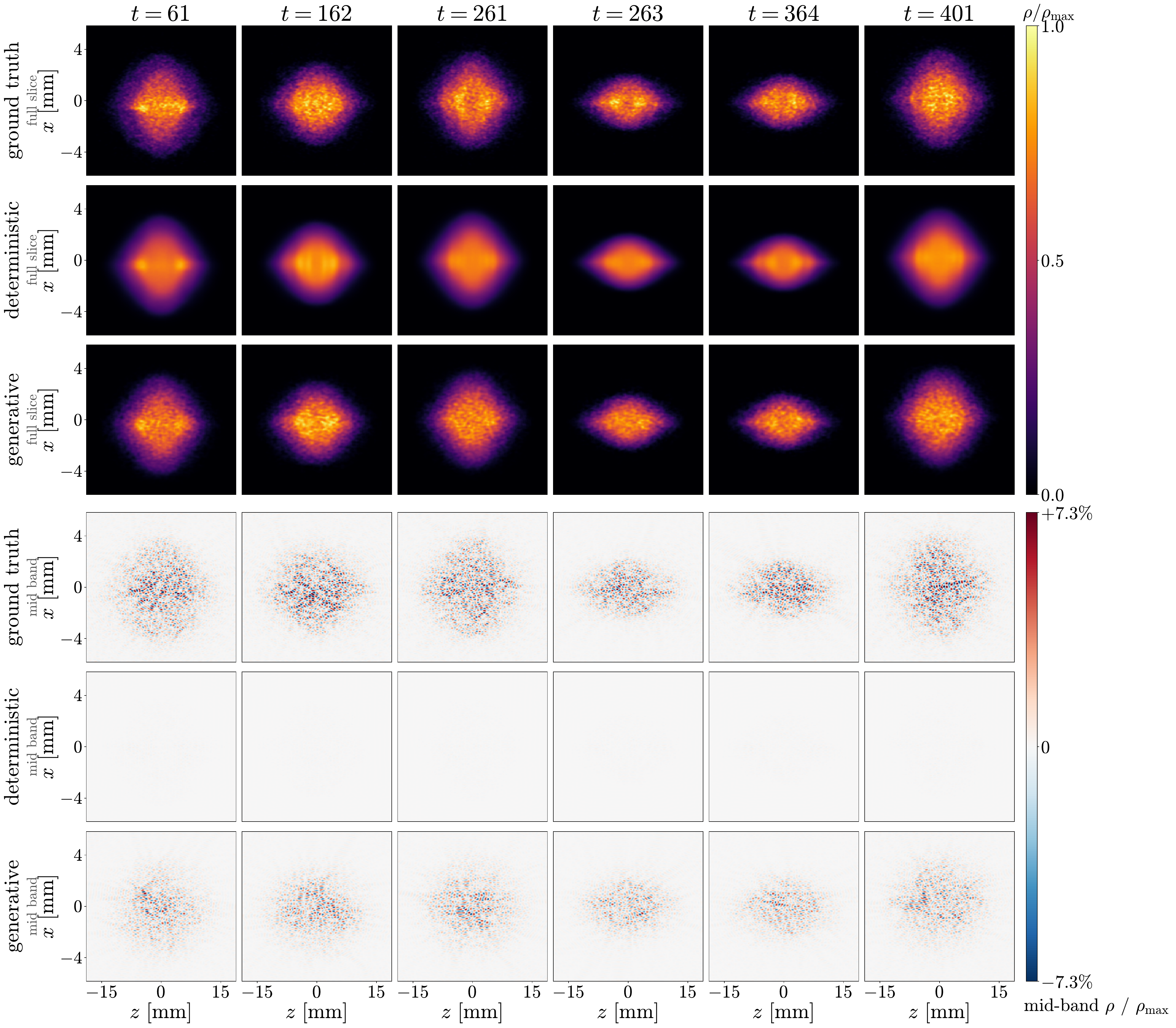}
    \caption{\small
    Bunched-beam Transport Run 87: fine-scale structure in the $y=0$ longitudinal slice.
    The deterministic prediction captures the bulk density evolution,
    while generative refinement recovers the realization-dependent
    mid-band structure observed in WarpX.}
    \label{fig:bunched_run87_texture}
\end{figure*}

\paragraph{Run 87: high-charge, longitudinally extended case.}
Run 87, visualized in Figures~\ref{fig:bunched_run87_transverse},~\ref{fig:bunched_run87_longitudinal}, and~\ref{fig:bunched_run87_texture}, has the largest bunch charge and weakest RF confinement among the selected trajectories, producing the longest bunch ($\sigma_z\approx4.5\,\mathrm{mm}$) and stronger space-charge effects. Despite this more challenging operating point, GLOBE tracks the transverse oscillations and longitudinal extent accurately. The projected densities remain accurate throughout the rollout. As in the lower-charge cases, generative refinement restores realization-dependent mid-band structure that is largely absent from the deterministic prediction.

\begin{figure*}[h]
    \centering
    \includegraphics[width=\textwidth]{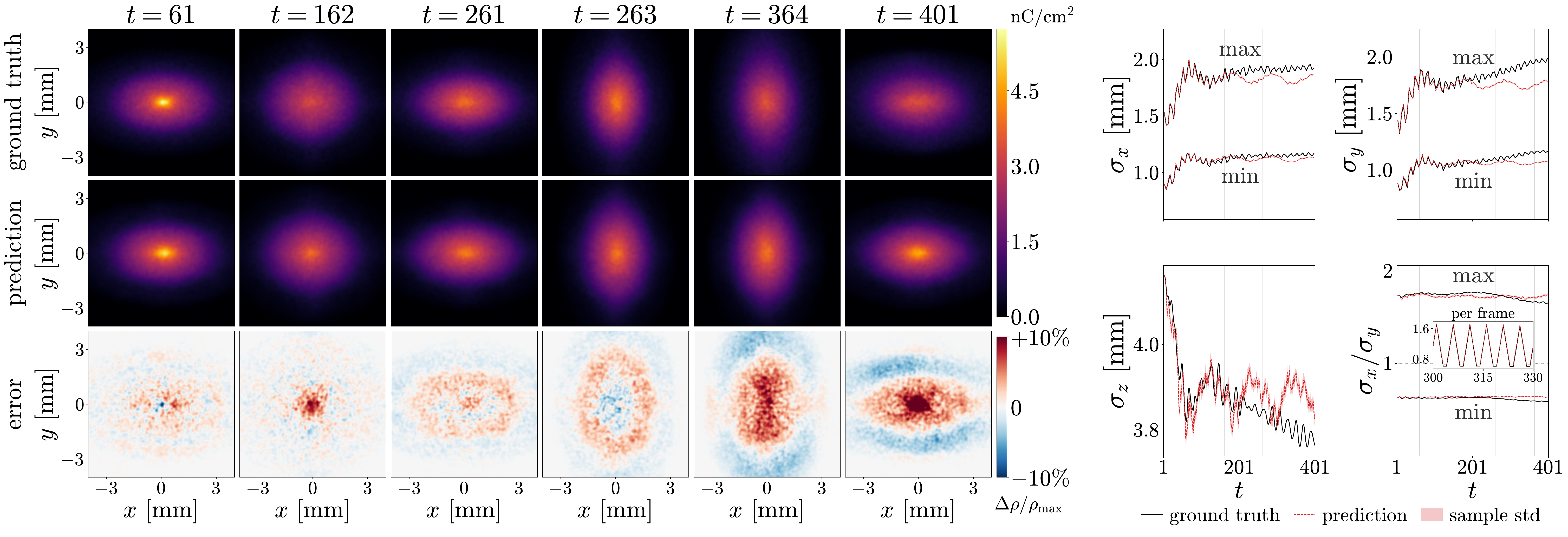}
    \caption{\small
    Bunched-beam Transport Run 343: transverse beam evolution. Ground-truth and predicted density projections are shown together with the evolution of
    $\sigma_x$, $\sigma_y$, $\sigma_z$, and $\sigma_x/\sigma_y$.
    }
    \label{fig:bunched_run343_transverse}
\end{figure*}

\begin{figure*}[h]
    \centering
    \includegraphics[width=0.85\textwidth]{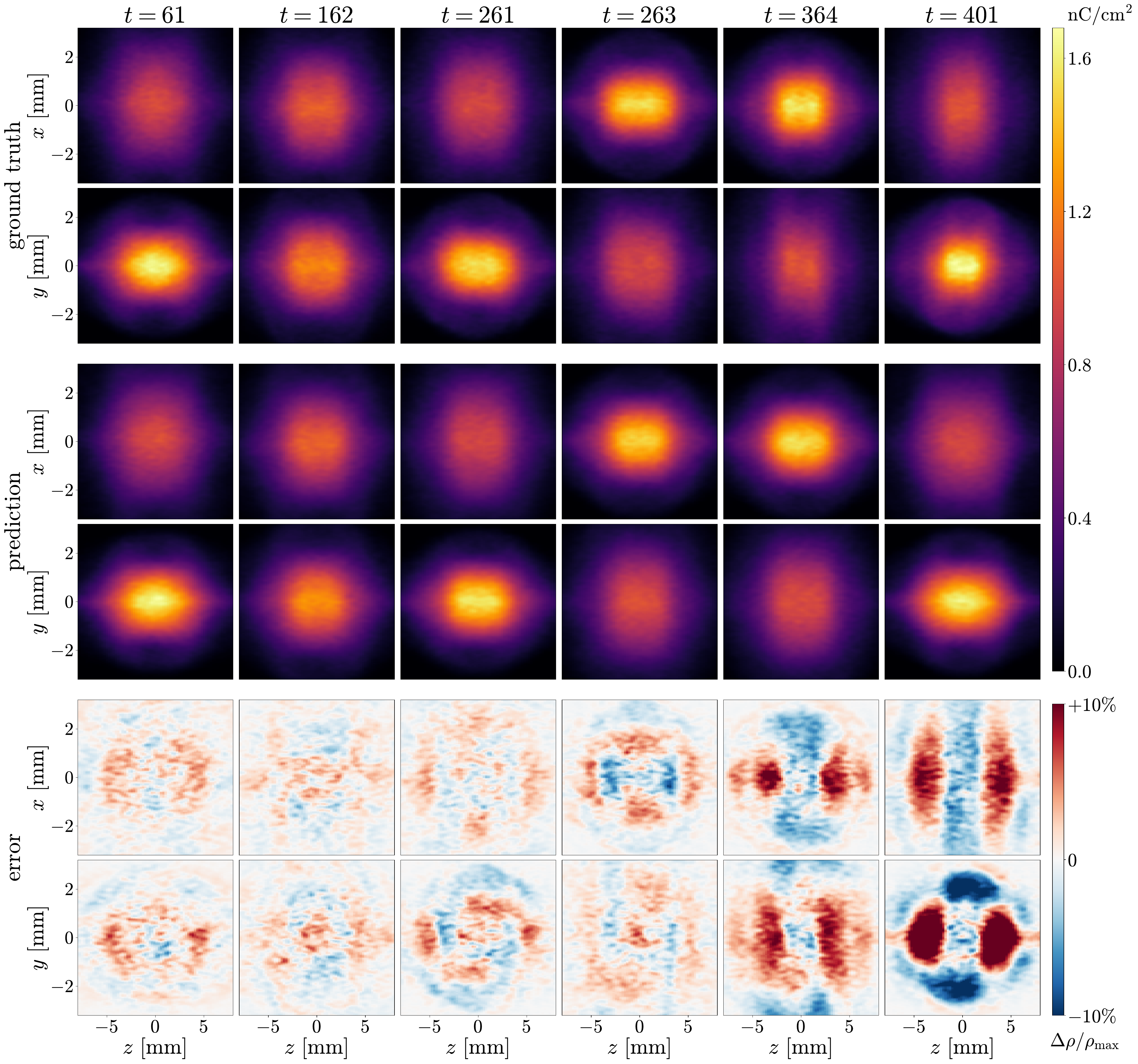}
    \caption{\small
    Bunched-beam Transport Run 343: longitudinal beam evolution. Ground-truth and predicted
    longitudinal projections are compared over the rollout, with relative density errors shown below.}
    \label{fig:bunched_run343_longitudinal}
\end{figure*}

\begin{figure*}[h]
    \centering
    \includegraphics[width=0.85\textwidth]{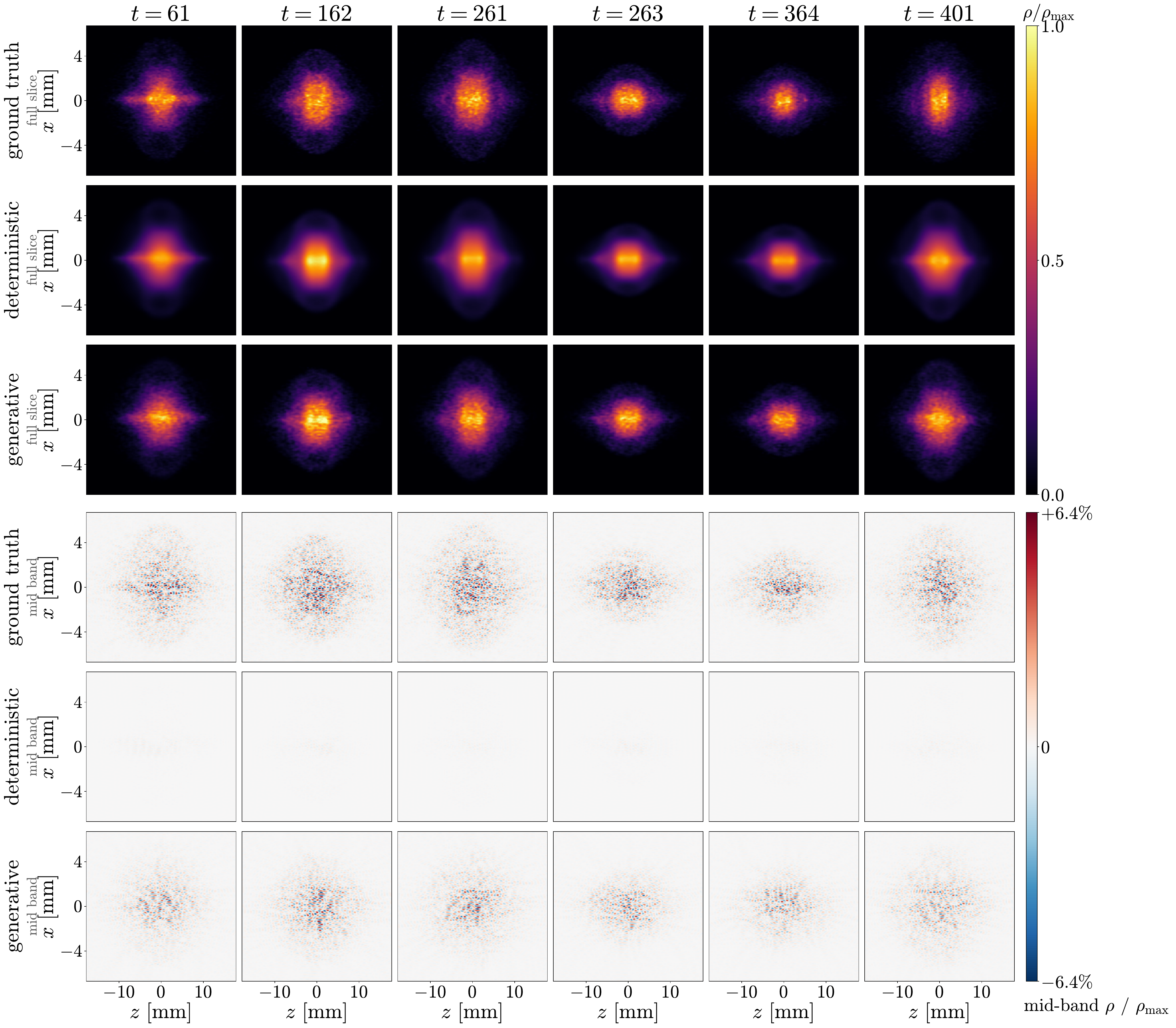}
    \caption{\small
    Bunched-beam Transport Run 343: fine-scale structure in the $y=0$ longitudinal slice.
    The deterministic prediction captures the bulk density evolution,
    while generative refinement recovers the realization-dependent
    mid-band structure observed in WarpX.}
    \label{fig:bunched_run343_texture}
\end{figure*}

\paragraph{Run 343: strong space-charge case.}
Run 343, visualized in Figures~\ref{fig:bunched_run343_transverse},~\ref{fig:bunched_run343_longitudinal}, and~\ref{fig:bunched_run343_texture} is the most challenging selected trajectory. Strong space charge drives approximately $35\%$ transverse envelope growth together with $8.5\%$ longitudinal compression, producing substantial redistribution among the beam dimensions. GLOBE captures the overall expansion, compression, and alternating transverse geometry, although deviations accumulate at later times. The density errors are consequently larger and more spatially organized than in the preceding cases. Nevertheless, the generative residual continues to recover a substantial fraction of the mid-band spatial structure that is suppressed by the deterministic forecast, showing that the refinement remains effective even in the strongest space-charge regime considered.

\subsection{PWFA Visualization}
\label{app:PWFA-visualization}

We also present additional PWFA test trajectories to complement the main-text example. For each case, we provide three complementary visual diagnostics. First, particle-and-wake snapshots together with trajectory-level observables evaluate the evolution of the accelerating field and bubble geometry. Second, at a representative intermediate frame, we compare the ground truth, prediction, and error for $E_z$, $E_x$, $cB_y$, and the derived transverse wakefield $W_x=E_x-cB_y$. While $E_z$ determines longitudinal acceleration, $W_x$ is the effective transverse field acting on an ultra-relativistic witness and therefore directly probes whether the model captures the focusing structure of the wake. Finally, channel-wise relative-error maps at representative times show how prediction errors evolve across the dominant field and density variables. Together, these figures illustrate how GLOBE captures wake formation, beam loading, transverse focusing, and bubble evolution across regimes ranging from quasi-linear operation to strongly nonlinear blowout.

\begin{figure*}[t]
    \centering
    \includegraphics[width=\textwidth]{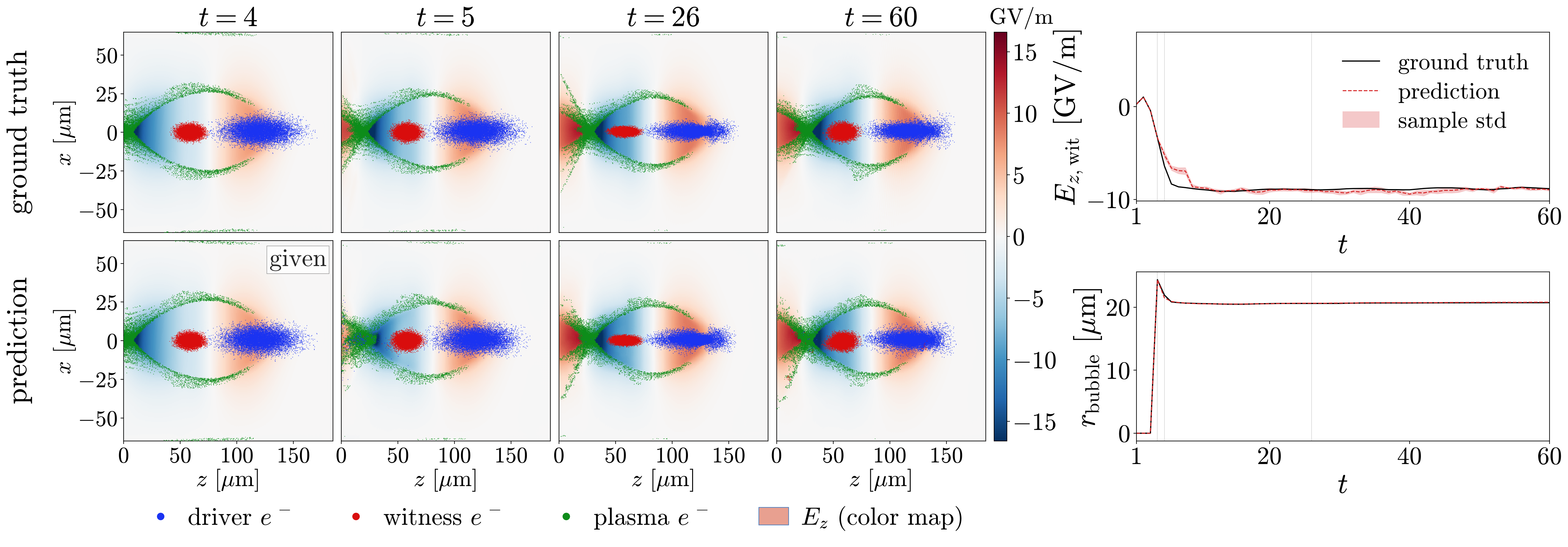}
    \caption{\small PWFA Run 312: wake evolution in the $x$--$z$ plane together with trajectory-level observables. The color map shows $E_z$; particles are overlaid by species. Right: predicted and reference observables with the ensemble sample deviation.}
    \label{fig:pwfa_run312_diag}
\end{figure*}

\begin{figure*}[t]
    \centering
    \includegraphics[width=0.88\textwidth]{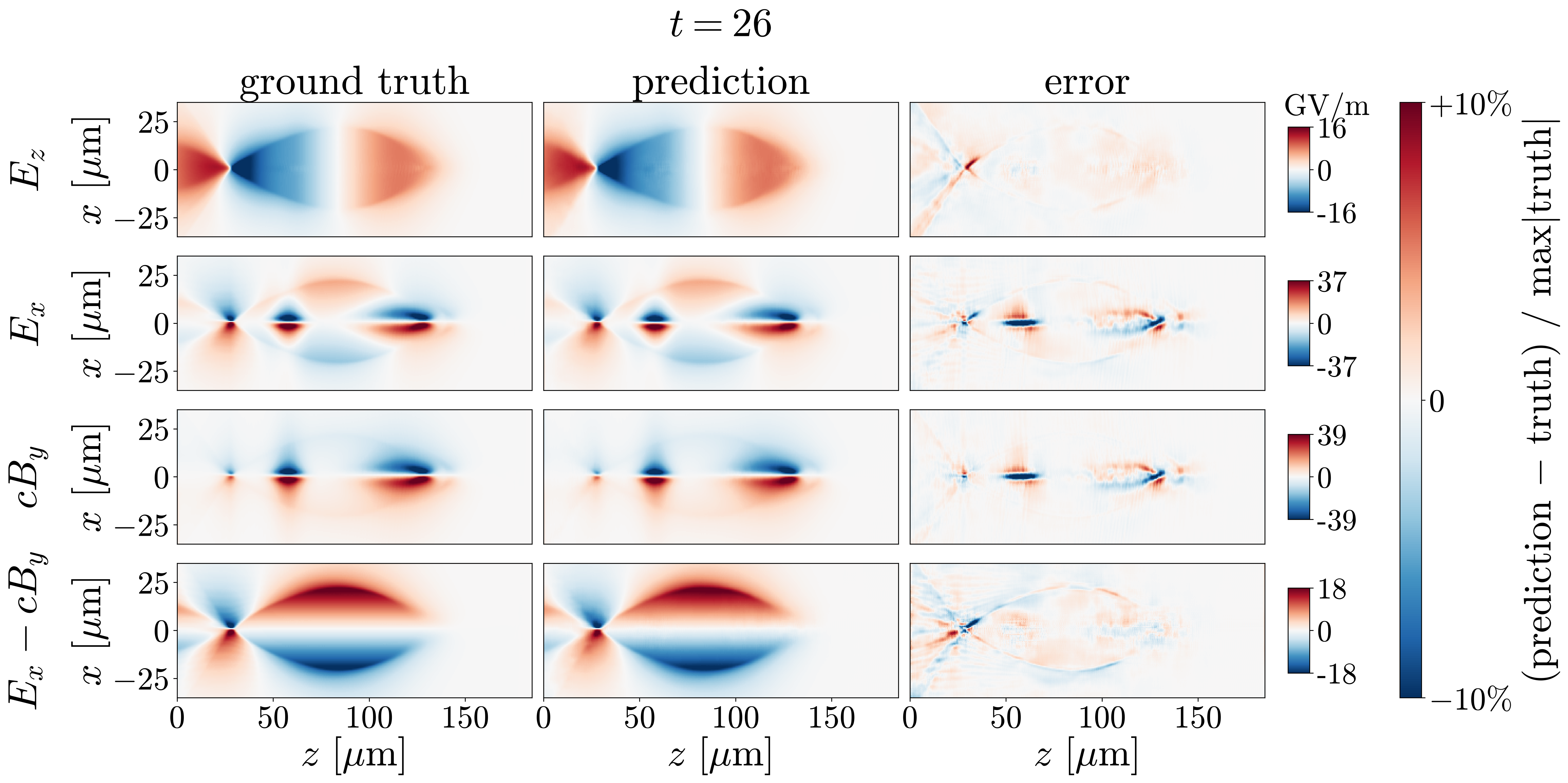}
    \caption{\small PWFA Run 312: detailed comparison at $t=25$ for representative field quantities. Left to right: ground truth, prediction, and error.}
    \label{fig:pwfa_run312_field}
\end{figure*}

\begin{figure*}[t]

    \centering
    \includegraphics[width=\textwidth]{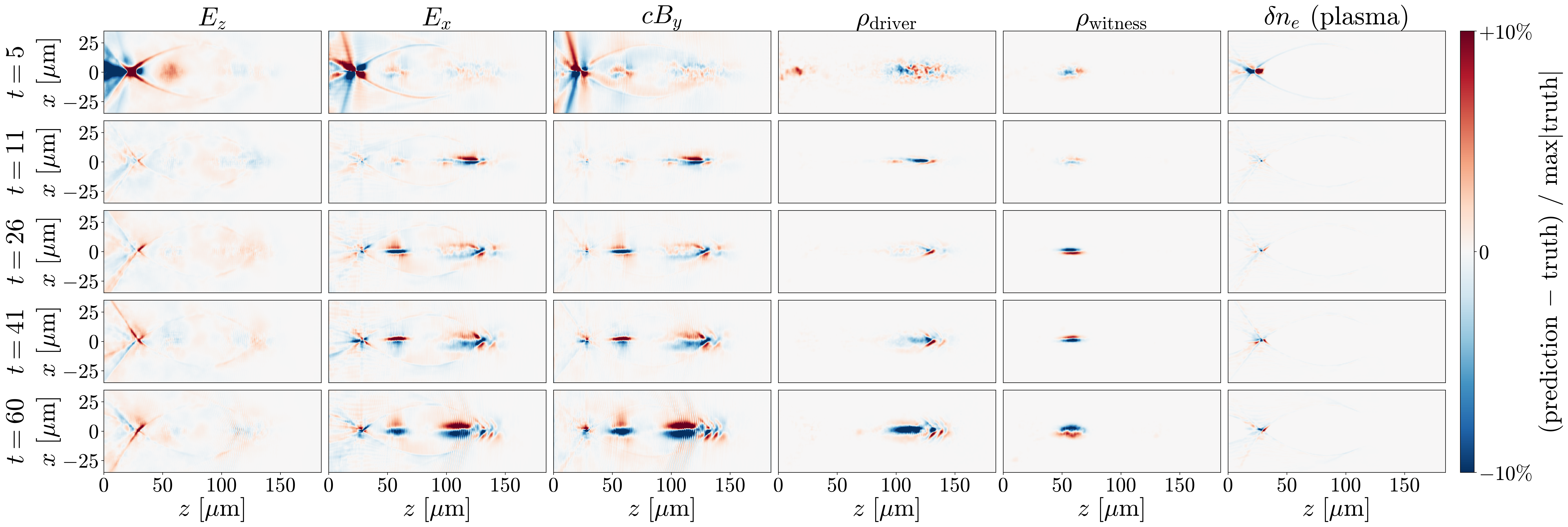}
    \caption{\small PWFA Run 312: relative-error maps for representative channels at selected times. Colors show $(\mathrm{prediction}-\mathrm{truth})/\max|\mathrm{truth}|$.}
    \label{fig:pwfa_run312_err}
\end{figure*}

\paragraph{Run 312 (main-text case).}
Run~312, visualized in Figures~\ref{fig:pwfa_run312_diag},~\ref{fig:pwfa_run312_field}, and~\ref{fig:pwfa_run312_err}, is the main-text reference case, with a moderately nonlinear blowout wake, near-optimal beam loading ($\eta_{\mathrm{load}}=1.03$), and a slightly offset, relatively wide driver. The particle-and-wake rollout shows close agreement in the evolving bubble geometry and witness location. $E_{z,\mathrm{wit}}$ follows the WarpX trajectory after the initial transient, while $r_{\mathrm{bubble}}$ remains particularly accurate. At $t=25$, GLOBE reproduces the large-scale structure of $E_z$, $E_x$, and $cB_y$, as well as their derived combination $W_x=E_x-cB_y$, indicating that both the accelerating field and transverse focusing structure are captured. The channel-wise error maps show that discrepancies remain concentrated around the driver, witness, and thin plasma-sheath structures rather than the bulk wake, even at late times. Over the rollout, the whole-field Rel-$L_2$ errors are $0.129/0.132/0.144$ for the deterministic, ensemble-mean, and sampled predictions.

\begin{figure*}[t]
    \centering
    \includegraphics[width=\textwidth]{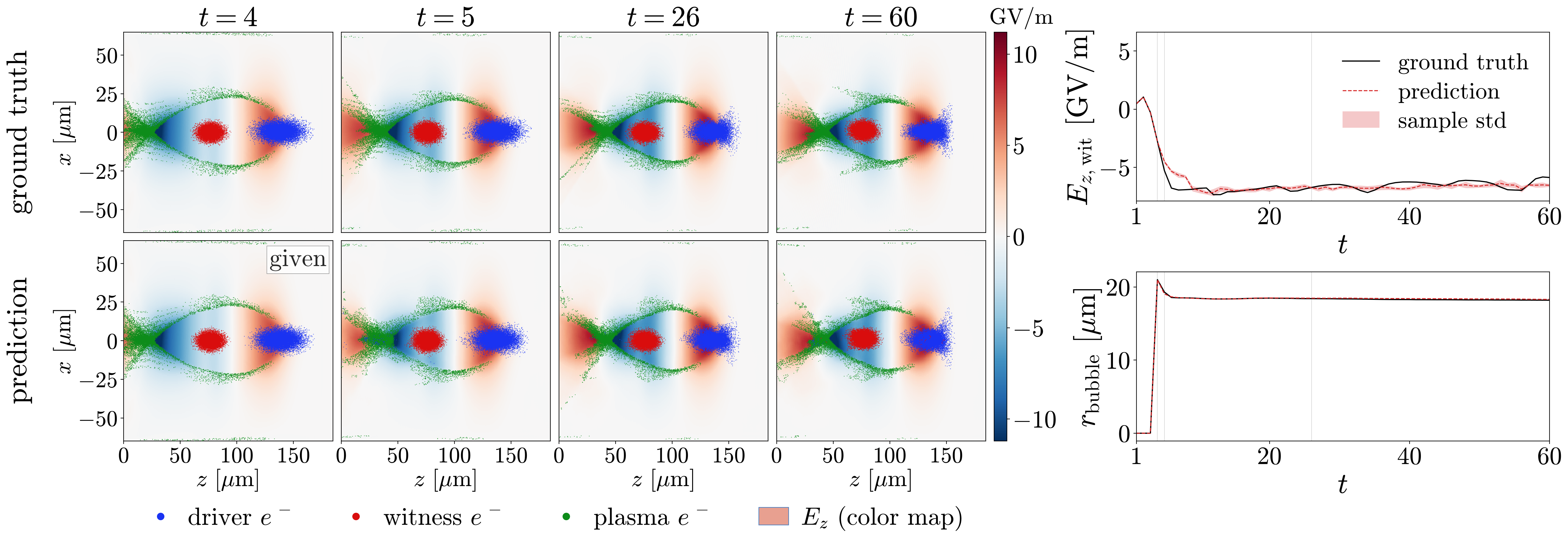}
    \caption{\small PWFA Run 355: wake evolution in the $x$--$z$ plane together with trajectory-level observables. The color map shows $E_z$; particles are overlaid by species. Right: predicted and reference observables with the ensemble sample deviation.}
    \label{fig:pwfa_run355_diag}
\end{figure*}

\begin{figure*}[t]
    \centering
    \includegraphics[width=0.88\textwidth]{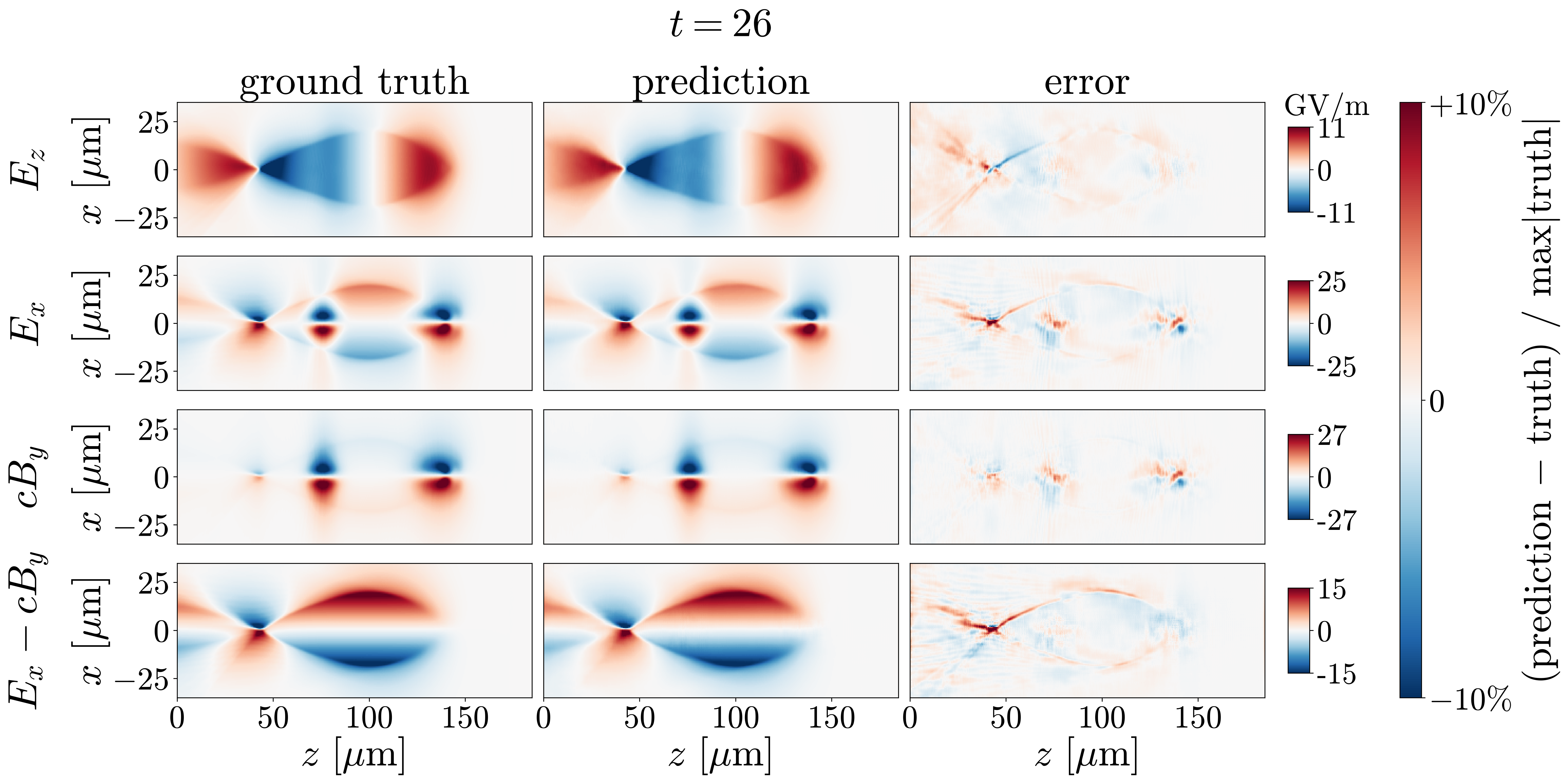}
    \caption{\small PWFA Run 355: detailed comparison at $t=25$ for representative field quantities. Left to right: ground truth, prediction, and error.}
    \label{fig:pwfa_run355_field}
\end{figure*}

\begin{figure*}[t]

    \centering
    \includegraphics[width=\textwidth]{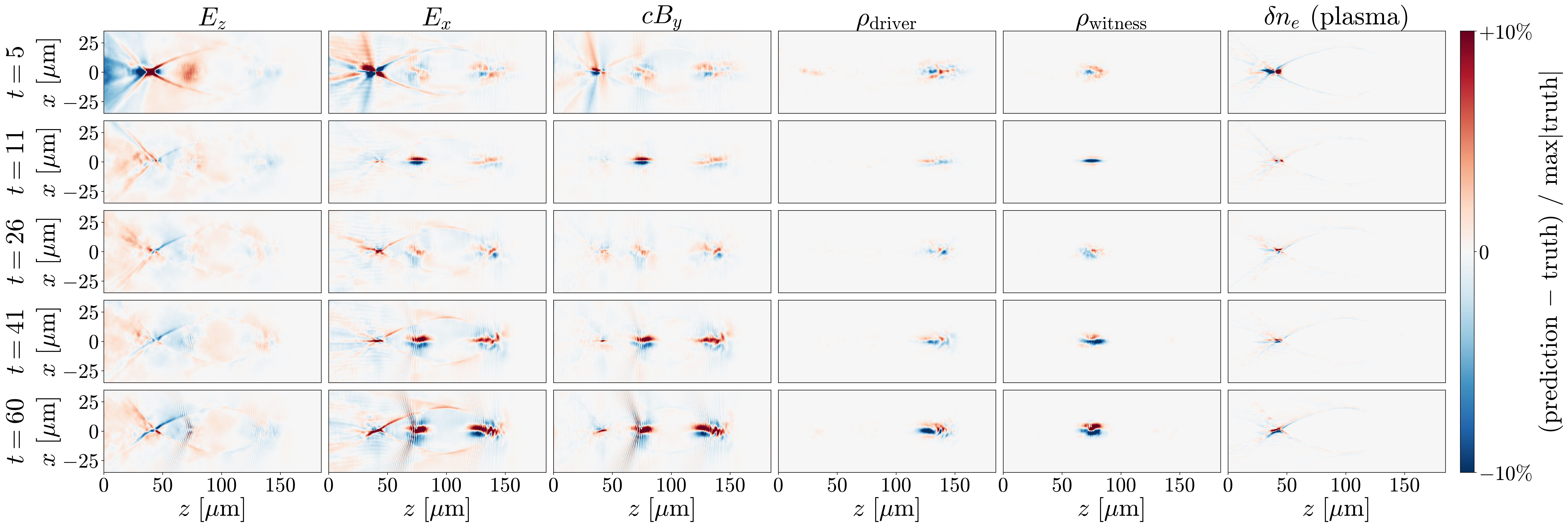}
    \caption{\small PWFA Run 355: relative-error maps for representative channels at selected times. Colors show $(\mathrm{prediction}-\mathrm{truth})/\max|\mathrm{truth}|$.}
    \label{fig:pwfa_run355_err}
\end{figure*}

\paragraph{Run 355: overloaded, strongly depleted driver case.}
Run 355, visualized in Figures~\ref{fig:pwfa_run355_diag},~\ref{fig:pwfa_run355_field}, and~\ref{fig:pwfa_run355_err}, combines an overloaded witness ($\eta_{\mathrm{load}}=1.23$) with the lowest driver energy among the selected cases, producing the strongest driver depletion (about $14\%$). The macroparticle-and-wake evolution shows that GLOBE captures the loaded cavity, witness acceleration, and progressive driver evolution, with only modest drift in $E_{z,\mathrm{wit}}$ and little error in the bubble radius. At $t=25$, the predicted $E_z$, $E_x$, and $cB_y$ retain the correct phase and spatial organization, while $W_x=E_x-cB_y$ confirms that the transverse focusing structure is preserved despite the stronger beam loading. The channel-wise error maps show that the largest discrepancies develop near the rear of the cavity and around the witness, where the loading most strongly reshapes the wake.

\begin{figure*}[t]
    \centering
    \includegraphics[width=\textwidth]{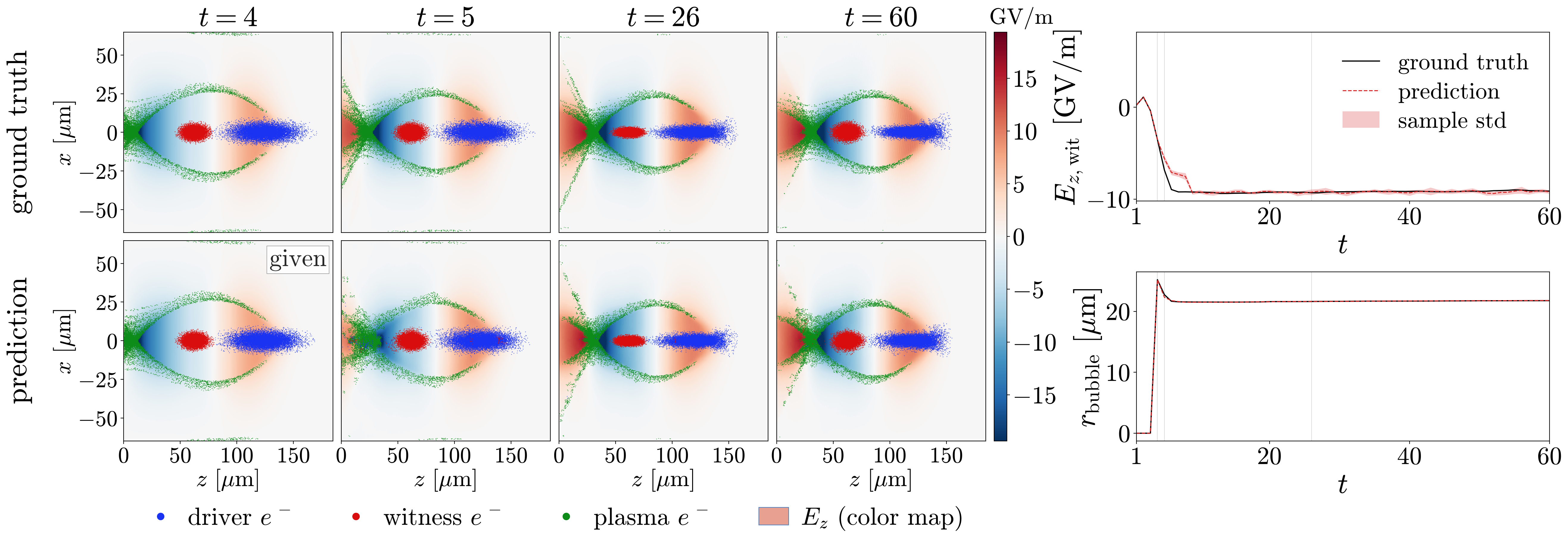}
    \caption{\small PWFA Run 271: wake evolution in the $x$--$z$ plane together with trajectory-level observables. The color map shows $E_z$; particles are overlaid by species. Right: predicted and reference observables with the ensemble sample deviation.}
    \label{fig:pwfa_run271_diag}
\end{figure*}

\begin{figure*}[t]
    \centering
    \includegraphics[width=0.88\textwidth]{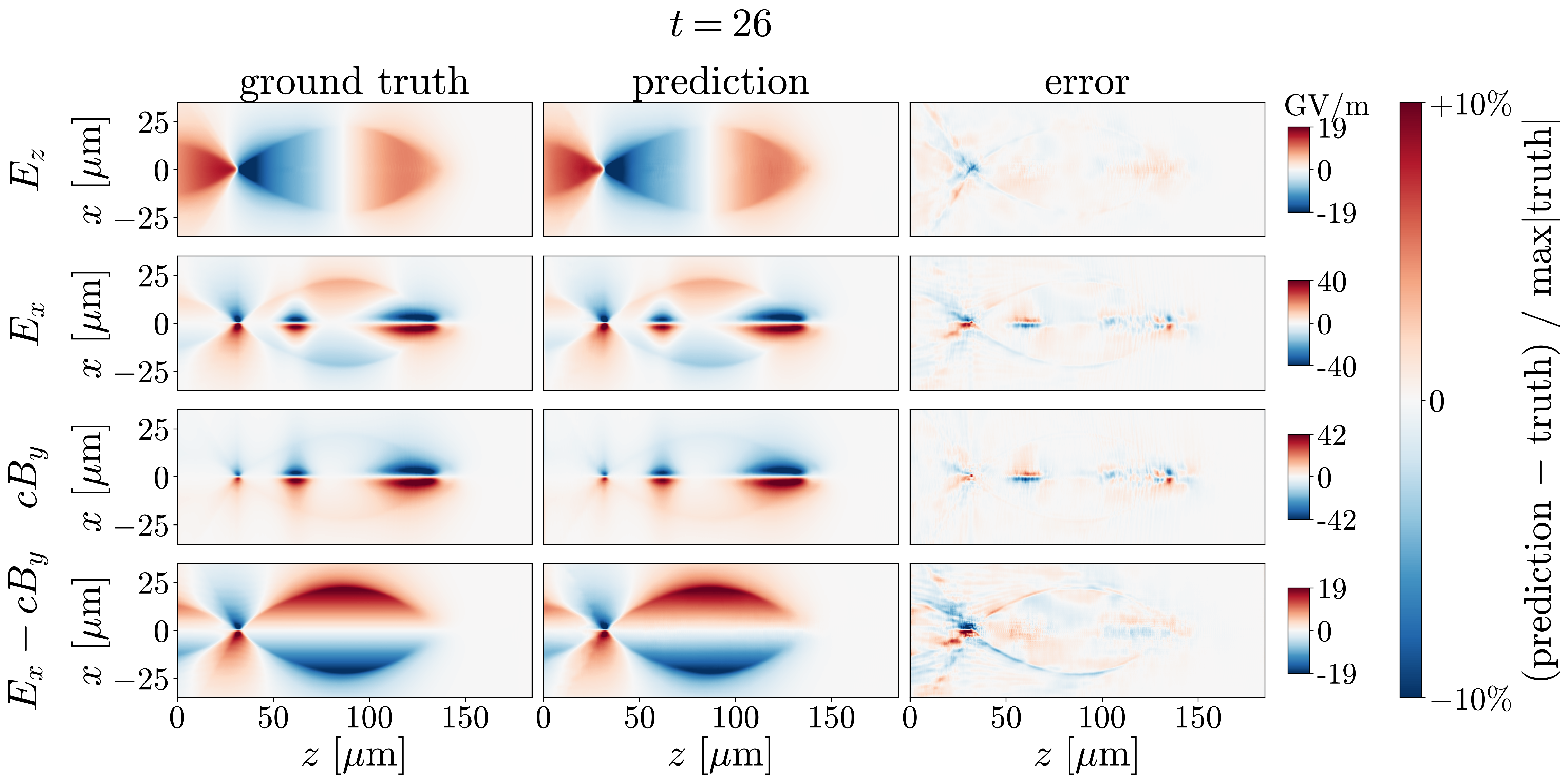}
    \caption{\small PWFA Run 271: detailed comparison at $t=25$ for representative field quantities. Left to right: ground truth, prediction, and error.}
    \label{fig:pwfa_run271_field}
\end{figure*}

\begin{figure*}[t]

    \centering
    \includegraphics[width=\textwidth]{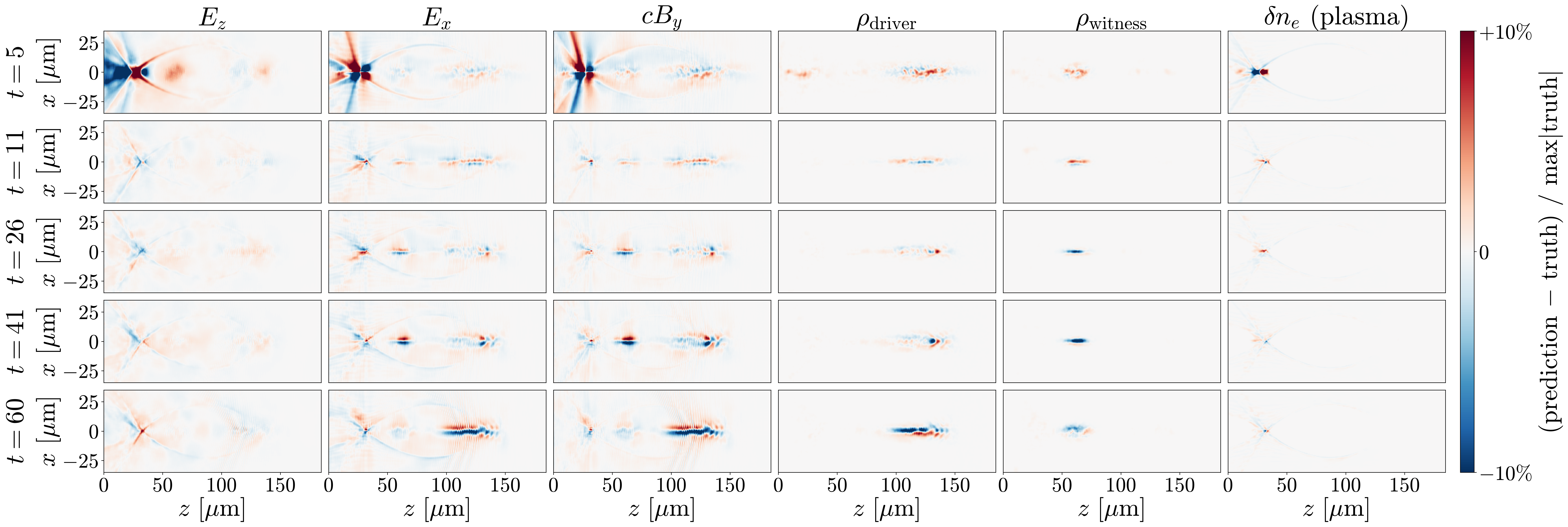}
    \caption{\small PWFA Run 271: relative-error maps for representative channels at selected times. Colors show $(\mathrm{prediction}-\mathrm{truth})/\max|\mathrm{truth}|$.}
    \label{fig:pwfa_run271_err}
\end{figure*}

\paragraph{Run 271: high-charge, underloaded axisymmetric wake case.}
Run 271, visualized in Figures~\ref{fig:pwfa_run271_diag},~\ref{fig:pwfa_run271_field}, and~\ref{fig:pwfa_run271_err}, has the largest normalized driver charge ($\Lambda=0.40$), weak transverse offset, and an underloaded witness ($\eta_{\mathrm{load}}=0.72$), producing the strongest peak field among the selected cases. The macroparticle-and-wake rollout shows a clean, nearly axisymmetric blowout, with GLOBE closely reproducing the driver, witness, plasma sheath, and bubble evolution throughout propagation. At $t=25$, the model accurately captures the strong accelerating $E_z$ field and the opposing $E_x$ and $cB_y$ contributions; their combination $W_x$ preserves the transverse focusing structure experienced by the witness. The relative-error maps remain localized mainly near the driver and rear sheath, while the dominant cavity structure is maintained through the rollout.

\begin{figure*}[t]
    \centering
    \includegraphics[width=\textwidth]{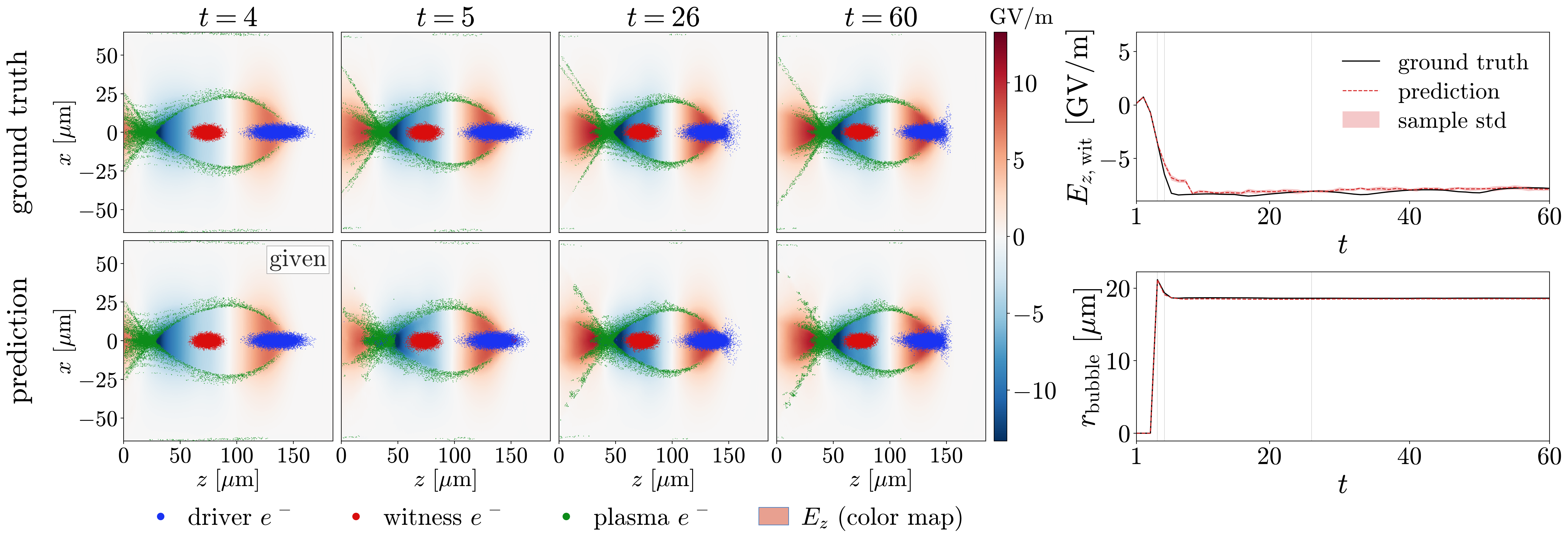}
    \caption{\small PWFA Run 27: wake evolution in the $x$--$z$ plane together with trajectory-level observables. The color map shows $E_z$; particles are overlaid by species. Right: predicted and reference observables with the ensemble sample deviation.}
    \label{fig:pwfa_run27_diag}
\end{figure*}

\begin{figure*}[t]
    \centering
    \includegraphics[width=0.88\textwidth]{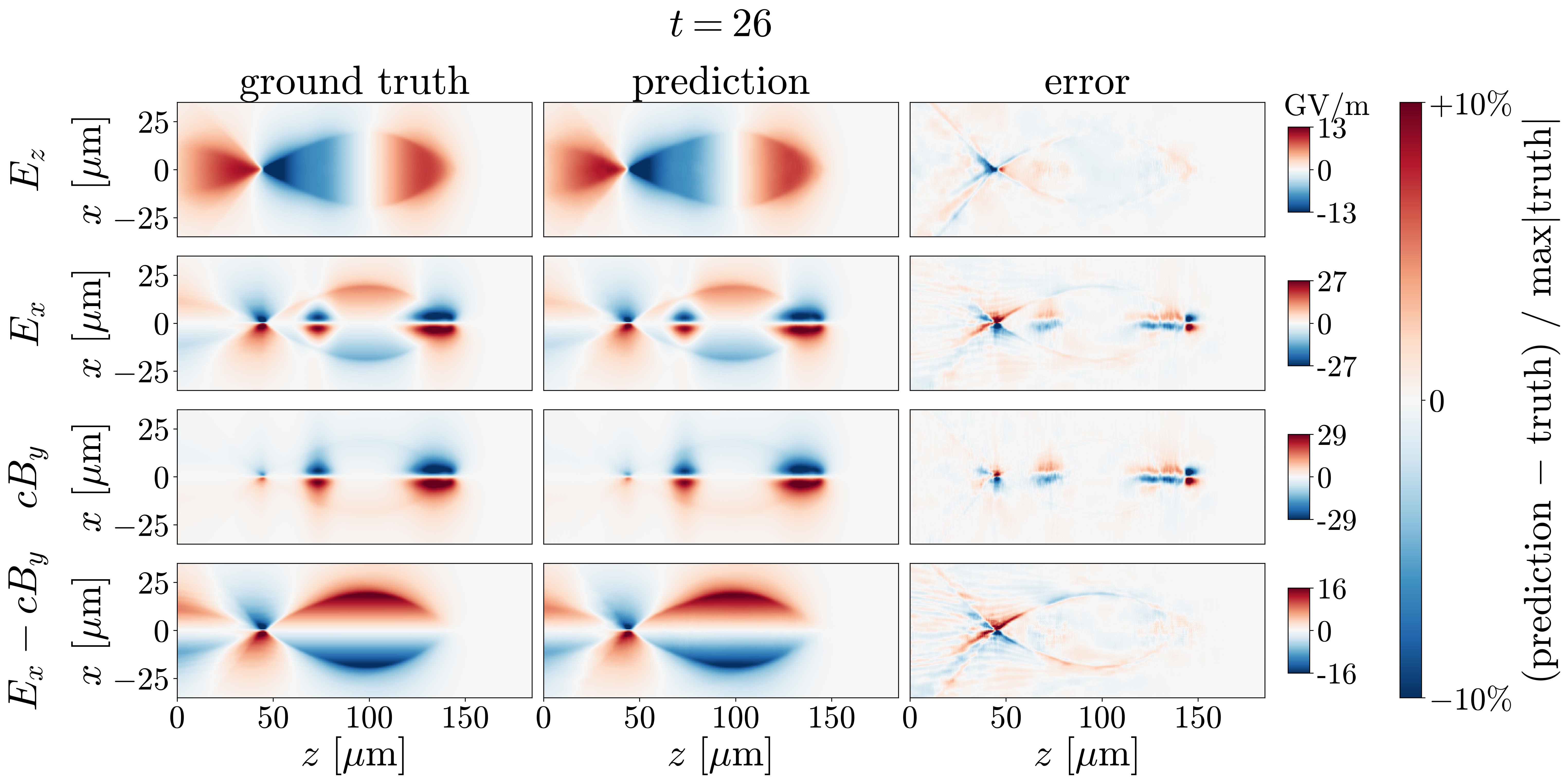}
    \caption{\small PWFA Run 27: detailed comparison at $t=25$ for representative field quantities. Left to right: ground truth, prediction, and error.}
    \label{fig:pwfa_run27_field}
\end{figure*}

\begin{figure*}[t]

    \centering
    \includegraphics[width=\textwidth]{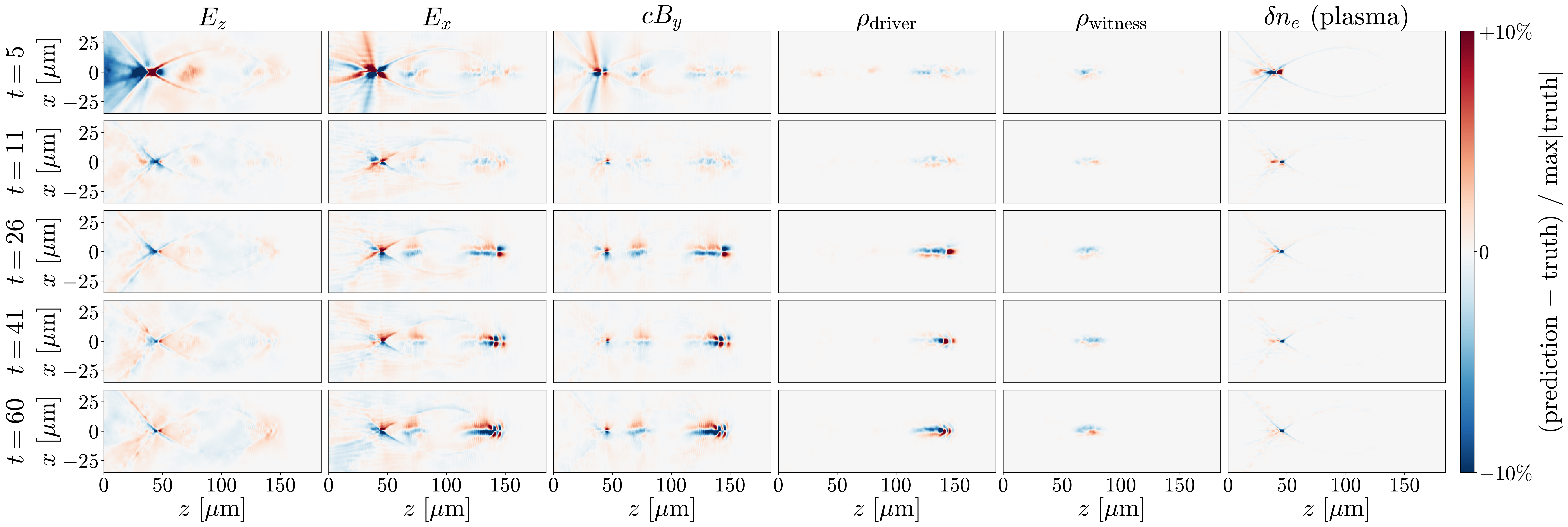}
    \caption{\small PWFA Run 27: relative-error maps for representative channels at selected times. Colors show $(\mathrm{prediction}-\mathrm{truth})/\max|\mathrm{truth}|$.}
    \label{fig:pwfa_run27_err}
\end{figure*}

\paragraph{Run 27: high-energy, lightly loaded wake case.}
Run 27, visualized in Figures~\ref{fig:pwfa_run27_diag},~\ref{fig:pwfa_run27_field}, and~\ref{fig:pwfa_run27_err}, uses the highest driver energy in the selected set together with a narrower-than-matched driver and relatively light witness loading ($\eta_{\mathrm{load}}=0.86$). The macroparticle-and-wake sequence shows stable propagation of the bubble and witness, with the predicted $E_{z,\mathrm{wit}}$ following the reference trend and the bubble radius remaining nearly indistinguishable from WarpX. At $t=25$, GLOBE reproduces the longitudinal accelerating structure in $E_z$ as well as the transverse $E_x$ and $cB_y$ fields; their combination $W_x$ remains close to the reference, indicating that the focusing force is reproduced in addition to the accelerating field. The multi-time error maps show that the remaining discrepancies are confined primarily to the narrow driver/witness regions and fine plasma-sheath structure. 

\begin{figure*}[t]
    \centering
    \includegraphics[width=\textwidth]{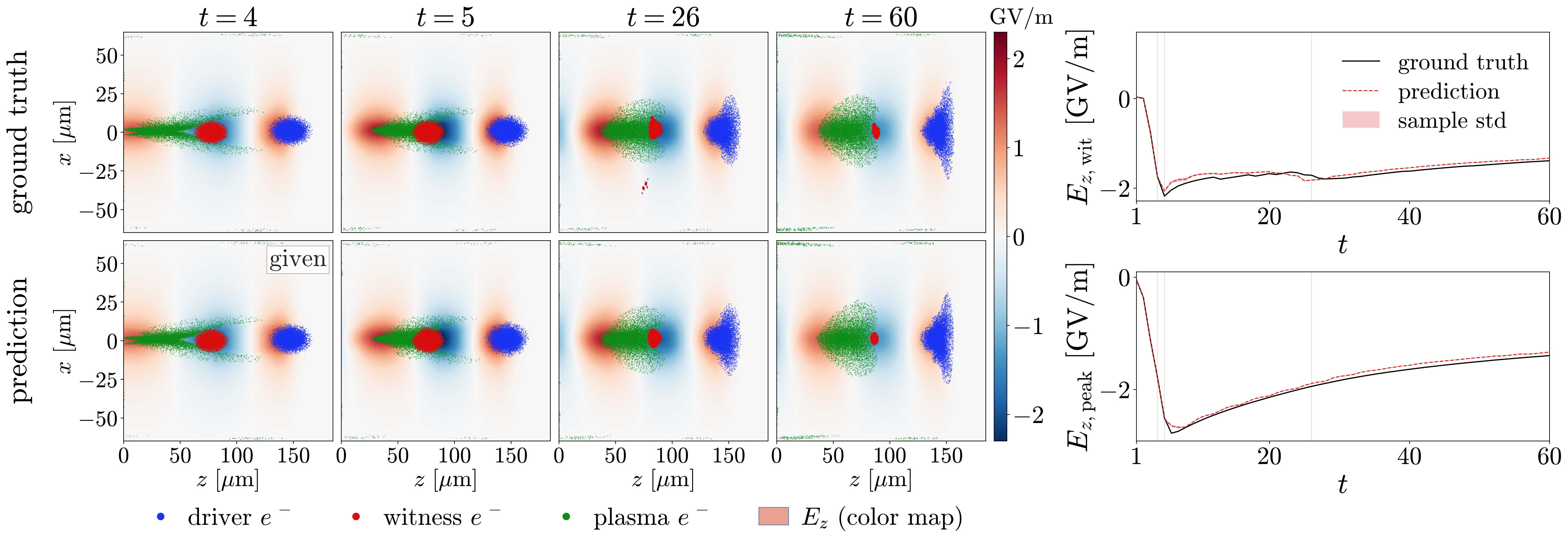}
    \caption{\small PWFA Run 118: wake evolution in the $x$--$z$ plane together with trajectory-level observables. The color map shows $E_z$; particles are overlaid by species. Right: predicted and reference observables with the ensemble sample deviation.}
    \label{fig:pwfa_run118_diag}
\end{figure*}

\begin{figure*}[t]
    \centering
    \includegraphics[width=0.88\textwidth]{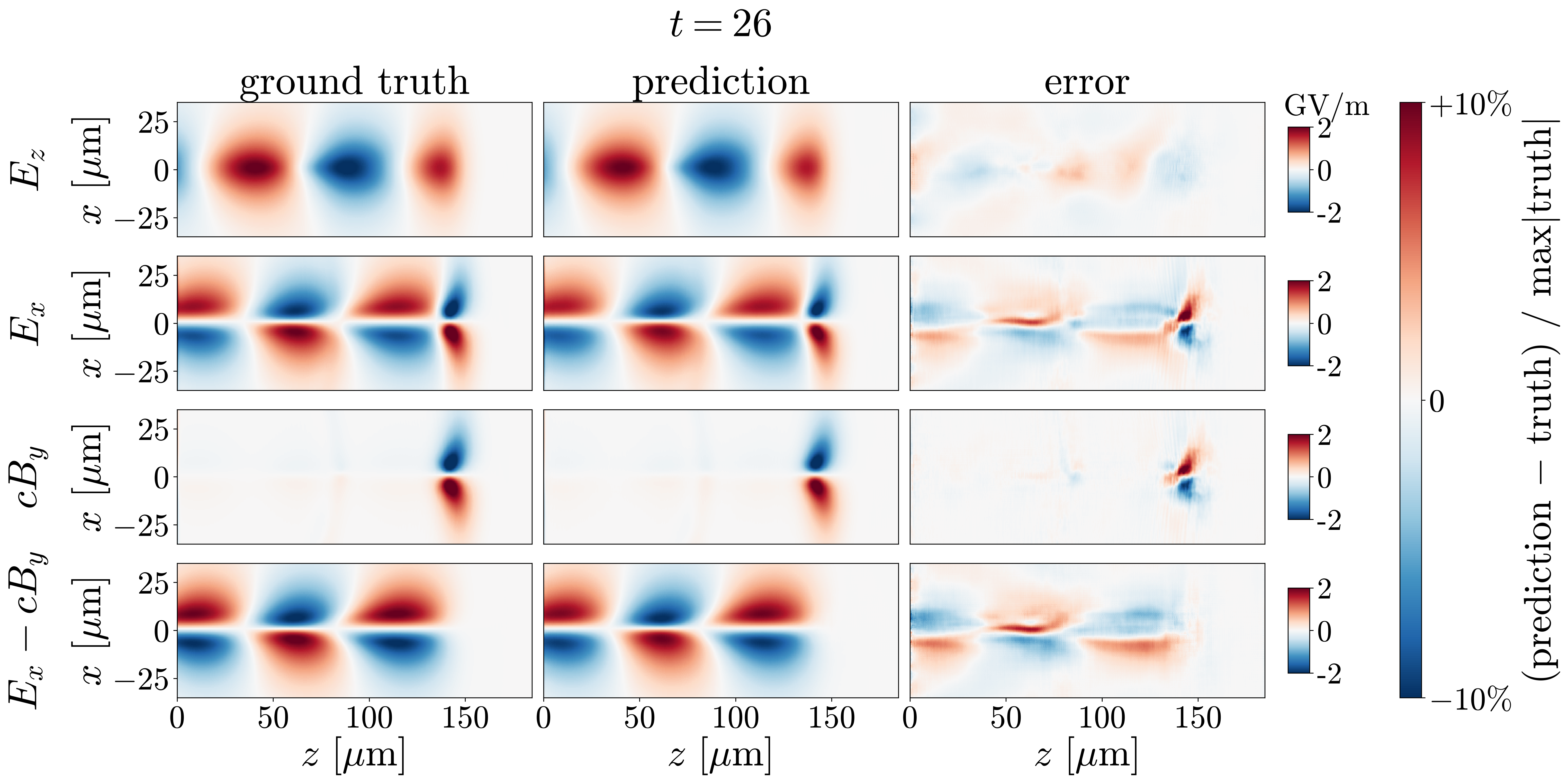}
    \caption{\small PWFA Run 118: detailed comparison at $t=25$ for representative field quantities. Left to right: ground truth, prediction, and error.}
    \label{fig:pwfa_run118_field}
\end{figure*}

\begin{figure*}[t]

    \centering
    \includegraphics[width=\textwidth]{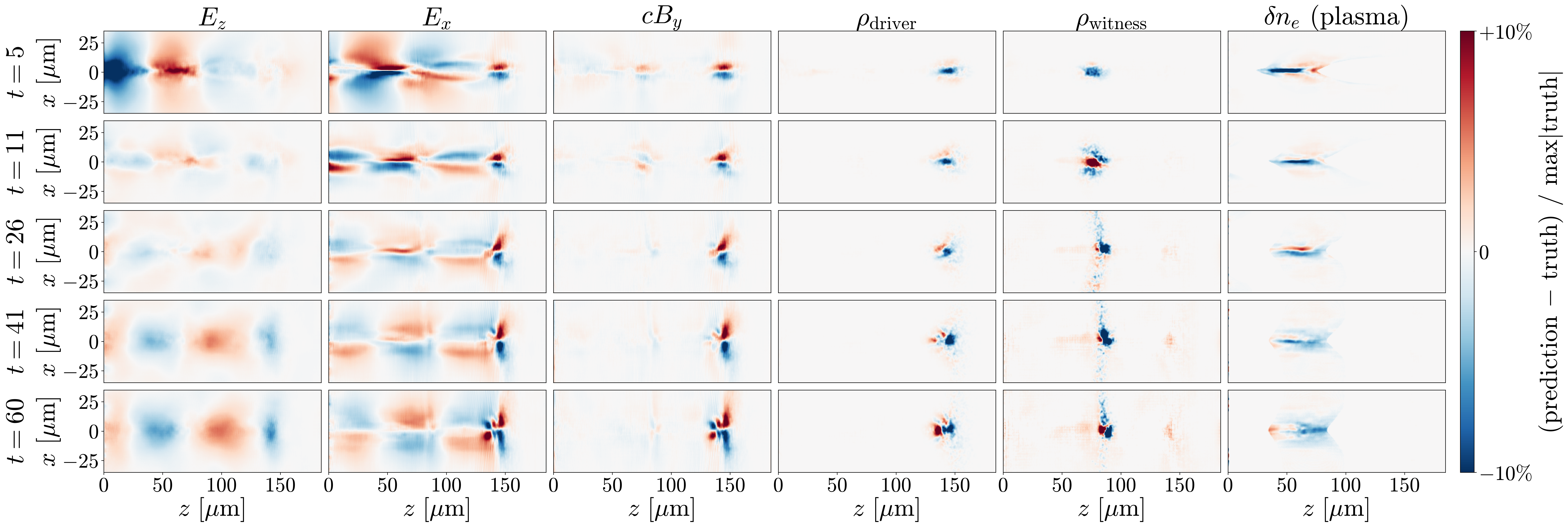}
    \caption{\small PWFA Run 118: relative-error maps for representative channels at selected times. Colors show $(\mathrm{prediction}-\mathrm{truth})/\max|\mathrm{truth}|$.}
    \label{fig:pwfa_run118_err}
\end{figure*}

\paragraph{Run 118: quasi-linear wake case.}
Run 118, visualized in Figures~\ref{fig:pwfa_run118_diag},~\ref{fig:pwfa_run118_field}, and~\ref{fig:pwfa_run118_err}, lies in the quasi-linear regime, with a much smaller normalized driver charge ($\Lambda=0.07$), a $\sim4~\mathrm{GV/m}$-class wake, and the largest relative transverse offset among the selected cases. The macroparticle-and-wake rollout shows that GLOBE follows the weak plasma response and witness evolution through the full horizon despite the absence of a well-developed blowout cavity. At $t=25$, the model reproduces the smoother $E_z$, $E_x$, and $cB_y$ patterns and their derived transverse field $W_x$, demonstrating that the same model captures both acceleration and focusing outside the bubble regime. The multi-time error maps show substantially weaker field and density structures than in the nonlinear cases, with the predicted perturbations remaining aligned with the reference as the wake propagates. The witness carries only $7$~pC and lies at the diagnostic floor, so its transverse-size error is not reported.

\end{document}